\documentclass[useAMS,usenatbib]{mn2e}

\usepackage{txfonts}
\usepackage[dvips]{graphicx}
\usepackage{epsfig}
\usepackage{natbib}

\usepackage{multirow}
\usepackage{url}
\bibpunct{(}{)}{;}{a}{}{,}
\usepackage{color}
\usepackage{listings}

\usepackage{color}

\title[The Evolution of Assembly Bias]%
{The Evolution of Assembly Bias}

\author[Contreras et al.]  {
S.~Contreras$^{1,2,3}$, I.~Zehavi$^{4}$, N.~Padilla$^{2,5}$, C.~M.~Baugh$^{6}$, 
E.~Jim\'enez$^{2}$, I.~Lacerna$^{7,8}$  
\\
 $^{1}$Centro de Estudios de F\'{\i}sica del Cosmos de Arag\'on (CEFCA), 
Plaza San Juan 1, Planta-2, Teruel, E-44001, Spain \\
 $^{2}$Instit\'uto Astrof\'{\i}sica, Pontifica
Universidad Cat\'olica de Chile, Santiago, Chile \\
 $^{3}$Donostia International Physics Center (DIPC),
Manuel Lardizabal pasealekua 4, 20018 Donostia, Basque Country, Spain \\
 $^{4}$Department of Physics, Case Western Reserve University, Cleveland, OH
44106, USA \\  
 $^{5}$Centro de Astro-Ingenier\'{\i}a, Pontificia Universidad Cat\'olica de 
Chile, Santiago, Chile \\
 $^{6}$Institute for Computational Cosmology, Department of Physics, 
Durham University, South Road, Durham, DH1 3LE, UK \\
 $^{7}$Instituto de Astronom\'ia, Universidad Cat\'olica del Norte, Av. Angamos 0610, Antofagasta, Chile\\
 $^{8}$Instituto Milenio de Astrof\'{\i}sica, Av.\ Vicu\~na Mackenna 4860, 
Macul, Santiago, Chile
}

\def\LaTeX{L\kern-.36em\raise.3ex\hbox{a}\kern-.15em
    T\kern-.1667em\lower.7ex\hbox{E}\kern-.125emX}

\begin{document}

\pagerange{\pageref{firstpage}--\pageref{lastpage}} \pubyear{2019}

\maketitle

\label{firstpage}

\begin{abstract}
We examine the evolution of assembly bias using a semi-analytical model of 
galaxy formation implemented in the Millennium-WMAP7 N-body simulation. We consider 
fixed number density galaxy samples ranked by stellar mass or star formation 
rate. We investigate how the clustering of haloes and their galaxy content 
depend on halo  formation time and concentration, and how these relationships 
evolve with redshift. At $z=0$ the dependences of halo clustering on halo 
concentration and formation time are similar. At higher redshift,  
halo assembly bias weakens for haloes selected by age, and reverses and 
increases for haloes selected by concentration, consistent with previous studies. The variation of the halo 
occupation with concentration and formation time is also similar at $z=0$ and 
changes at higher redshifts. Here, the occupancy variation with halo 
age stays mostly constant with redshift but decreases for concentration.
Finally, we look at the evolution of assembly bias reflected in the galaxy
distribution by examining the galaxy correlation functions relative to those 
of shuffled galaxy samples which remove the occupancy variation. 
This correlation functions ratio monotonically decreases with larger redshift and for lower number 
density samples, going below unity in some cases, leading to reduced galaxy clustering.
While the halo occupation functions themselves vary, the assembly bias 
trends are similar whether selecting galaxies by stellar mass or star 
formation rate. Our results provide further insight into the origin and 
evolution of assembly bias. Our extensive occupation function measurements and 
fits are publicly available and can be used to create realistic mock catalogues.
\end{abstract} 

\section{Introduction}
\label{Sec:Intro}
Cosmic structure evolves hierarchically in the cold dark matter model. 
Density fluctuations grow by gravitational instability and form dark matter 
haloes, 
which evolve via accretion and mergers with other haloes \citep{Press:1974}. 
\cite{WhiteRees:1978} formulated the basis of modern galaxy formation theory 
starting from this concept, postulating that galaxies form inside dark matter 
haloes via the cooling of gas, star formation and mergers of galaxies. This 
framework is the basis of semi-analytic models of galaxy formation (SAMs; see, 
e.g., \citealt{Baugh:2006,Benson:2010} for  reviews). These models use the 
merger histories of dark matter haloes as the starting point to model galaxy 
formation. The first SAMs used merger trees constructed using Monte-Carlo 
approaches based on the extended Press-Schechter theory (e.g., 
\citealt{Lacey:1993,Cole:1994,Kauffmann:1993}), while modern SAMs use merger 
trees extracted from high-resolution dark matter simulations (e.g., 
\citealt{Kauffmann:1999,Bower:2006,DeLucia:2007,Lagos:2008,Benson:2012,Jiang:2014,Croton:2016,Lagos:2018,Stevens:2018}). This opens up the prospect of studying environmental 
influences on the formation histories and properties of dark matter haloes 
and the impact on the galaxies they host. 

The framework that led to SAMs also underpins the development of the halo 
occupation distribution (HOD) approach as an empirical description of galaxy 
clustering  (e.g., \citealt{Peacock:2000,Berlind:2002,Cooray:2002,Zheng:2005}).
The HOD formalism characterizes the relationship  between galaxies and dark 
matter haloes in terms of the probability distribution that a halo of virial 
mass $M_h$ contains $N$ galaxies of a given type, together with the spatial and 
velocity distribution of galaxies inside haloes.  An assumed cosmology and
a specified shape of the HOD then allows us to predict any galaxy clustering 
statistic. The HOD approach is a powerful way to interpret observed galaxy
clustering measurements, essentially transforming correlation function
measurements to the relationship connecting  galaxies with haloes (e.g., 
\citealt{Zehavi:2011,Coupon:2012} and references therein).  It is also a useful
method to characterize the predictions of galaxy formation models in a 
concise form that allows us to quantify the galaxy-halo relation 
(e.g., \citealt{Zheng:2005,C13,C17}). 
Another important application of the HOD approach is to facilitate the 
generation of realistic galaxy mock catalogues by populating dark matter 
haloes from an N-body simulation with galaxies that reproduce a particular 
target clustering measurement. This method has become increasingly popular due 
to the growing demand for such catalogues for planning for and interpreting 
the results from large galaxy surveys and due to its good performance and low 
computational cost (e.g., \citealt{Manera:2015,Zheng:2016}). 
In the standard HOD framework mass is the only halo property that plays a role.
This foundation of the HOD method  has its origins in the Press-Schechter 
formalism and the uncorrelated nature of the random walks used to describe halo assembly 
in excursion set theory.  This leads to the prediction that the halo 
environment is correlated with halo mass but not with how the halo is assembled 
\citep{Bond:1991,Lemson:1999,White:1999}. This is, however, not the case for 
haloes in $N$-body simulations  where halo populations of the same mass but 
with a different `secondary property' display different clustering, an effect 
that is now generally termed {\sl (halo) assembly bias}. This was convincingly 
demonstrated in the Millennium N-body simulation of \cite{Springel:2005} by 
\cite{Gao:2005} who showed the age-dependence of the clustering of haloes of 
the same mass (see also \citealt{Sheth:2004}); this dependence of halo 
clustering on secondary properties besides mass  was later extended to, e.g., 
concentration, spin, substructure 
(e.g., \citealt{Wechsler:2006,Gao:2007,Jing:2007,Lacerna:2012,Xu:2017,Villarreal:2017, Mao:2018}). 

\cite{Croton:2007} used a SAM applied to the Millennium Simulation to show 
that halo assembly bias also impacts the clustering of galaxies, an effect 
that is now commonly referred to as {\sl galaxy assembly bias}, namely halo 
assembly bias as reflected in the galaxy distribution (see also  
\citealt{Zhu:2006,Zu:2008,Lacerna:2011,Chaves-Montero:2016}). This can potentially have 
important implications for interpreting galaxy clustering using the HOD 
framework (e.g., \citealt{Zentner:2014}). 
Detecting galaxy assembly bias has proven challenging and controversial. 
Despite some studies which claim to have uncovered the existence of assembly 
bias in the observable Universe (e.g., 
\citealt{Berlind:2006,Yang:2006,Cooper:2010,Wang:2013b, Lacerna:2014,Hearin:2015,Miyatake:2016,Saito:2016}) others argue that the impact of assembly is small 
\citep{Abbas:2006,Blanton:2007,Tinker:2008,Tinker:2011,Lin:2016,Zu:2016a,Dvornik:2017} or that the assembly bias signal could be a result of different systematics (e.g. \citealt{Campbell:2015b,Zu:2016b,Zu:2017,Busch:2017,Sin:2017,Tinker:2017a,Lacerna:2018}). 

This is the latest in a series of papers examining the spatial distribution of 
galaxies predicted by SAMs. \cite{C13} examined the clustering and HOD 
predicted by SAMS from different groups and found that the models give robust 
clustering predictions when the galaxies are selected by properties that scale 
with the halo mass (such as stellar mass). \cite{C15} studied how predicted 
galaxy properties (such as stellar mass, cold gas mass, star formation rate, 
and black hole mass) correlate with their host halo mass in different SAMs. 
\cite{C17} examined how the predicted HOD form evolves with redshift in SAMs. 
We proposed a parametric form for the evolution of the HOD fitting parameters 
that can be used when constructing mock galaxy catalogs or for consistently 
fitting clustering measurements at different epochs. 

Finally, in \cite{Zehavi:2017} (hereafter Z18) we use SAMs to investigate 
how the galaxy content of dark matter haloes is influenced by the large-scale
environment and halo age at $z=0$, 
for galaxy samples selected by their stellar mass, finding distinct variations 
of the halo occupation functions. We show that haloes which form early have 
more massive central galaxies, and thus start hosting them at lower halo mass, 
and fewer satellite galaxies, compared to late-forming haloes. We also find 
similar results in hydrodynamical simulations \citep{Artale:2018}. These 
{\sl occupancy variations}, namely the dependence of the HOD on halo properties
other than mass, are intimately related to assembly bias, as it is their effect
combined with {\sl halo} assembly bias that gives rise to {\sl galaxy} 
assembly bias.

Here, we build on our previous studies and 
investigate the evolution of assembly bias and specifically the occupancy 
variations in SAMs. We extend the analysis of Z18 in a number of ways: 
1) we study a wide range of redshifts between $z=0$ and $z=3$; 
2) we explicitly examine separately the different manifestations of assembly
bias, namely halo assembly bias, occupancy variation, and galaxy assembly bias;
3) we consider galaxy samples constructed using two properties, stellar mass 
and star formation rate (SFR); 
and 4) we select haloes using two  secondary  parameters, halo formation 
time and concentration. 
We use the \cite{Guo:2013a} SAM which is a recent galaxy formation model from 
the Munich group implemented in a Millennium class N-body simulation with a 
WMAP-7 cosmology. 

\cite{Wechsler:2006} and \cite{Gao:2007} study the evolution of halo 
assembly bias in large $N$-body simulations using a mark-correlation statistic
and the large-scale bias of the mass-halo cross-correlation, respectively.
\cite{Hearin:2016} examine the redshift dependence of assembly bias in the
context of an extension of the HOD framework that incorporates assembly bias 
(the so-called decorated HOD), finding that the impact of assembly bias on
galaxy clustering weakens at higher redshift for samples with fixed stellar 
mass.
We aim to comprehensively investigate the evolution of galaxy assembly bias 
using a physical galaxy formation model. We focus here on galaxy assembly bias 
as reflected in the halo occupation and galaxy clustering.  To our knowledge 
this is the first work that explicitly examines the evolution of the occupancy variation, 
and as a consequence, of galaxy assembly bias.
Our aim is to investigate the origin and evolution of assembly bias. This will 
enable the development of more sophisticated tests to search for assembly bias 
in the observable Universe. Our results will also help shape the design of new 
mock galaxy catalogues, which are necessary for the next generation of galaxy 
surveys.

The outline of this paper is as follows: in Section 2 we introduce the SAM 
used and describe the different galaxy and halo samples employed in this work. 
Section 3 shows our results regarding the evolution of halo assembly bias,
while Section 4 presents our main results regarding the evolution of the
occupancy variation.  In Section 5 we study the impact of assembly bias on 
galaxy clustering and the evolution of galaxy assembly bias. Finally, in 
Section 6 we summarise our results and present our conclusions. We describe 
our publicly available occupancy variation measurements and parametric fits 
in the appendix.
Throughout the paper masses are measured in $h^{-1}\, {\rm M_{\odot}}$, the SFR is measured
in $\rm M_{\odot}/yr$ 
and distances are measured in $h^{-1}\, {\rm Mpc}$ and are in comoving units.

\section{Theoretical background and sample definition} 
\label{Sec:GC}

In this section we describe the dark matter simulation and the semi-analytic 
model used in this paper. We also present the different galaxy and halo 
samples we utilize. Finally, we describe the techniques used to characterise 
the galaxy and halo samples.

\subsection{The semi-analytic model}
\label{Sec:Sam}
Semi-analytical modelling (SAMs) is one of the main tools used to 
study galaxy formation (see \citealt{Baugh:2006,Lacey:2016} for reviews). 
These models aim to follow the main physical processes involved in the formation and evolution of galaxies. Some of the processes modelled by the SAM are (i) the collapse and merging of dark matter haloes (ii) shock heating and radiative cooling of gas (iii) star formation (iv) supernovae, AGN, and photoionization feedback (v) chemical enrichment of gas and stars (vi) disc instabilities and (vii) galaxy mergers. 

The SAM used here is that of Guo et al. (2013; hereafter G13). This 
model is a version of {\tt L-GALAXIES} the SAM code developed by the Munich group \citep{DeLucia:2004,Croton:2006,DeLucia:2007,Bertone:2007,Guo:2011,Henriques:2013,Henriques:2015}. For an extended description of this model and its performance we refer the reader to Guo et~al. (2013, see also \citealt{Guo:2016b} and \citealt{C17}). The outputs are publicly available from the Millennium Archive\footnote{\url{http://gavo.mpa-garching.mpg.de/Millennium/}}.
G13 is the latest publicly available SAM of the Munich group that makes use of the Millennium-WMAP7 dark matter simulation.  
We will explore other SAMs in future work, but do not expect our conclusions to change.

\subsection{N-body Simulation}
\label{Sec:Mill}

The G13 model is implemented in the Millennium-WMAP7 N-body simulation \citep{Guo:2013a,Gonzalez:2014, Lacey:2016}. This simulation  has similar specifications to the original Millennium simulation of \cite{Springel:2005} but uses a WMAP7 cosmology (ie. $\Omega_{\rm \Lambda0}$ = 0.728, $\Omega_{\rm m0}$= $ \Omega_{\rm dm0}$+$\Omega_{\rm b0}$ = 0.272, $\Omega_{\rm b0}$ = 0.0455, $\sigma_8$ = 0.81, $n_{\rm s}$ = 0.967, $h$ = 0.704.). 
The simulation uses $2160^3$ particles in a periodic box of comoving volume $(\,500\, h^{-1}\, {\rm~Mpc})^3$  corresponding to a particle mass of $ 9.31\times10^8 h^{-1} \, {\rm M_{ \odot}}$ and a softening value of $5\, h^{-1}\, {\rm~kpc}$. There are 61 simulation snapshots output between $z=50$ and $z=0$.

Halo merger trees are constructed from the simulation outputs. Haloes are identified using the friends-of-friends ({\tt FoF}) group finding algorithm \citep{Davis:1985} at each snapshot of the simulation, using a minimum of 20 particles per halo (equivalent to a mass of $ 1.86\times10^{10} h^{-1} \, {\rm M_{ \odot}}$). {\tt SUBFIND} is then run on these groups to identify subhaloes \citep{Springel:2001}. Merger trees are constructed by linking a subhalo in one snapshot to a single descendant subhalo in the subsequent output, i.e., a subhalo merger tree. The semi-analytical code uses these merger trees as the starting point to build its galaxy catalogue. Here, the mass of a dark matter halo, $\rm M_{h}$, is defined as the mass within the radius where the halo overdensity is 200 times the critical density of the simulation (referred to as ``m\_crit200'' in the public database).

\subsection{The galaxy and halo samples}
\label{Sec:Samples}

\subsubsection{Classifying samples by galaxy properties}

For the main part of our analysis we use samples defined by galaxy number 
density. To do this we rank the model galaxies either by stellar mass or SFR 
and include all galaxies above a particular value of the stellar mass or SFR 
threshold that provides the desired number density. We construct galaxy 
samples for three different number densities, $n = 0.01, 0.00316$ and $ 0.001 
\,h^{3}\, {\rm Mpc}^{-3}$, and for a wide range of redshifts: $z=0, 0.5, 1, 1.5,
2, 2.5$ and $ 3$. The samples are chosen to be evenly spaced in logarithmic 
number density with differences of half a decade in log abundance. 

The cumulative comoving number density of galaxies ranked by stellar mass is 
often used to link galaxy populations across cosmic time (e.g., 
\citealt{Padilla:2010,Leja:2013,Mundy:2015,Torrey:2015,C17}). This type of 
selection is preferred over using a constant stellar mass cut to select 
galaxies at different epochs since it mitigates the need to assume a specific 
evolution model for the stellar mass, is insensitive to systematic shifts in 
the calculation of stellar masses and can be readily applied to observations. 
It also facilitates the comparison with galaxy samples selected using different
properties (here, e.g., with galaxies selected by their SFR). \cite{C13} also 
showed that the HOD predictions for samples defined in this way are robust 
among different SAMs at a  fixed redshift.

\begin{figure}
\includegraphics[width=0.50\textwidth]{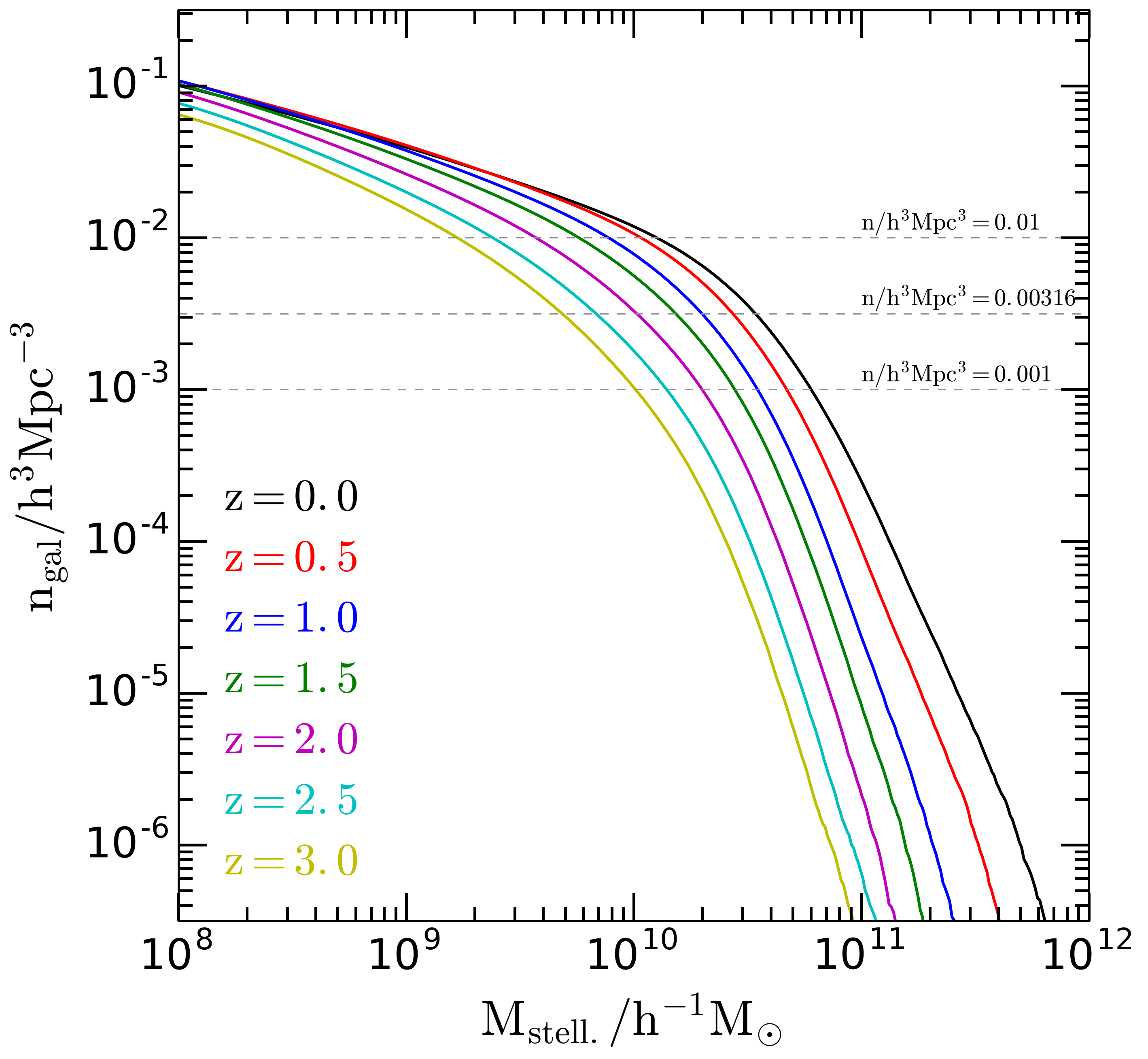}
\includegraphics[width=0.48\textwidth]{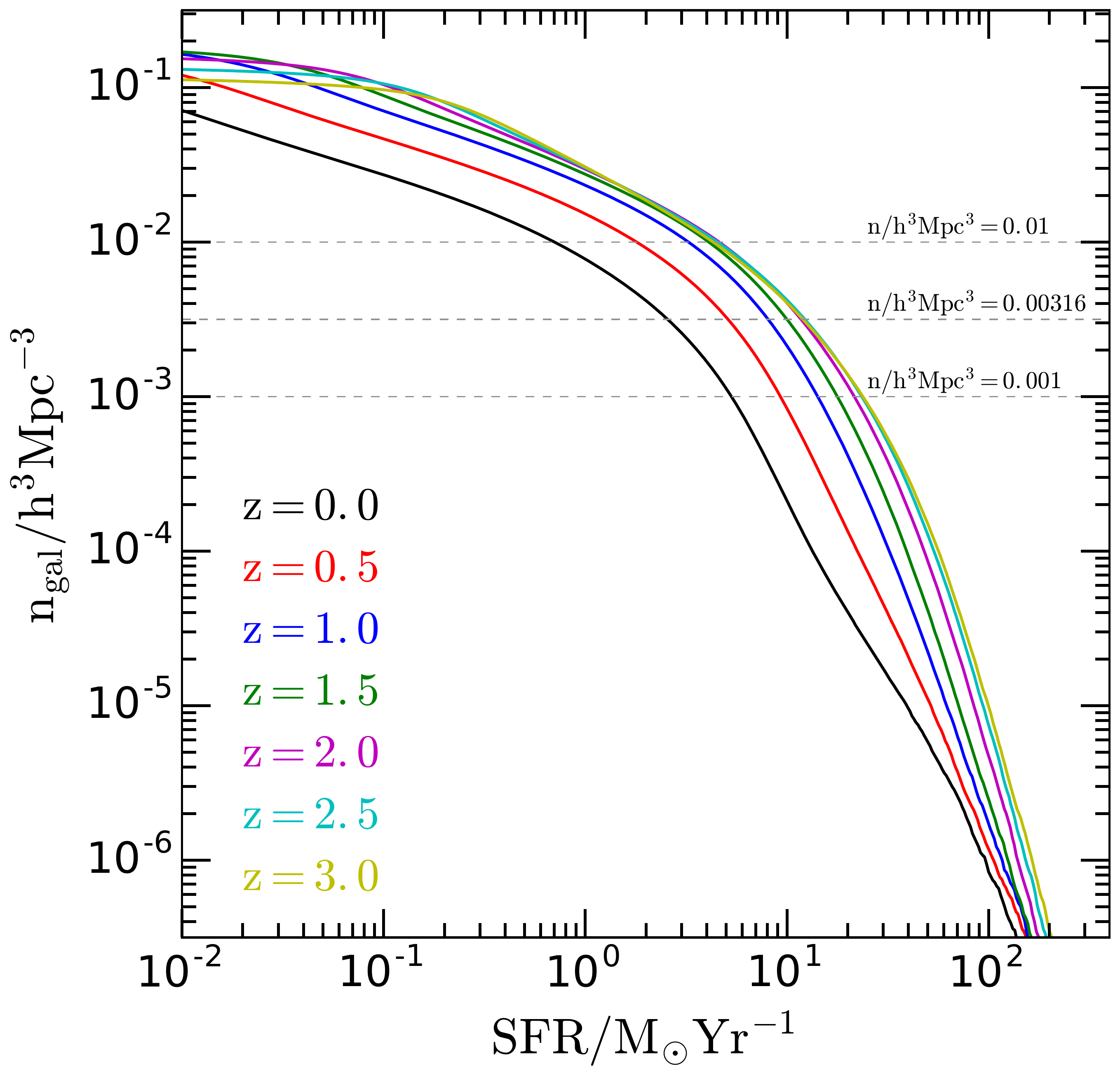}
\caption{The cumulative stellar mass function (top panel) and the cumulative SFR function (bottom panel) predicted by the G13 SAM for different redshifts (as marked). The dashed horizontal lines indicate the number densities of the samples used in this work.}
\label{Fig:AMF}
\end{figure}

Fig.~\ref{Fig:AMF} shows the cumulative stellar mass function (top panel) and 
SFR function (bottom panel) for all redshifts studied here. The horizontal 
dashed lines show  the different number density cuts we consider. The galaxies 
selected in each case are those to the right of the intersection with their 
associated dashed line. The top panel exhibits the expected growth of the 
galaxy stellar mass with time, while the bottom panel shows that there are 
fewer star forming galaxies at low redshifts than at high redshift.  

\subsubsection{Classification by halo properties}

To investigate assembly bias we define subsets of the fixed number density 
galaxy samples by selecting haloes using two different intrinsic or secondary 
properties: formation time (age) and concentration.

We define the formation time of a halo as the redshift when its main 
progenitor reaches half of the halo's present-day mass for the first time. This 
definition is commonly used in the study of assembly bias (e.g., 
\citealt{Gao:2004,Gao:2005,Croton:2007}, Z18).  We note that the formation 
time of a halo is calculated at each redshift independently.
We calculate the formation time using the merger trees available in the 
database and linearly interpolate the mass of the haloes between snapshots.

The other halo property we consider is the concentration. The halo concentration
characterizes the density profile. It is canonically defined as 
$c_{\rm vir} = r_{\rm vir}/r_{\rm s}$, where $r_{\rm vir}$ is the virial radius of 
the halo and $r_{\rm s}$ is the inner ``transitional'' radius appearing in 
the \citet{NFW:1996} profile, at which the density profile changes slope. 
It is often alternatively defined via the rotation curve of the halo, as the 
ratio between $V_{\rm max}$ and $V_{\rm vir}$, where $V_{\rm max}$ is the peak value 
of the rotation curve, $V^2_{\rm c} = G\ M(r)/r$, and $V_{\rm vir}$ the virial 
velocity of the halo, $V_{\rm vir} \equiv V_c(r_{\rm vir})$ 
\citep{Bullock:2001,Gao:2007}. We utilize the latter definition here, which is 
directly calculable from simulation data and does not require any model fitting.

In order to explore the variation in clustering and halo occupation with halo 
age and concentration, following Z18, we rank the haloes by these properties 
and identify the 20 per cent oldest and youngest haloes (based on their 
formation time)  and (separately) the 20 per cent of haloes that are  most or 
least concentrated. These divisions are made in 0.1 dex bins of halo mass, so 
as to factor out the influence of the changing halo mass function on these 
properties; the 20 per cent extremes of the distribution set up in this way 
effectively have the same mass function as the overall sample. We also tested 
using binnings of 0.05 and 0.2 dex in halo mass finding no difference in our 
main results.

\subsection{The HOD and the correlation function}
\label{Sec:HOD_CF}

To study the impact of assembly bias on galaxies we measure the halo 
occupation functions and the correlation functions for the various 
halo and galaxy samples. 

The HOD formalism describes the ``bias'' relation between galaxies and
mass at the level of individual haloes allowing us to characterize the 
galaxy-halo connection. 
The key ingredient is the halo occupation function, $\langle N(M_h) \rangle$, 
which represents, for a given galaxy sample, the average number of galaxies 
per halo as a function of halo mass (loosely referred to here also as the HOD). 
The commonly assumed shape for the halo occupation function is motivated by 
predictions of physical models such as SAMs and hydrodynamic simulations
\citep{Berlind:2003,Zheng:2005}.  When inferring the HOD it is often useful
to consider separately the contribution from central galaxies 
and that of the additional satellite galaxies populating the halo 
\citep{Kravtsov:2004,Zheng:2005}. For stellar mass (or luminosity) threshold 
galaxy samples, the expected form of the central galaxies occupation function 
is a smoothed step function and roughly a power-law for the satellites. 
For samples defined by SFR or color, the shape of the halo occupation
function is more complex to account for the paucity of blue/star forming
centrals in massive haloes (e.g., 
\citealt{Zehavi:2005,Geach:2012,C13,Gonzalez:2018}). We emphasize that the 
HODs presented in this work are all calculated directly from the SAMs, rather 
than inferred from the clustering, as is commonly done with observational data.

The correlation function (CF) is the most fundamental measure of the spatial 
distribution of haloes and galaxies.  It is defined as the excess probability 
of finding a pair of objects at a given separation compared to a random 
distribution. Following \cite{Gao:2007} and Z18, whenever we measure the CF 
for the full galaxy sample we calculate the auto CF (the correlation of a 
given sample of objects with respect to the same sample). In contrast, when 
we measure the CF of a subsample of galaxies (e.g., the ones associated with 
the 20 per cent earliest-formed haloes) we measure the cross CF between this 
sample and the full galaxy sample. As explained in Z18 (see specifically their 
Appendix B) using the cross CF increases the signal-to-noise of the 
measurements and facilitates the interpretation of the results compared 
with the use of the auto CF of the subsamples. 

\section{The evolution of halo assembly bias}
\label{Sec:HAB}

There are two basic ingredients necessary for galaxy assembly bias: (i) halo 
assembly bias, namely the dependence of halo clustering on halo properties 
other than mass, and (ii) the variation in the galaxy content of haloes with 
these properties, which we refer to as the occupancy variation (see Z18).
Galaxy assembly bias requires both effects to be present. In this paper, we 
study how halo assembly bias, the occupancy variation and the resulting galaxy 
assembly bias evolve with time.  This will provide further insight into the
nature and origin of assembly bias and may guide attempts to detect it in 
observational galaxy samples.
We show the evolution of halo assembly bias in this Section. The evolution
of the occupancy variation is discussed in Section~\ref{Sec:OV}, and the 
evolution of galaxy assembly bias is presented in Section~\ref{Sec:GAB}.

\begin{figure*}
\includegraphics[width=0.47\textwidth]{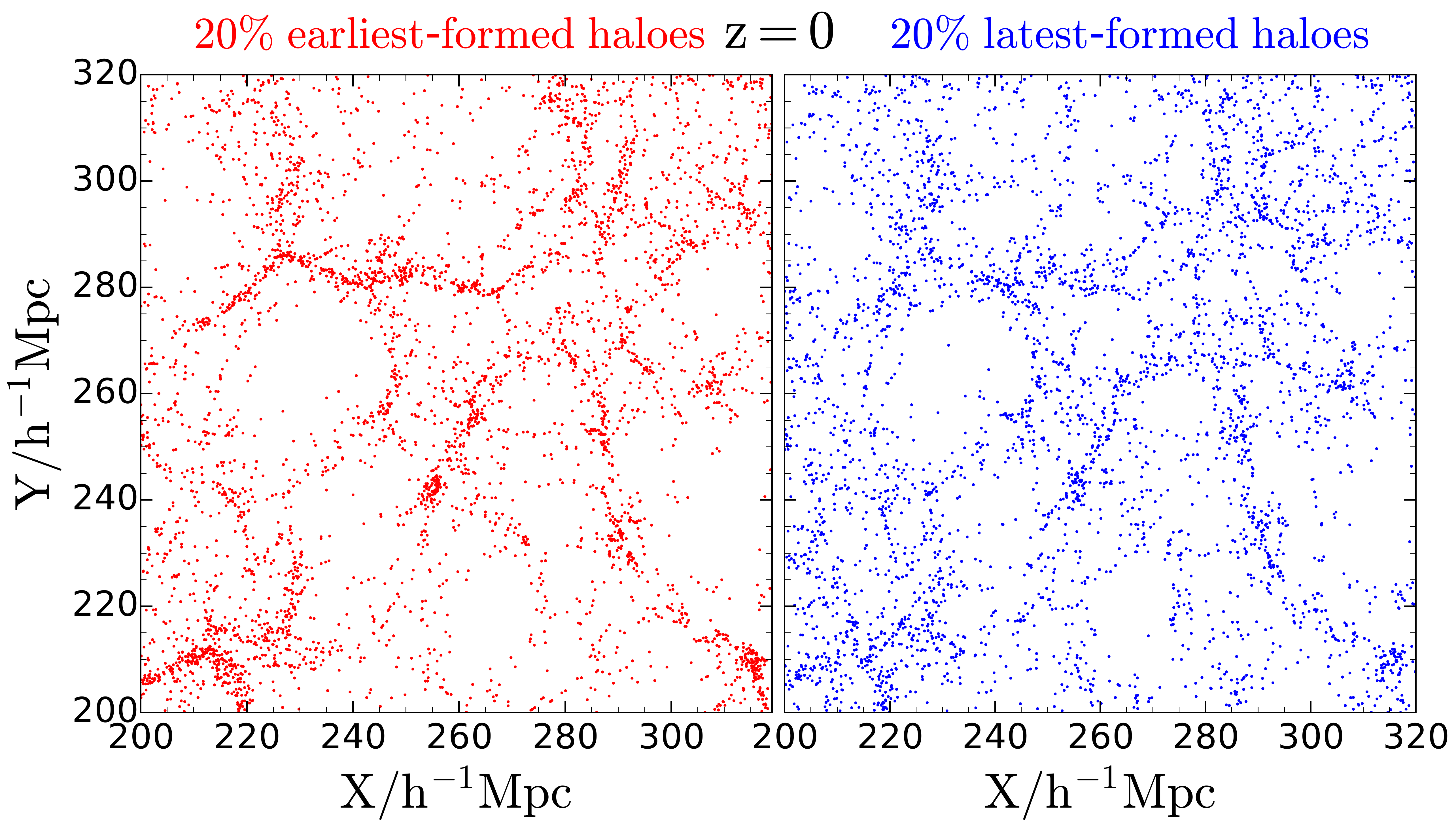}
\includegraphics[width=0.47\textwidth]{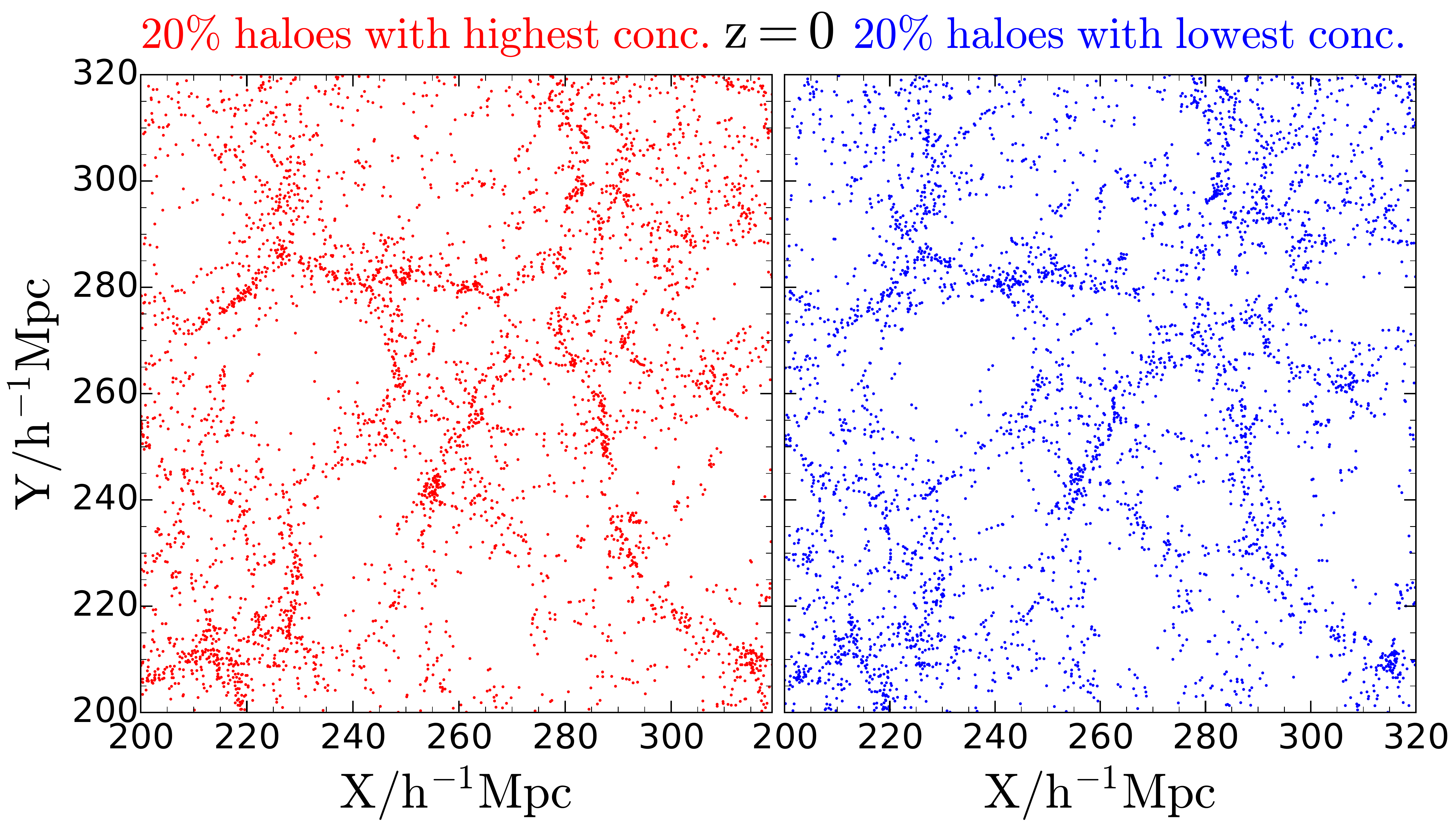}
\includegraphics[width=0.47\textwidth]{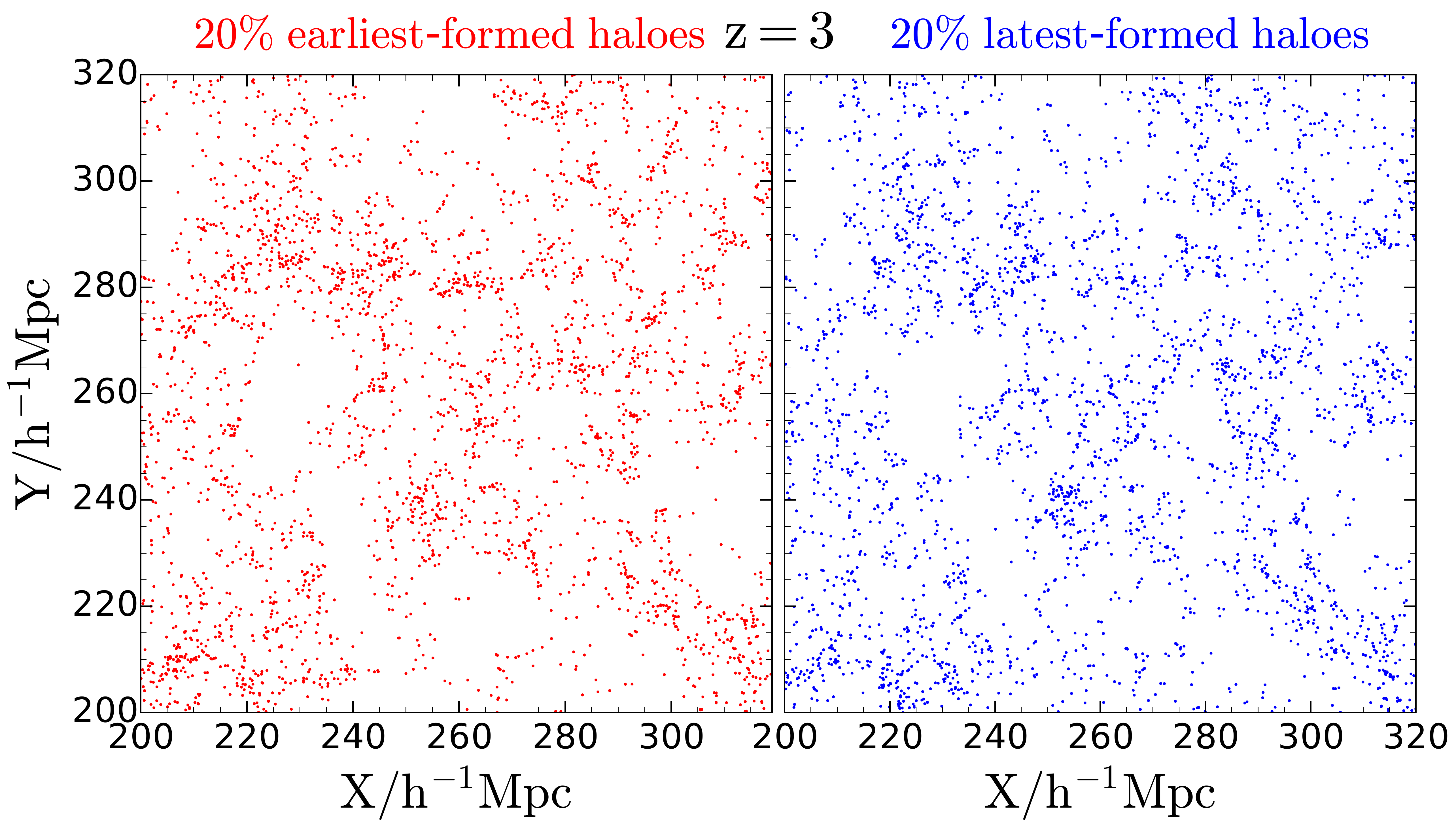}
\includegraphics[width=0.47\textwidth]{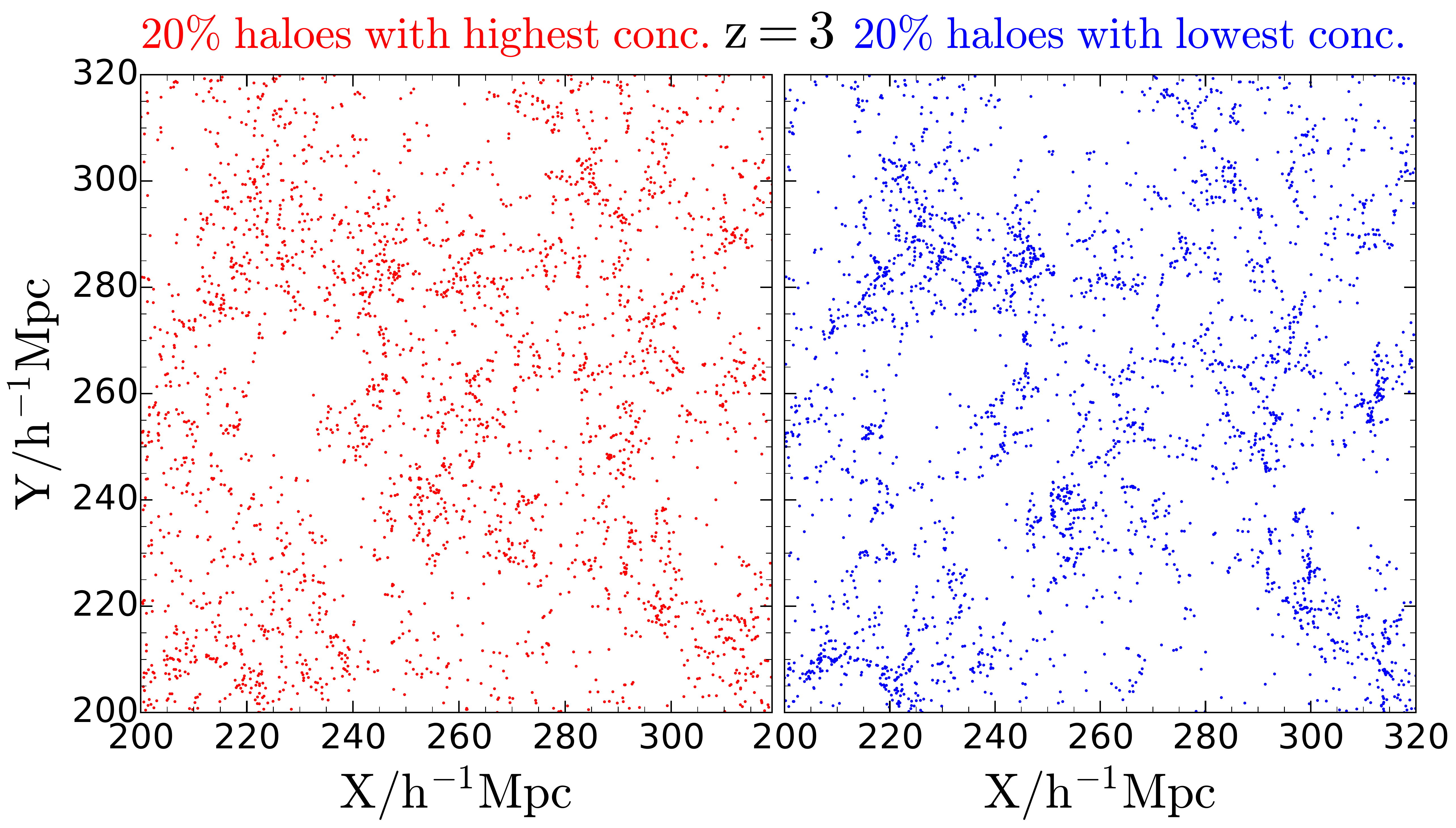}
\caption{Left panels: A slice of $120 x 120 x 10  h^{-1}\, {\rm Mpc}$ from the Millennium-WMAP7 simulation at $z=0$ (top panels) and $z=3$ (bottom panels). The red dots show the 20 per cent of haloes with the highest formation redshifts while the blue dots show the 20 per cent with the lowest formation redshifts from the full halo sample. Right panels: the distribution of haloes using red for the 20 per cent with the highest concentrations and blue for the 20 per cent with the lowest concentrations. 
}
\label{Fig:CW}
\end{figure*}

First we look at the evolution of halo assembly bias in the dark matter-only 
$N$-body simulation without reference to the SAM galaxies. We begin with
a visual inspection of the distribution of haloes in the simulation.
Fig.~\ref{Fig:CW} shows haloes in a slice of the Millennium-WMAP7 simulation 
at $z=0$ and $z=3$,  distinguishing between those with early and late formation
times and also those with high and low concentrations. Starting with the
halo age dependence at $z=0$ (Fig.~\ref{Fig:CW}, top-left double panels), we 
see that while both early-formed and late-formed haloes trace the same cosmic 
web, the early-formed haloes present a sharper view of the web and appear 
somewhat more clustered. The view of the cosmic web when highlighting the extremes of halo concentration
at $z=0$ (Fig.~\ref{Fig:CW}, top-right panels) is reminiscent of that using 
halo formation time, though the clustering differences are slightly less 
apparent in this case. 
The bottom half of Fig.~\ref{Fig:CW} shows the distribution of haloes chosen
similarly, but now at $z=3$.  As expected, the haloes overall appear
less clustered than at $z=0$.  The differences between the early-formed and
late-formed haloes (bottom-left panels) are much smaller in this case, and
interestingly for the concentration (bottom-right panels), it appears that
the haloes with low concentration are in fact now more clustered than the ones
with high concentration.  We quantify all of the trends discussed above shortly
below using the CF.

These results are in agreement with those of  \cite{Gao:2007}, who found that the halo assembly bias signal does not depend on redshift when the halo samples are selected using a fixed cut in peak height ($\nu = \delta_c/\sigma(M_h)$, where  $\sigma(M_h)$ is the root mean square linear overdensity within a sphere with mean mass $M_h$, and $\delta_c(z)$ is the linear overdensity threshold for collapse at redshift z). For an increase in the peak height from $\nu = 0.5$ to $\nu = 1.55$ (the peak height values of the minimum halo mass of our simulation at $z=0$ and $z=3$, respectively), \cite{Gao:2007} found that the difference in the clustering signal between early and late-forming haloes decreases (with  early formed haloes being more clustered at low $\nu$). For haloes selected by their  concentration, they showed that at low peak height, high concentration haloes are more correlated than late forming,  low concentration haloes, and for high peak height, low concentration haloes are more correlated than high concentration haloes. These results are equivalent to the redshift evolution trend found for a fixed halo mass cut. The same can be concluded if a fixed cut in the nonlinear mass for collapse is used \citep{Wechsler:2006}. 

\begin{figure*}
\includegraphics[width=0.47\textwidth]{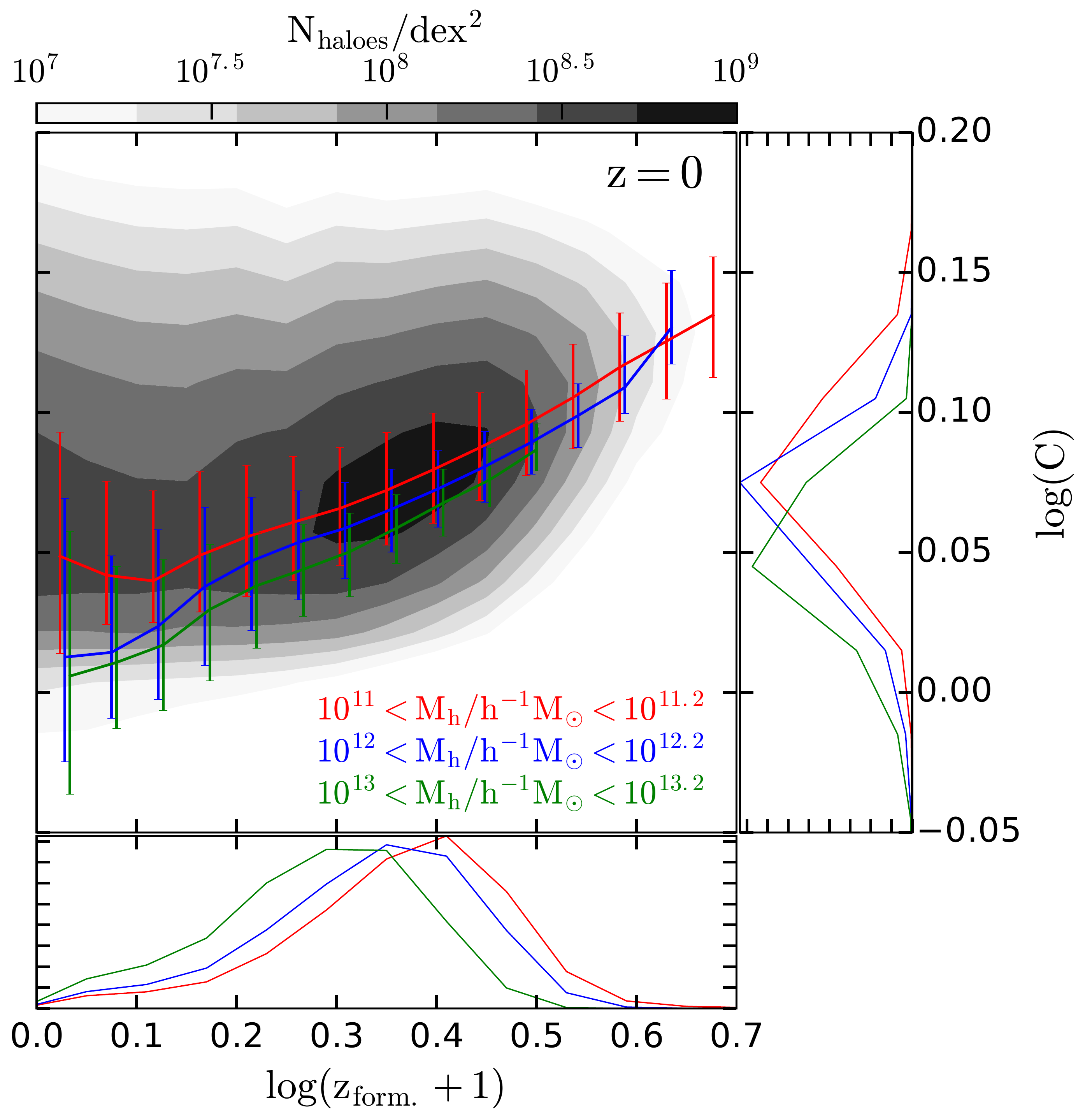}
\includegraphics[width=0.47\textwidth]{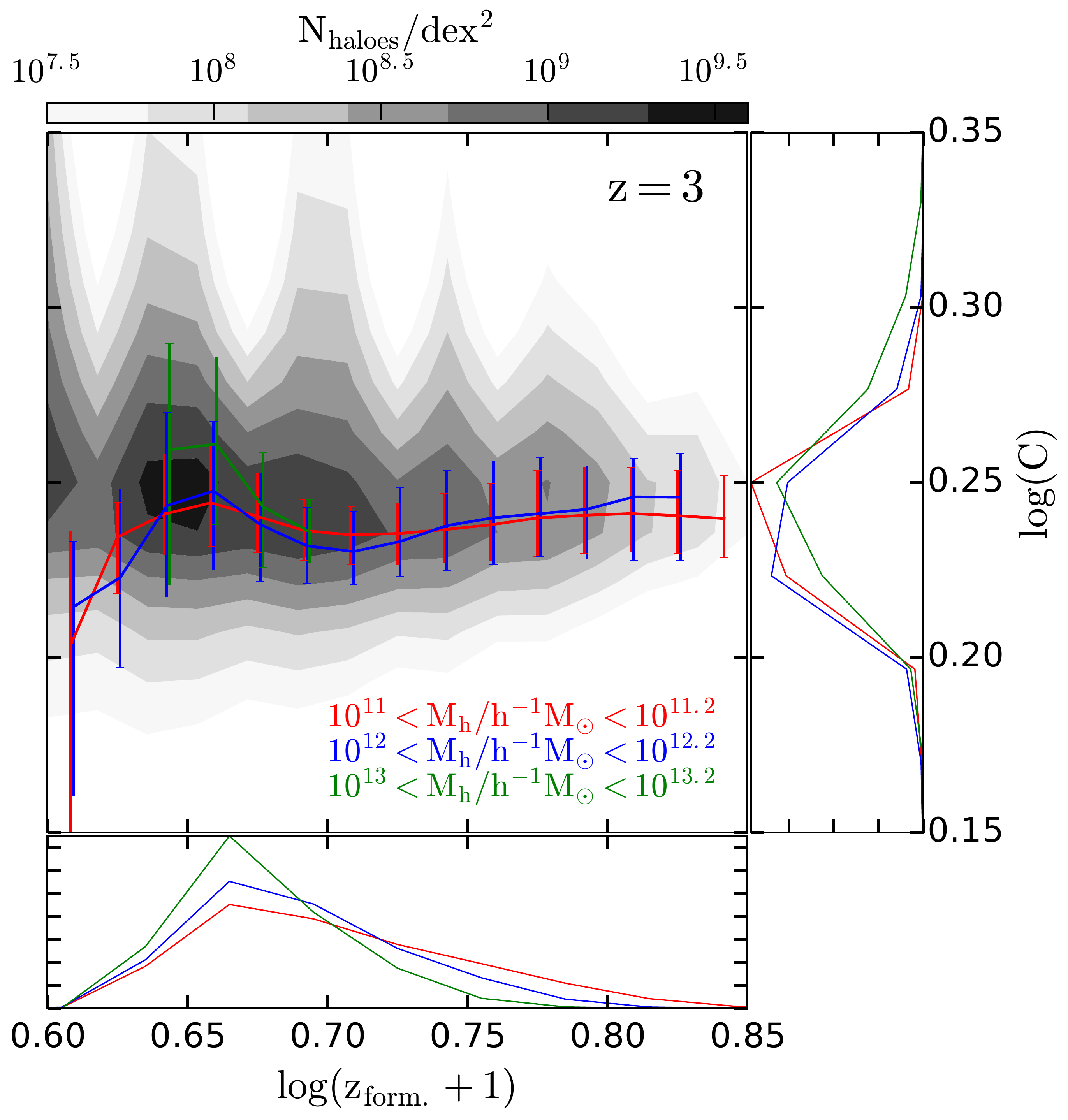}
\caption{Halo concentration plotted as a function of formation redshift for haloes in the Millennium-WMAP7 simulation at $z=0$ (left) and $z=3$ (right). The contours represent the density of haloes as labelled by the key at the top of each panel. The solid red, blue and green lines represent the median concentration as a function of formation redshift for halo masses in the ranges $10^{11} h^{-1} {\rm M_{\odot}}\ -\ 10^{11.2} h^{-1} {\rm M_{\odot}}$ (red line), $10^{12} h^{-1} {\rm M_{ \odot}}\ -\ 10^{12.2} h^{-1} {\rm M_{ \odot}}$ (blue line), and $10^{13} h^{-1} {\rm M_{\odot}}\ -\ 10^{13.2} h^{-1} {\rm M_{ \odot}}$ (green line). The shadowed region is computed using all the haloes in the simulation. Haloes with fewer than 200 particles ($10^{11.26}\ h^{-1}M_{\odot}$, ie. the red line) might underestimate the halo concentration due to resolution effects (see text for more details). The bars indicate the $20$ per cent $-$ 80 per cent range of the distribution. The lower panels in both cases show the distribution of haloes as a function of formation redshift and the side panels show the distribution as a function of concentration.} 
\label{Fig:CP}
\end{figure*}

To better understand how the age and concentration of haloes correlate with 
one another at different redshifts, Fig.~\ref{Fig:CP} shows the joint 
distribution of halo concentration and formation time at $z=0$ (left) and 
$z=3$ (right). We show both the distribution of the full set of haloes 
(contours) and the median concentration as a function of halo age for three 
narrow bins of halo mass (lines and errorbars). The jags in the contours in 
the $z=3$ panel are artificial being  caused by the limited time resolution 
of the Millennium-WMAP7 simulation outputs at high redshift. At $z=0$ there 
is a clear trend of concentration increasing with formation redshift  (as shown by the solid lines). 
On the other hand, at $z=3$ there is little variation of halo concentration 
with formation redshift which suggests that the assembly bias effect with 
concentration and halo age might be different.   
\cite{Xu:2018} also look at the correlation of halo bias with different secondary halo properties (including formation time and concentration). They show that this correlation changes dramatically with halo mass. Therefore, it is unsurprising that the evolution of the clustering signal with formation time and concentration of the halo is different (see also \citealt{Mao:2018} who studied the dependence of clustering on several halo secondary properties).

The concentration of haloes with a small number of  particles (fewer than 200 particles or $M_h < 10^{11.26} h^{-1} {\rm M_{ \odot}}$, ie. the red lines in Fig.~\ref{Fig:CP}) could be underestimated due to resolution effects \citep{Trenti:2010, Paranjape:2017}. Most of the following analysis uses halo masses above this threshold and ranks samples in halo concentration at fixed mass, rather than using its actual value; we expect that our results should be unaffected by this possible source of systematics.

\begin{figure*}
\includegraphics[width=0.33\textwidth]{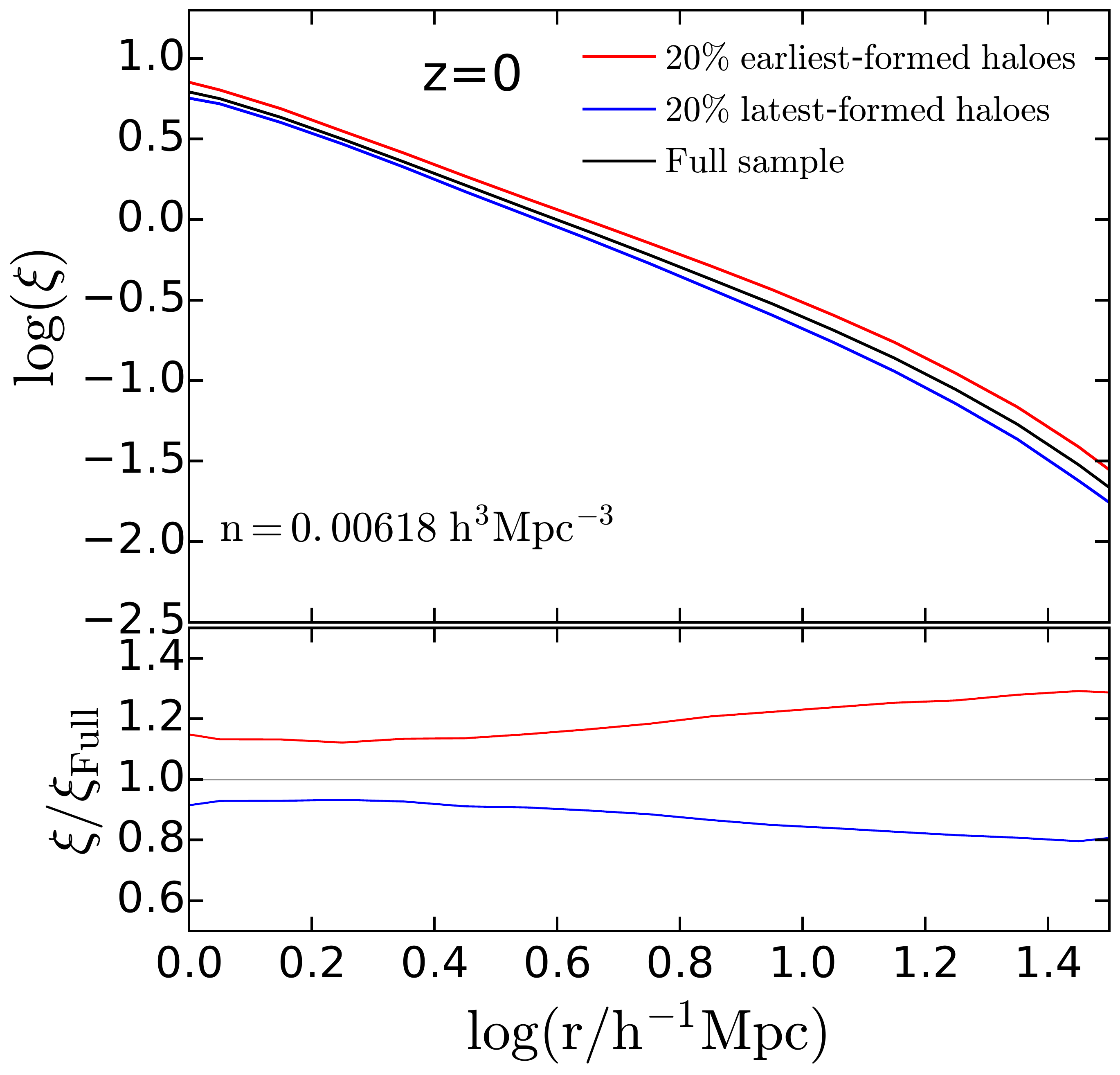}
\includegraphics[width=0.33\textwidth]{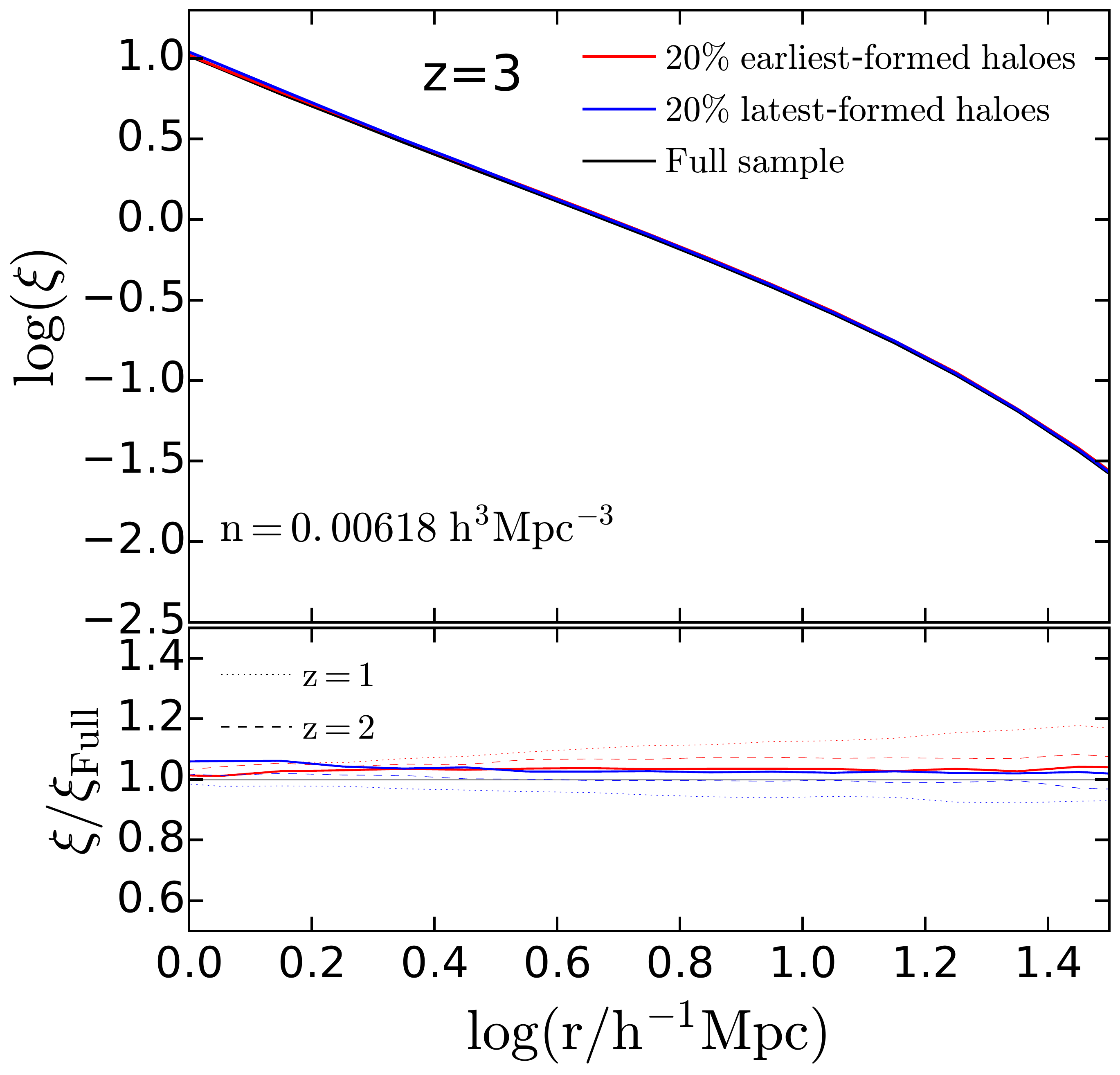}
\includegraphics[width=0.33\textwidth]{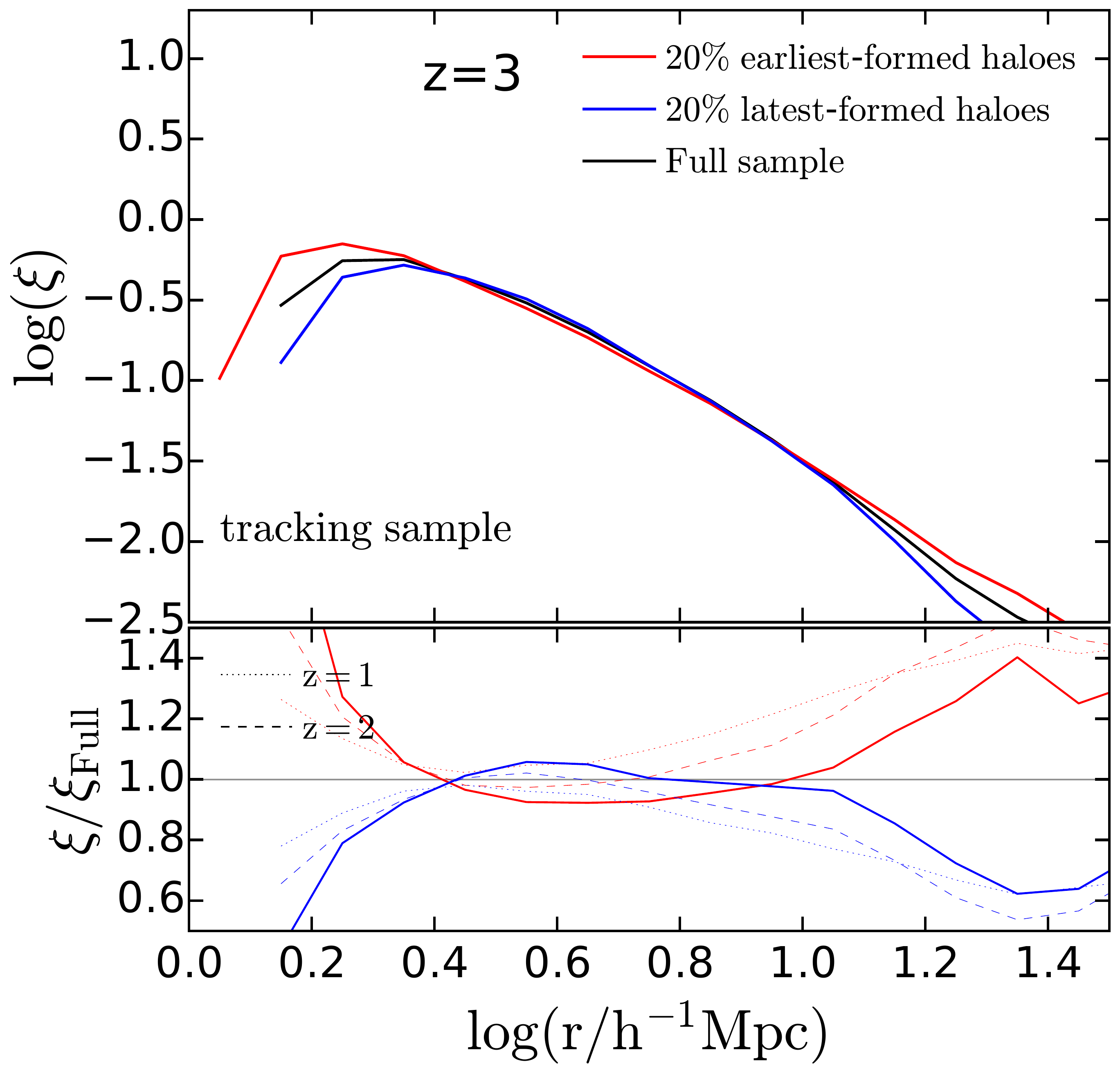}
\includegraphics[width=0.33\textwidth]{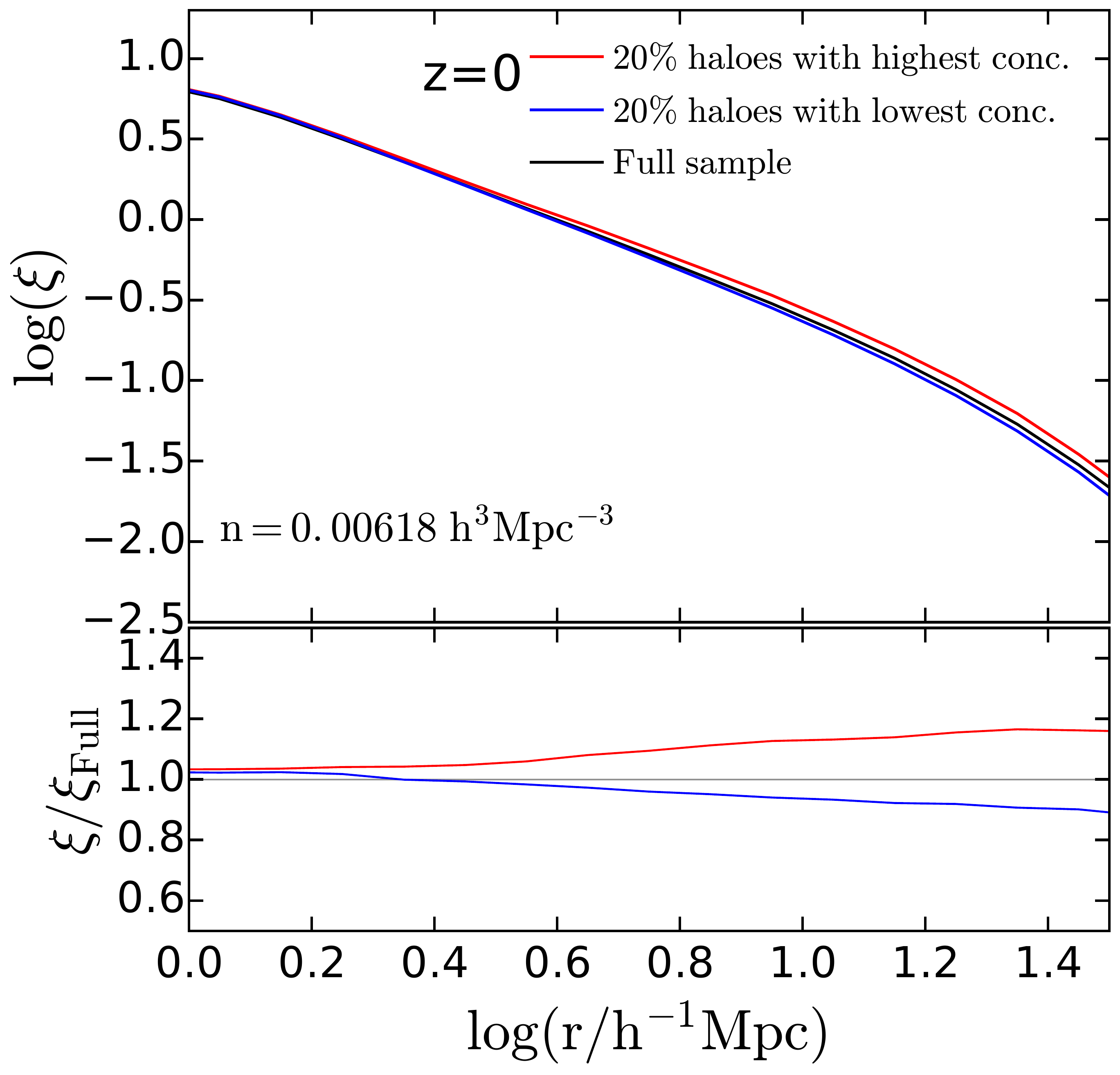}
\includegraphics[width=0.33\textwidth]{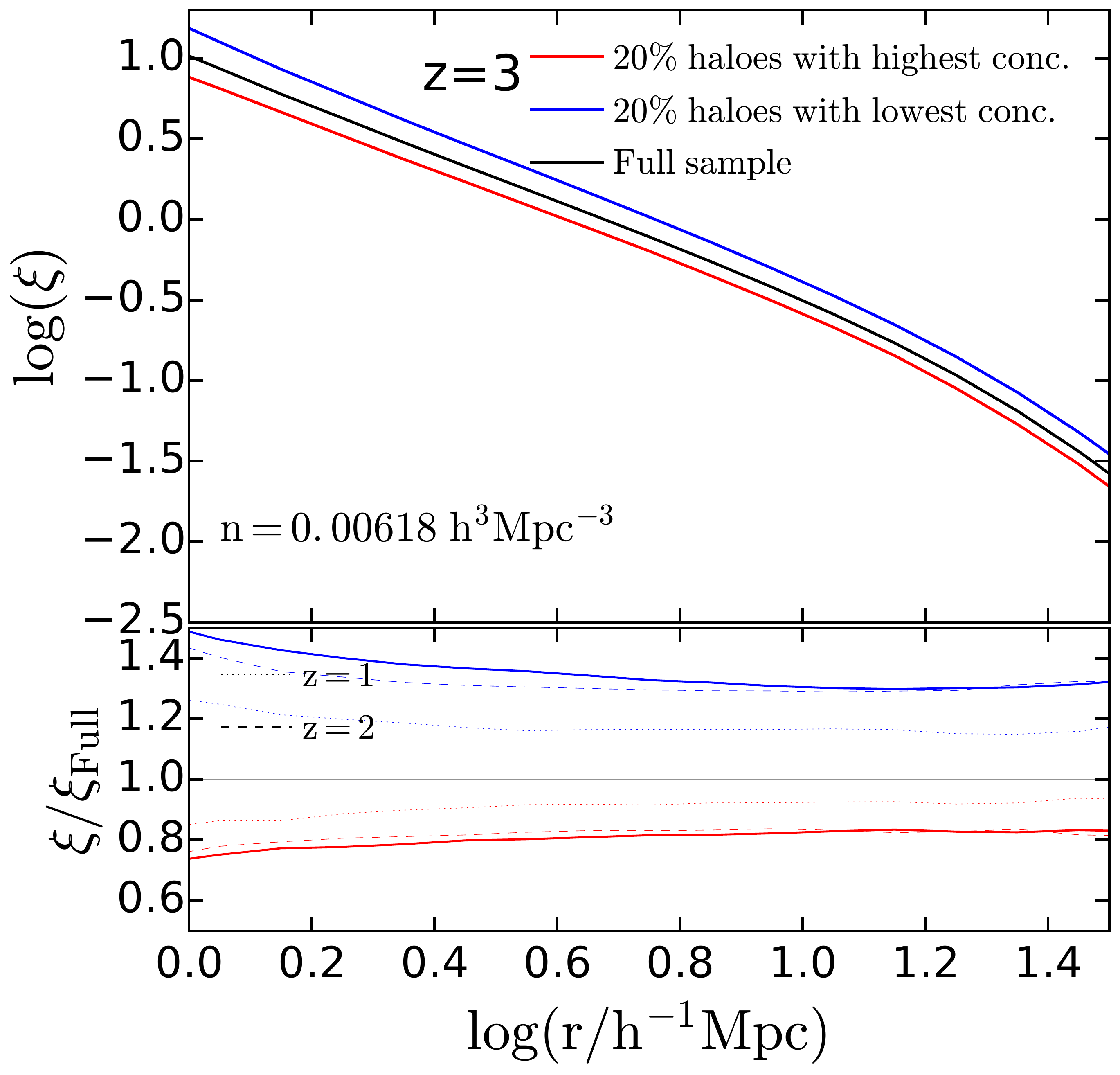}
\includegraphics[width=0.33\textwidth]{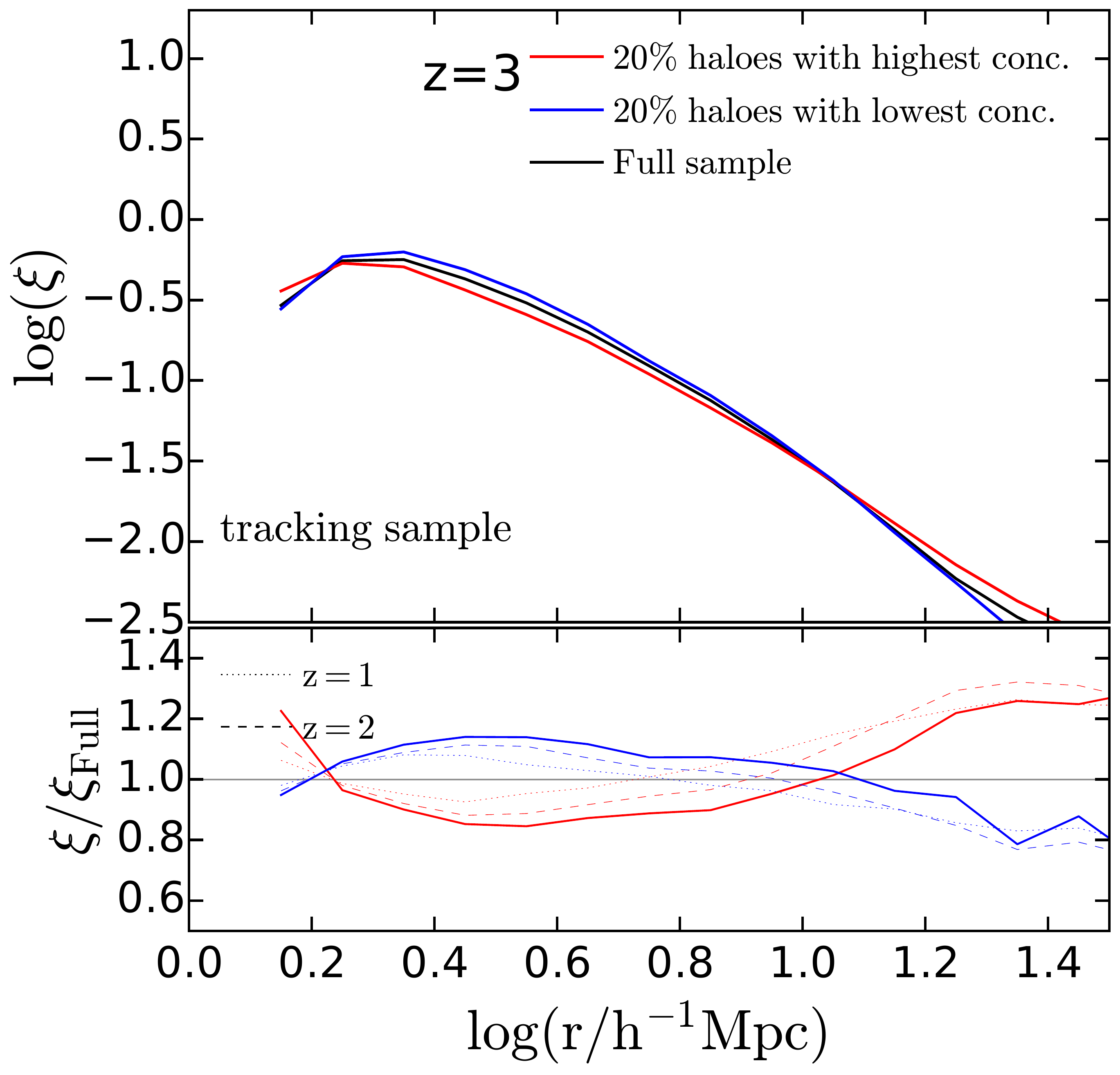}
\caption{(Left) The halo cross CF (coloured lines) and the auto CF (black lines) at $z=0$, for a halo number density of $0.00618\ h^3\, {\rm Mpc}^{-3}$. 
The top and bottom panels show the contribution of haloes with high and low formation redshifts and concentrations, respectively. (Middle) Same as the left panel, but for haloes at $z=1, 2, 3$. 
(Right) The halo cross CF and the auto CF of the main progenitors of the $z=0$ haloes at $z=1, 2$ and $3$. For this plot only, the classification in terms of age and concentration is made using the descendants of these haloes at $z=0$.}  
\label{Fig:CF_Ev}
\end{figure*}

We now look into the CF of halo samples with different concentrations and 
formation times for a fixed number density after rank ordering the haloes 
in decreasing mass at $z=0$ and $z=3$ (Fig.~\ref{Fig:CF_Ev}). 
We use a halo number density of $n = 0.00618 \, h^{3}\, {\rm Mpc}^{-3}$, which 
is comparable to the number density of central galaxies in the 
$n = 0.01 \, h^{3}\, {\rm Mpc}^{-3}$ galaxy sample (when selecting galaxies by 
their stellar mass). The equivalent halo mass (peak height) cut for these 
samples are $10^{11.75}\ h^{-1}M_{\odot}$ (0.76) for $z=0$ and $10^{11.24}\ h^{-1}M_{\odot}$ (2.01) for $z=3$.
In each  subplot of Fig.~\ref{Fig:CF_Ev}, the black line in the top panel denotes 
the auto CF of the full halo sample, while the red (blue) lines correspond to the cross 
CF of the full sample with the 20 per cent oldest (youngest) haloes in the top row of the figure, or the
20 per cent highest (lowest) concentration haloes in the bottom row (see \S~\ref{Sec:HOD_CF}).

We find that at $z=0$, haloes with early formation times and high 
concentrations are more clustered than ones with late formation times and low 
concentrations. This is the well-studied behavior of halo assembly bias (e.g., 
\citealt{Gao:2005,Wechsler:2006,Gao:2007}). The halo assembly bias effect,
as reflected by the clustering differences, is slightly stronger for the case 
of halo formation time than for halo concentration and extends to smaller 
separations. We note that as we are measuring here halo (instead of galaxy) 
clustering, the scales involved are all in the so-called 2-halo regime.
At higher redshift (e.g., $z=3$) 
there is no difference in the clustering measured for haloes at the extremes 
of the formation time distribution and low concentration haloes are more 
clustered than high concentration haloes, reversing the trend seen at the 
present day. 

We reach the same conclusions as already inferred from Fig.~\ref{Fig:CW}, 
namely that the halo assembly bias signal (i.e. the difference between the 
red and blue lines) decreases with increasing redshift 
for halo samples selected by age. 
For concentration, the evolution of halo assembly bias is stronger in the
sense that the clustering differences reverse at high redshift. 
These trends are in agreement with the evolution of the halo assembly bias 
signal found in the original Millennium simulation by \cite{Gao:2007} 
where they found the same trends when the peak height increases from 0.76 to 2.01, that is the increase of peak height from our samples between $z=0$ and $z=3$ (see \citealt{Wechsler:2006}  for a comparison using the nonlinear mass for collapse.) 

To understand the origin of this difference between using age and concentration
as the secondary parameter we show in the right panel of Fig.~\ref{Fig:CF_Ev} 
the CF of the main progenitors of the $z=0$ haloes selected at $z=1, 2$ and $3$.
We call this sample the `tracking sample'. For this sample only, the secondary 
property halo labels (i.e. in terms of the extremes of concentration or 
formation time) refer to the $z=0$ descendants. 
We find different trends for different scales.
At large scales ($>10\ h^{-1}{\rm Mpc}$), the tracking sample shows the same clustering trend as their descendants at $z=0$ but with a higher amplitude. 
The behaviour of the tracking sample cannot be easily related to the \cite{Wechsler:2006} or \cite{Gao:2007} results, since there is no fixed mass cut (or peak height or nonlinear mass for collapse)  in the tracking sample. We interpret this non-evolution in the halo clustering as a negligible change in the comoving position and  abundance of the haloes in this redshift range.
This means that the evolution of halo assembly bias is 
not caused by a change in the clustering of haloes with extreme values of 
formation time or concentration. Instead we attribute the evolution of the 
assembly bias signal at a fixed halo number density to a shift in the ranking 
of haloes according to their mass and secondary property. 
This means that, for example, haloes with the highest concentrations at $z=0$ 
are not necessarily the ones with the highest concentrations at $z=3$. 

We demonstrate this shift in the ranking of the haloes 
in Table~\ref{Table:SampleEv}. 
Here we show that fewer than $40$ per cent of the progenitors 
of $z=0$ haloes were part of the original sample at $z=1$. At $z=3$ this number 
decreases to $\sim 20$ per cent. This shift also explains the different 
evolution of halo samples selected by age and concentration. 
Table~\ref{Table:C_Age} shows that while at $z=0$ there is a $\sim 40$ per 
cent overlap between members of the early (late) formation time halo sample 
and the high (low) concentration halo sample
this number decreases to $\sim 10$ per cent at $z=3$. 

\begin{table}
\begin{center}
 \begin{tabular}{c c c c c} 
 \hline
   & z=0 & z=1 & z=2 & z=3 \\ 
  \hline
  Early forming & 100 & 29 & 23 & 22 \\
  Late forming & 100 & 18 & 18 & 17\\
  High conc. & 100 & 37 & 23 & 19 \\
  Low conc. & 100 & 22 & 14 & 11 \\
 \hline
  \end{tabular}
\end{center}
\caption{The progenitors of a 100 representative haloes from the $n = 0.00618\ h^3\,  {\rm Mpc}^{-3}$ sample at $z=0$, drawn from the early and late formation time, high and low concentration samples, that were part of those samples at $z=0,1,2$ and $3$. } 
\label{Table:SampleEv}
\end{table}

\begin{table}
\begin{center}
 \begin{tabular}{c c c c c} 
 \hline
    & z=0 & z=1 & z=2 & z=3 \\ 
  \hline
  Early forming - high conc. & 40 & 29 & 19 & 10 \\
  Late forming - low conc. & 41 & 27 & 17 & 13 \\
  \hline
  \end{tabular}
\end{center}
\caption{The haloes in common in a representative sample of 100 haloes for a fixed number density of  $n = 0.00618\ h^3\, {\rm Mpc}^{-3}$ in the early formation time  - high concentration samples and in the late formation time -low concentration samples. The numbers are calculated as the ratio between the number of haloes in the interception of the two samples ($\rm A \cap B $) and the number of haloes in the union of both samples ($A \cup B$).} 
\label{Table:C_Age}
\end{table}

Different trends are seen at intermediate and small separations in the right 
panels of Fig.~\ref{Fig:CF_Ev}. The progenitors of early formation time and 
high concentration haloes are more correlated on small scales and less 
correlated on intermediate scales (compared to haloes with late formation 
times and low concentrations). These are not the focus of our work presented
here, and we provide just some heuristic considerations. As early-formed haloes 
grow faster at higher redshifts it may be expected that they exhibit stronger 
clustering on small scales at $z=3$ (since they accrete mass from nearby 
structures). The stronger clustering on intermediate scales for haloes with 
late formation times may be explained in terms of these haloes accreting more 
mass at lower redshifts and the structures that will merge with these 
haloes being in their vicinity but not immediate proximity. 

Since at $z=0$ there is a $\sim 40$  per cent overlap between halo samples 
selected by age and concentration (Table~\ref{Table:C_Age}) we can assume 
this explanation is also valid for the main progenitors of the haloes selected 
by concentration. 

One might be concerned that the agreement between the correlation functions of 
early and late formation time haloes in the top-middle panel of 
Fig.~\ref{Fig:CF_Ev} could be an artifact of the limited time resolution of 
the Millennium-WMAP7 simulation at high redshifts. To check this we also 
calculated the correlation functions using the P-Millennium simulation \citep{Baugh:2019}, a dark matter only simulation with over four times as many 
snaphots as the Millennium-WMAP7 run and with a better mass resolution. We find the same trends as those 
presented in this work, confirming that our results are not a product of the 
finite time resolution of the dark matter simulation used.

\section{The occupancy variation evolution} 
\label{Sec:OV}

In this section we show the evolution of the occupancy variation in the SAM, 
i.e., how the dependence of the HOD on a secondary halo property 
varies with time. This may provide us more insight into the nature and 
origin of this phenomena. 

\begin{figure}
\includegraphics[width=0.49\textwidth]{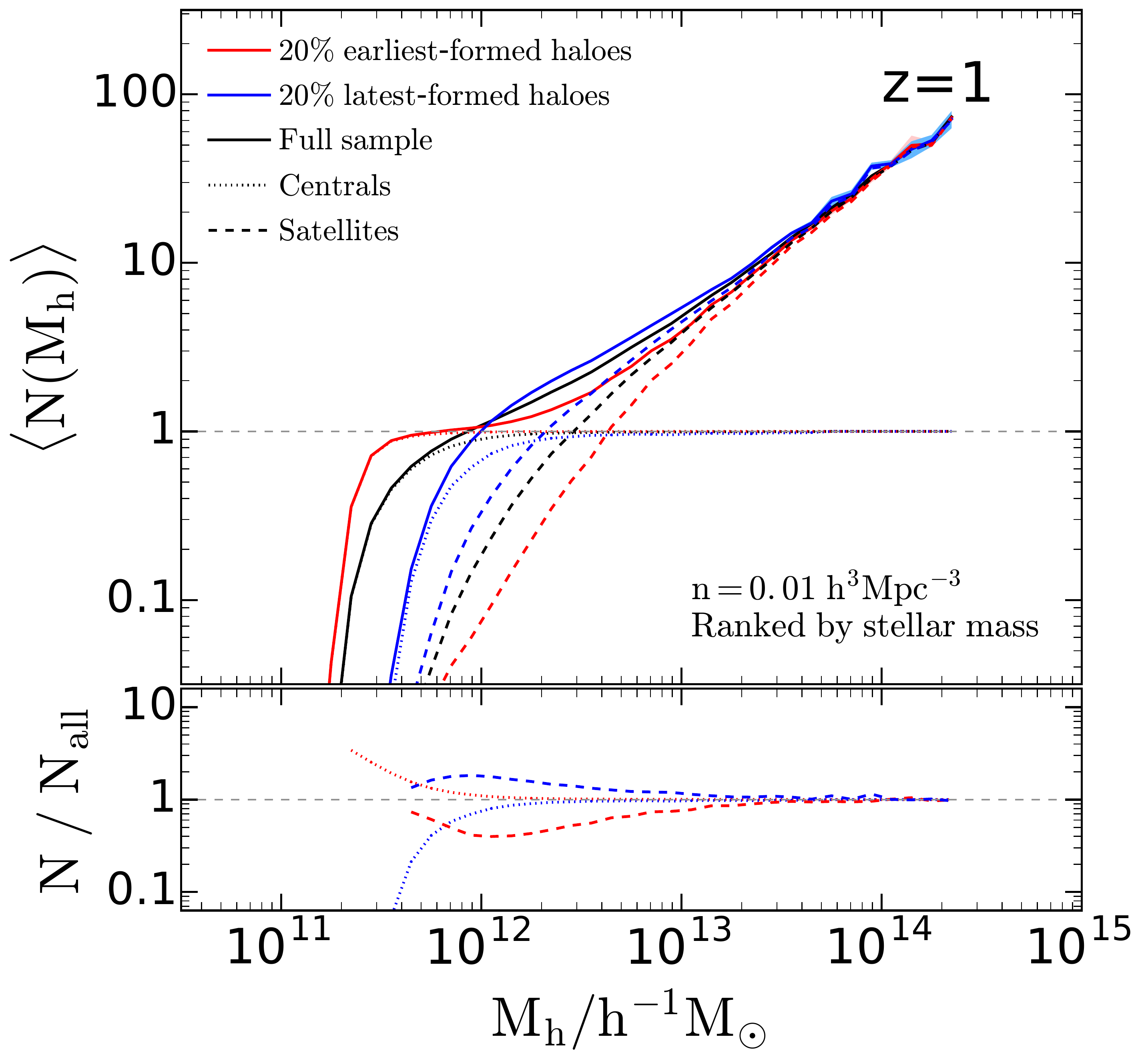}
\caption{(Top) The halo occupation distributions at $z=1$ for a galaxy sample with $n=0.01\ h^3\, {\rm Mpc}^{-3}$. The lines show the contribution of all galaxies (solid), centrals galaxies only (dotted) and satellite galaxies only (dashed). The red and blue lines show the halo occupation distributions for the $20$ per cent earliest and latest forming haloes, respectively. The shaded region represents the jackknife errors using 10 subsamples, and is effectively negligible over most of the range. 
(Bottom) The ratio between the different halo subsets (with the oldest haloes in red and youngest ones in blue) and the full sample HOD for central galaxies (dotted line) and satellites (dashed line). The lines are plotted only when there are at least 20 haloes per halo mass bin of 0.1 dex width.}
\label{Fig:HOD_single}
\end{figure}

Z18 showed that in SAMs, when selecting galaxies at $z=0$ by their stellar 
mass, the predicted HOD depends on halo formation time as well as halo mass. 
They found that haloes with early formation times tend to start being populated
by central galaxies (the main galaxy of a dark matter halo) at lower masses 
than those with late formation times, but they have a lower number of
satellites. \cite{Artale:2018} show that this is also the case in hydrodynamic 
simulations. 
We find that the above results also hold for other redshifts. This is 
shown, for example, in Fig.~\ref{Fig:HOD_single}, where we plot the HOD  at 
$z=1$ for $n = 0.01 \, h^3\, {\rm Mpc}^{-3}$  for galaxies ranked by their 
stellar mass. The occupancy variation for both central and satellite galaxies
is clearly evident.

\begin{figure}
\includegraphics[width=0.24\textwidth]{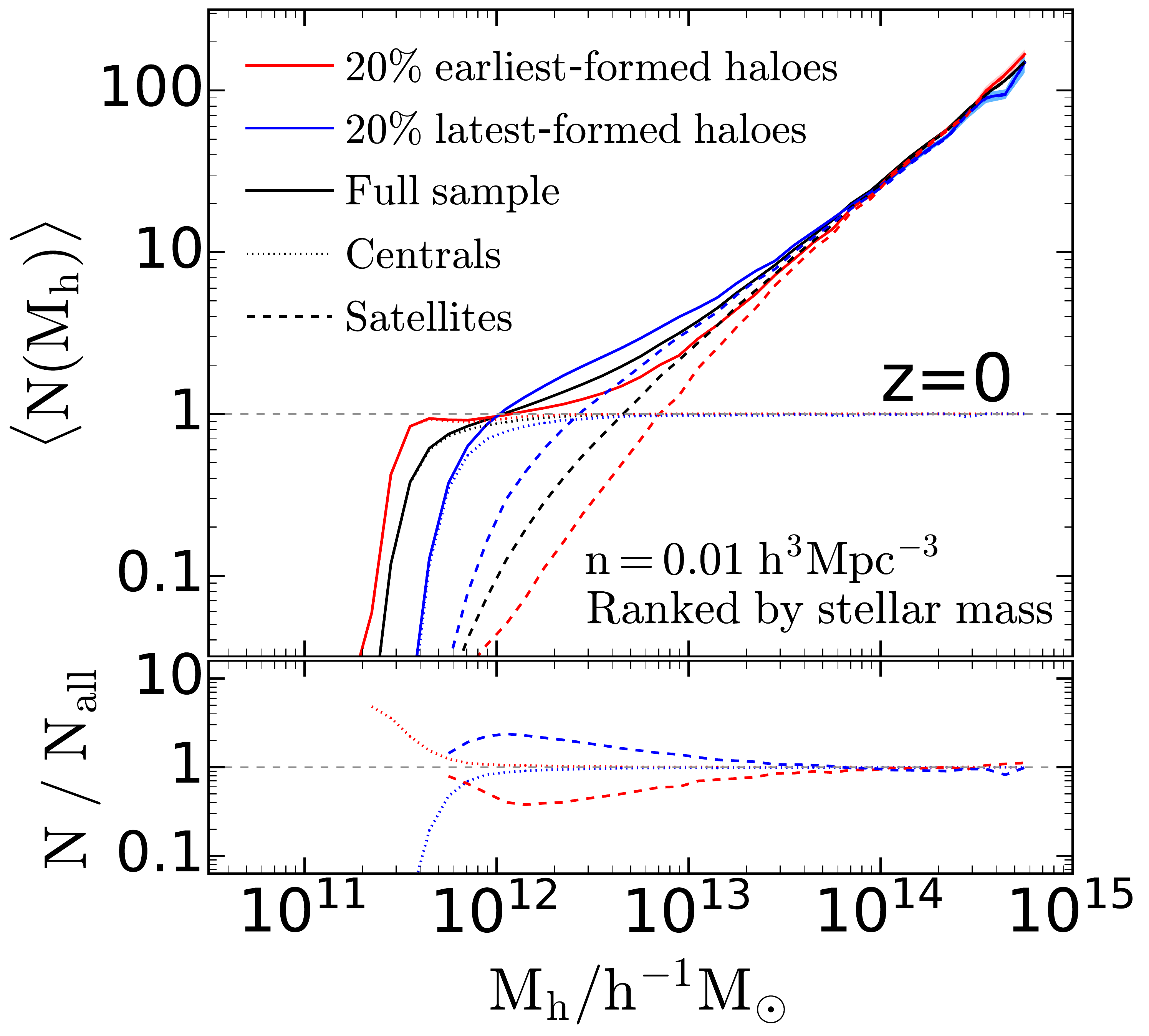}
\includegraphics[width=0.24\textwidth]{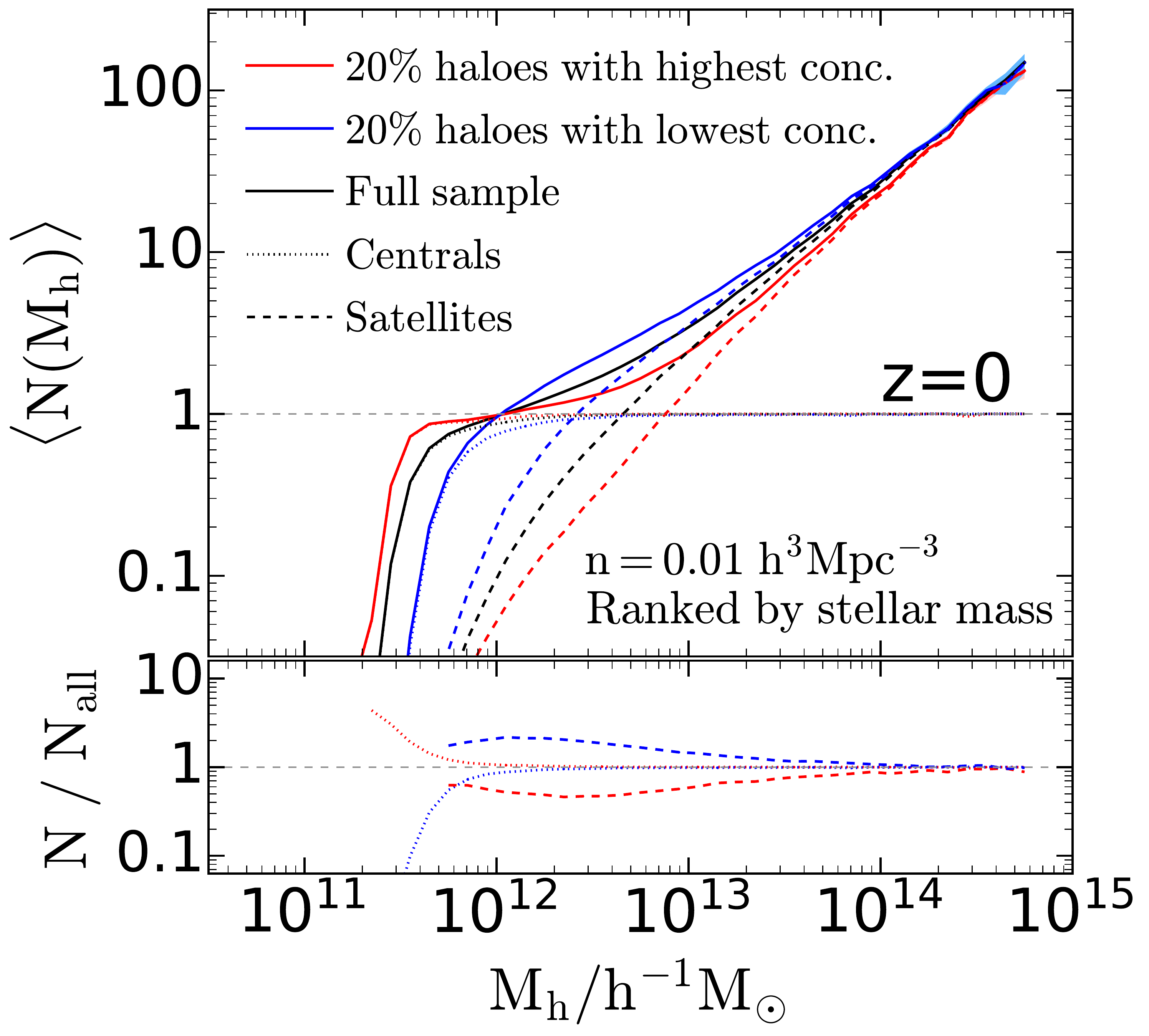}
\includegraphics[width=0.24\textwidth]{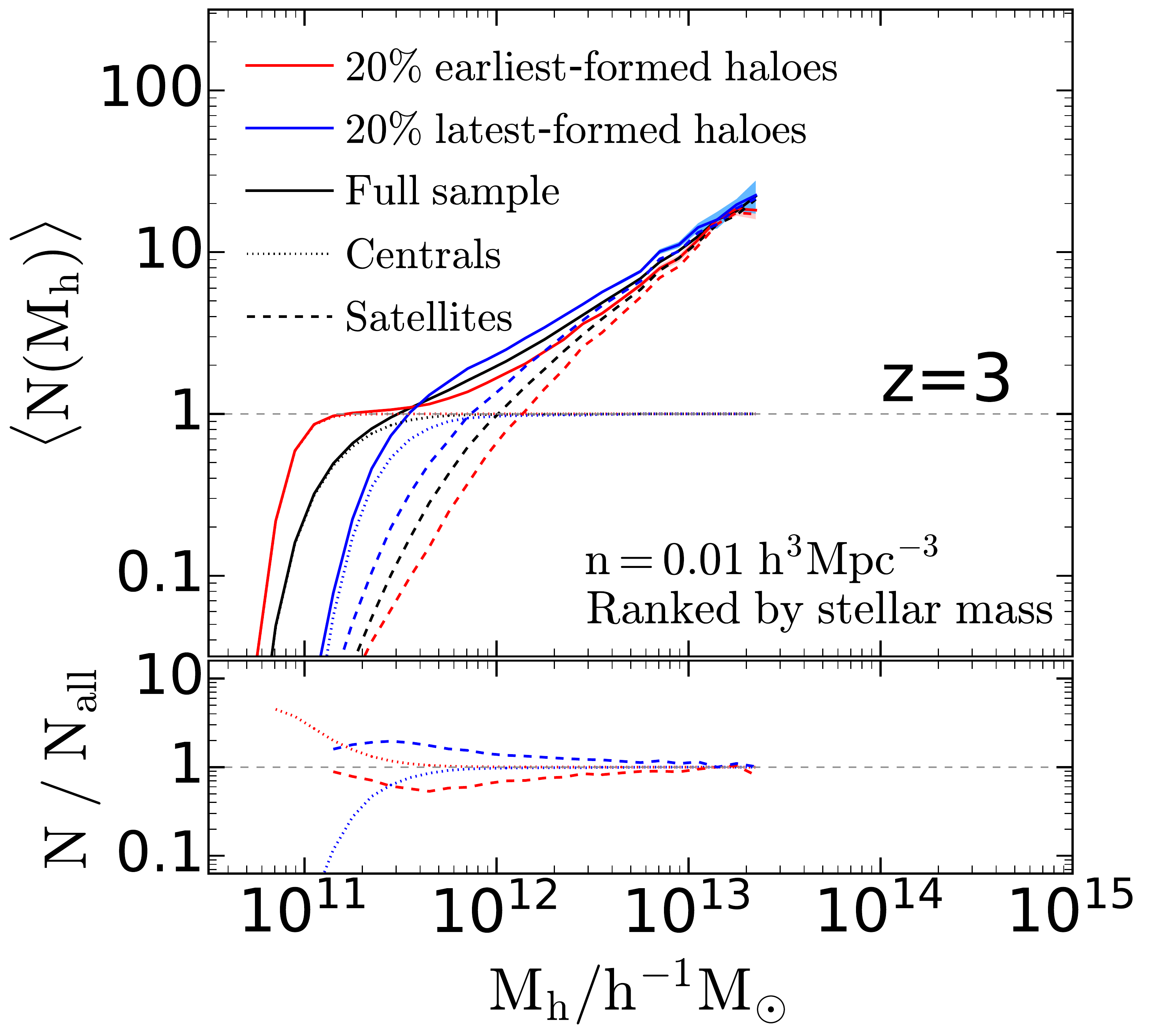}
\includegraphics[width=0.24\textwidth]{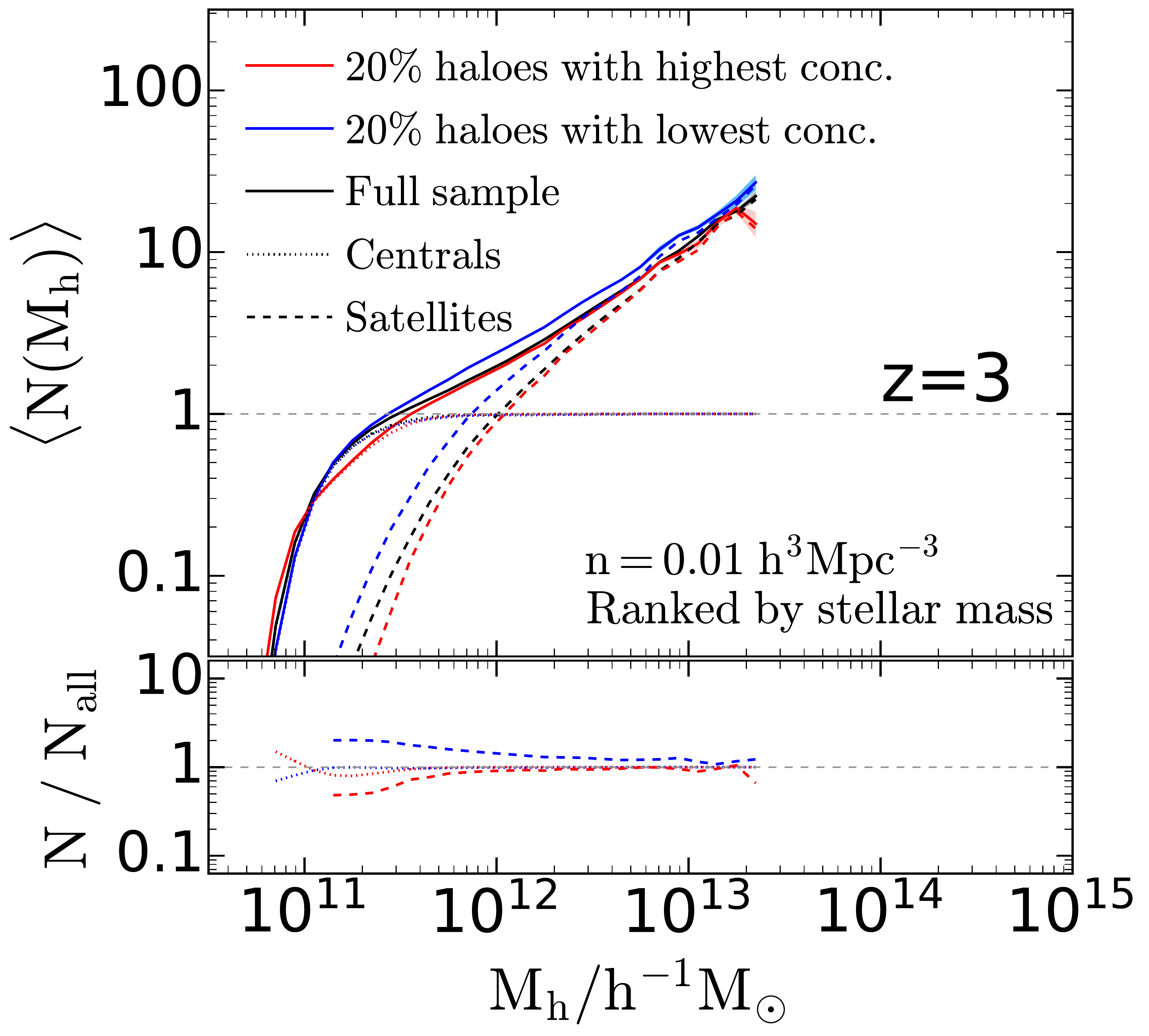}

\caption{(Left) Same as Fig.~\ref{Fig:HOD_single} but for $z=0$ (top) and $z=3$ (bottom). (Right) Same as the left panels, but for haloes selected by concentration instead of formation time.}
\label{Fig:HOD_1}
\end{figure}

Fig.~\ref{Fig:HOD_1}, shows the HOD for $z=0$ (top) and $z=3$ (bottom) for the 
same sample selection. 
In the left panels of Fig.~\ref{Fig:HOD_1} the lines represent the contribution
from the $20$ per cent of haloes with the earliest (red) and latest (blue) 
formation times, while the right panels show the contribution from the haloes 
with the $20$ per cent highest and lowest concentrations. We remind the reader 
that these halo subsamples are constructed by selecting the haloes in narrow 
bins of halo mass.
At $z=0$, the predictions for the high (low) concentration samples are similar
to those with early (late) formation times. This is consistent with what we 
found in Section~\ref{Sec:HAB}, that the behaviour of these samples in terms of clustering is similar at $z=0$,
but now extended to the halo occupation with galaxies. 
This similarity is no longer present at $z=3$. At this redshift, the occupancy 
variation for haloes selected by their age is qualitatively similar to that 
at $z=0$ (and at $z=1$; see Fig.~\ref{Fig:HOD_single}).  For haloes selected 
by concentration, the occupancy variation decreases somewhat for the 
satellite galaxies and it almost disappears for the central galaxies as we go 
to $z=3$. These trends also hold for other number density samples.

\begin{figure}

\includegraphics[width=0.24\textwidth]{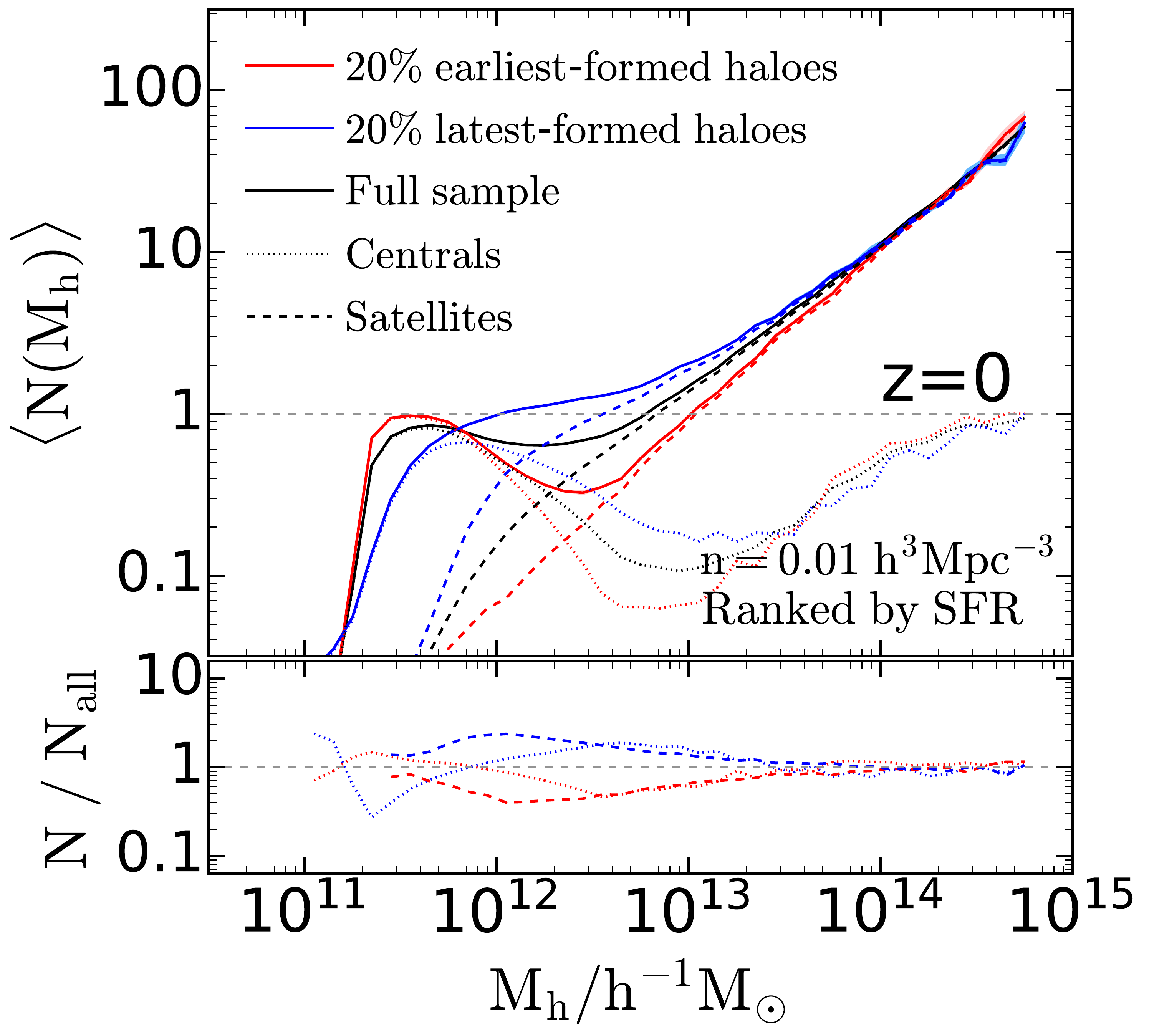}
\includegraphics[width=0.24\textwidth]{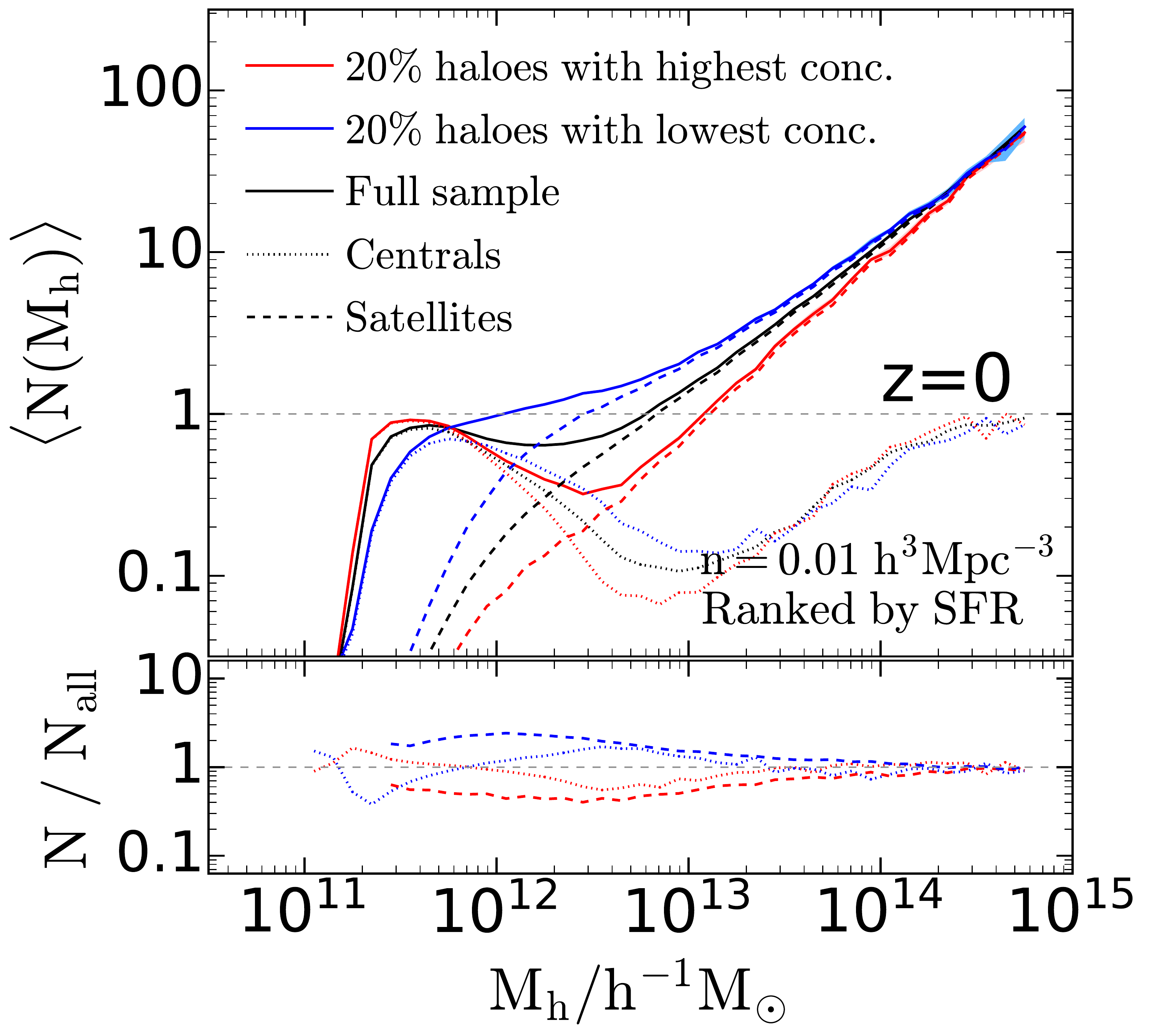}
\includegraphics[width=0.24\textwidth]{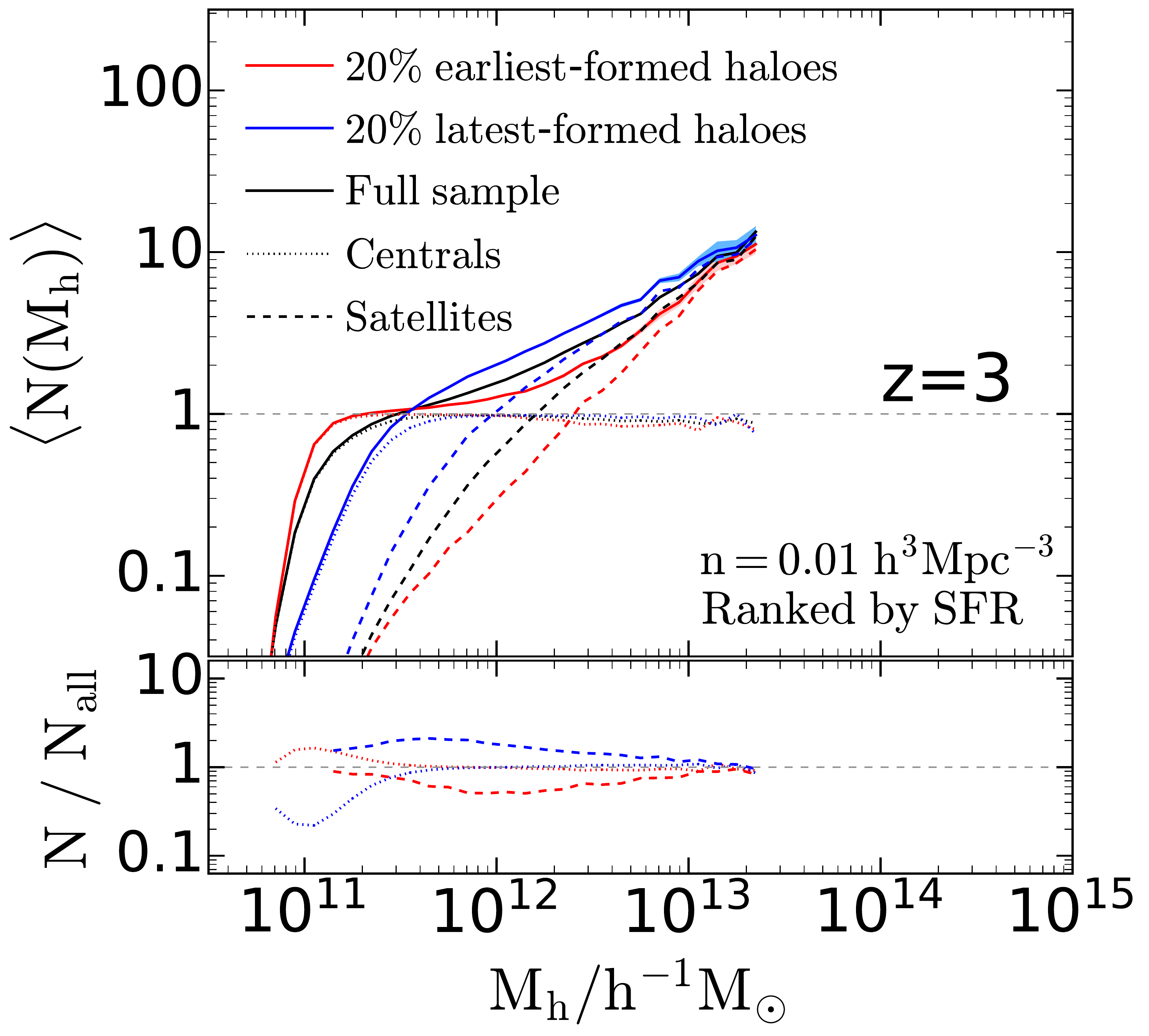}
\includegraphics[width=0.24\textwidth]{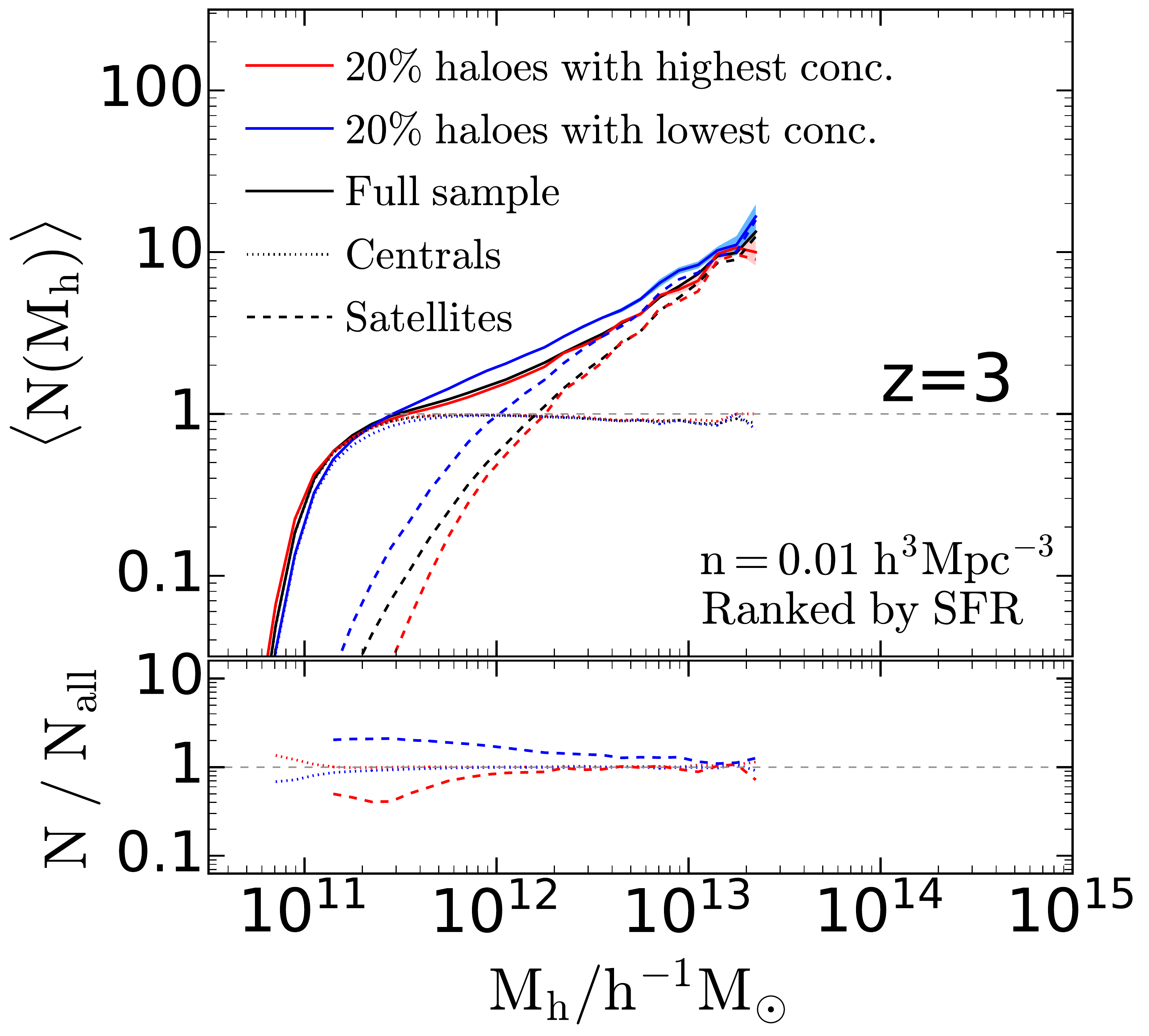}
\caption{Same as Fig.~\ref{Fig:HOD_1} but for galaxies selected by their SFR instead of stellar mass.} 
\label{Fig:HOD_2}
\end{figure}

We repeat the above analysis for galaxies selected by SFR in 
Fig.~\ref{Fig:HOD_2}. The overall shape of the HOD at $z=0$ for SFR-selected 
galaxies is different than for galaxies selected by stellar mass due to the 
tendency of high mass halos to host non star forming (red) central galaxies, as 
discussed in Section~\ref{Sec:HOD_CF}. Interestingly, the ``dip'' feature 
diminishes as one goes toward higher redshifts, possibly due
to having less time for quenching mechanisms to occur.
By $z=3$ the overall shape of the HOD, and in particular the central galaxies 
contribution, is very similar for the SFR-selected galaxy samples and the 
stellar mass selected samples (see also \citealt{Orsi:2008}).
We have verified that the transition in the shape of the HOD between $z=0$ and $z=3$ is smooth with increasing
redshift. A large set of HOD measurements for different 
redshifts and number densities is being released with this paper (see Appendix~A 
for more details).

At $z=0$, for the SFR-selected galaxy samples, early forming and high 
concentration haloes have a lower number of satellite galaxies compared to 
haloes with late formation times or low concentrations (same as for galaxies 
selected by stellar mass). For the central galaxies, at low halo masses, 
early forming and high concentration haloes have a larger number of central 
galaxies, while for higher halo masses they have a lower number of central 
galaxies compared to haloes with late formation times or low concentrations. 
The latter trend perhaps arises since the central galaxies in the 
early-formed high-mass halos have more time to be impacted by star formation 
quenching.
At $z=3$, the HODs for galaxies selected by SFR display the same trends as 
those for galaxies selected by stellar mass. The occupancy variation (i.e. 
the difference between the red and blue lines) stays roughly constant for 
haloes selected by age. The occupancy variation with halo concentration 
decreases with redshift for the HOD of the satellites and nearly diminishes for that of the
central galaxies.

\begin{figure*}
\includegraphics[width=0.45\textwidth]{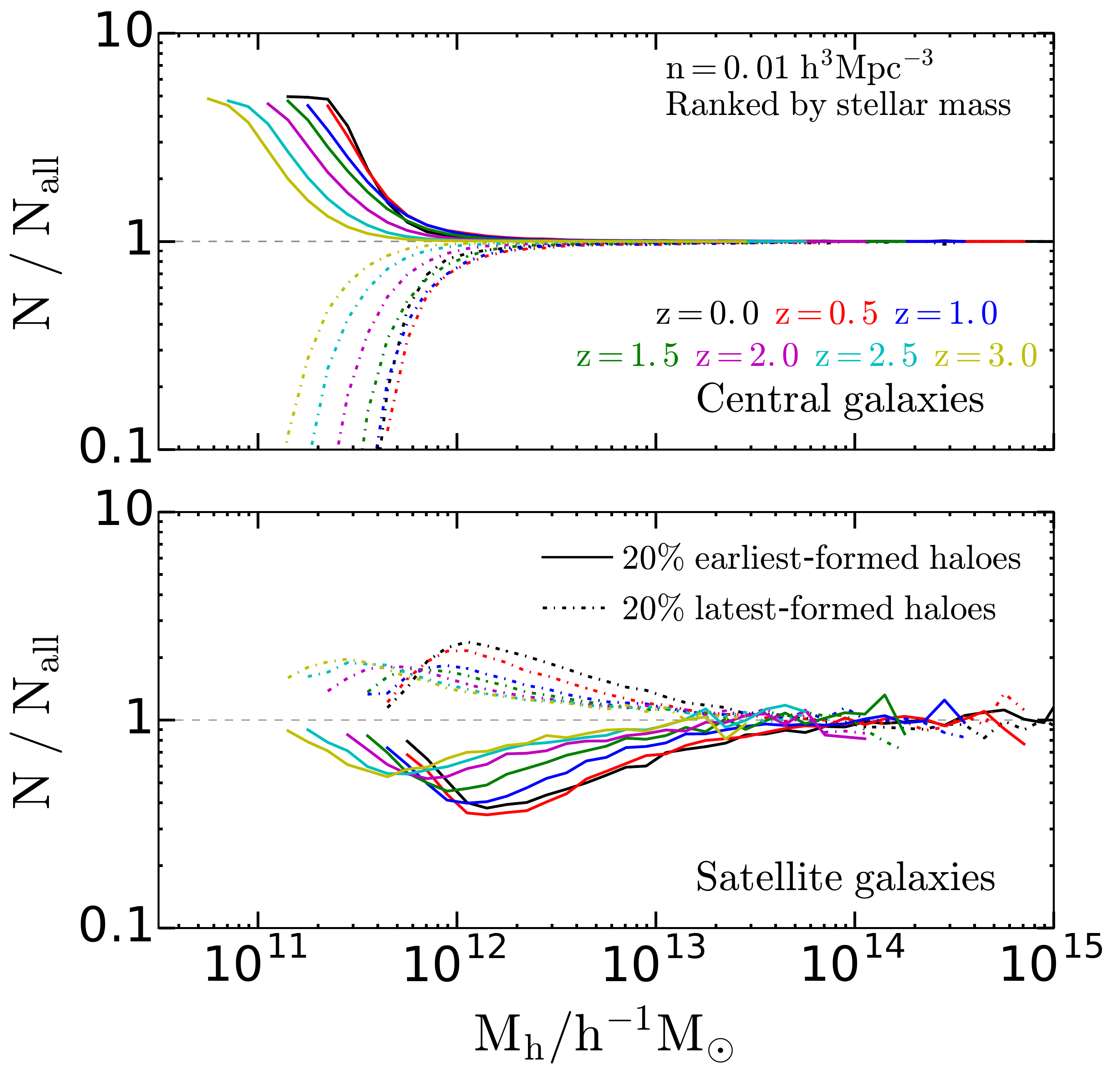}
\includegraphics[width=0.45\textwidth]{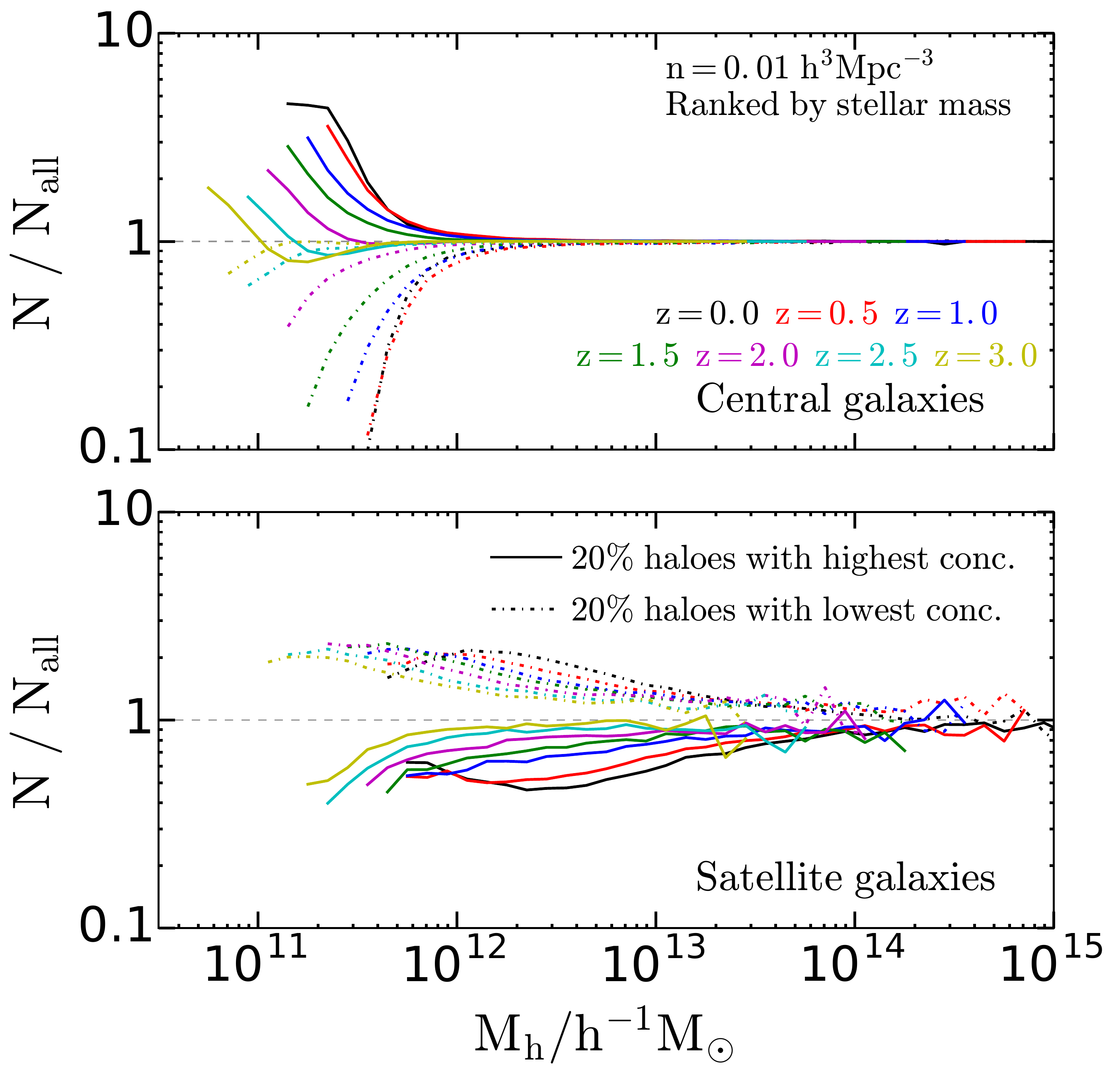}
\caption{Same as the bottom sub-panels of Fig.~\ref{Fig:HOD_single}, but now combined for $z=0, 0.5, 1, 1.5, 2, 2.5$ and $3$. The top panels presents the central galaxies contribution and the bottom panels is the satellites contribution. The two left panels show, as solid (dotted) lines, the ratio between the HOD of the $20$  per cent earliest (latest) formation time haloes and the full HOD, and in the right panels, the ratio of the HOD of the $20$ per cent high (low) concentration haloes and the full HOD, for galaxies selected by their stellar mass.} 
\label{Fig:HOD_ratio_1}
\end{figure*}

\begin{figure*}
\includegraphics[width=0.45\textwidth]{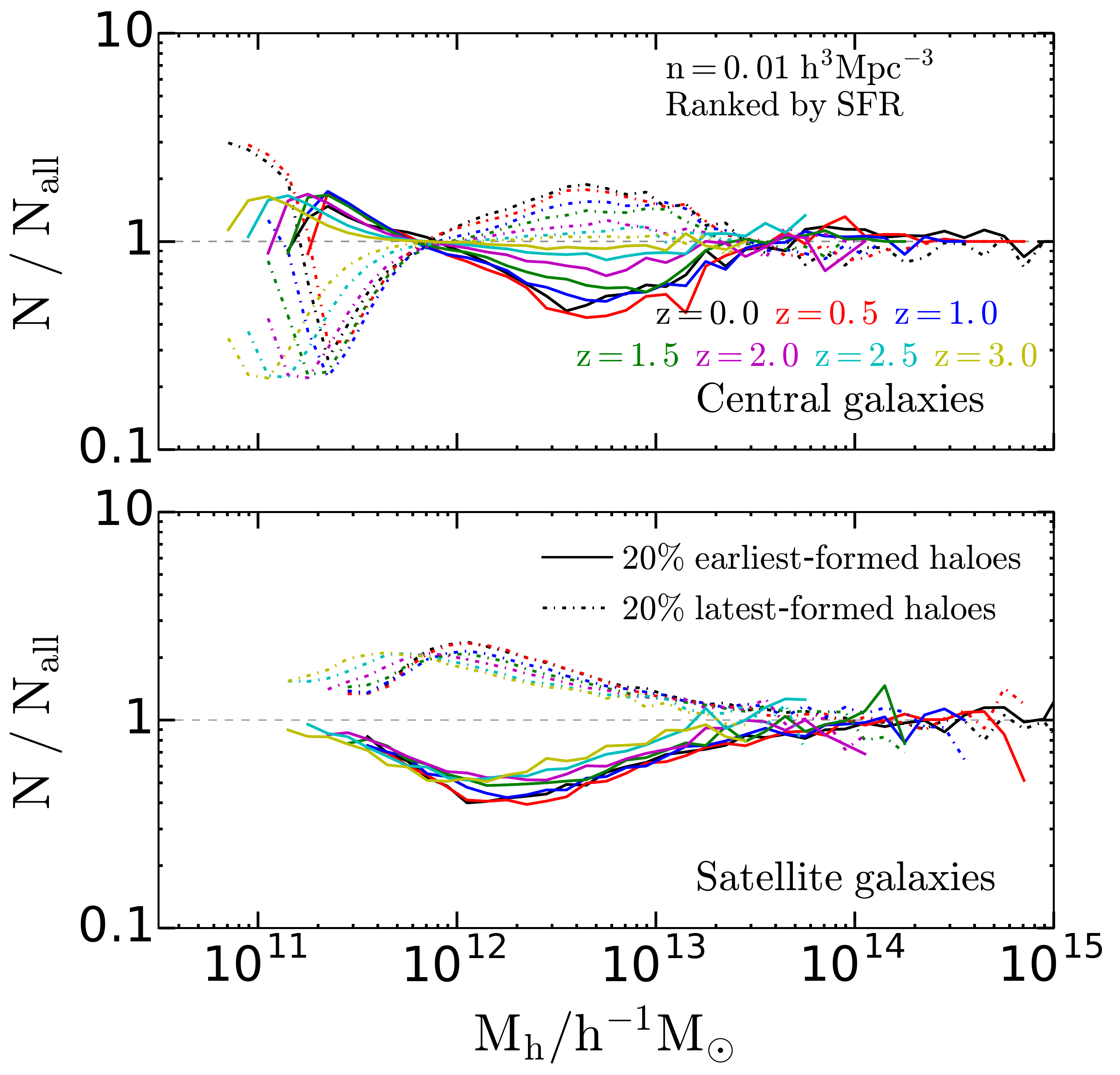}
\includegraphics[width=0.45\textwidth]{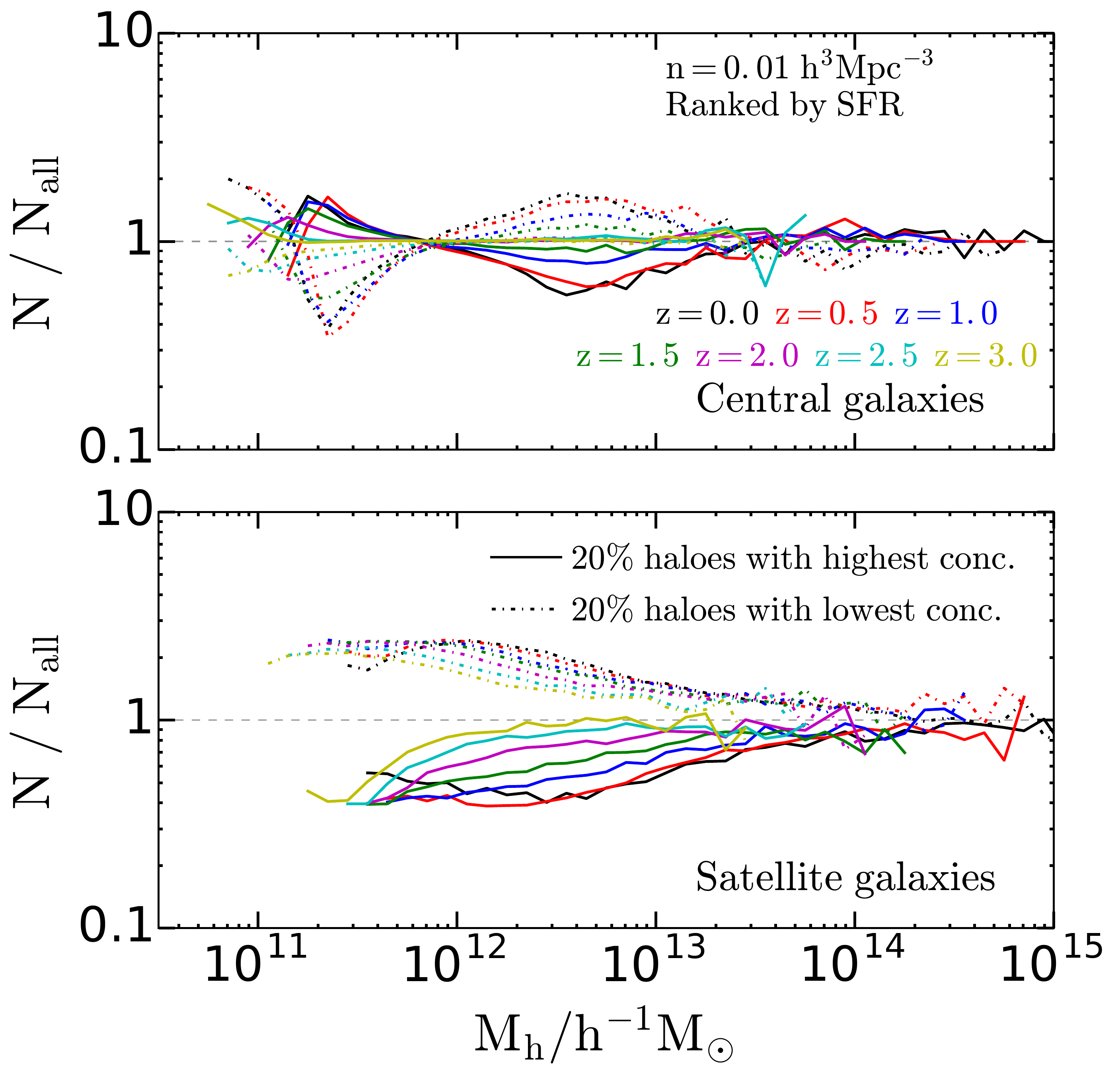}
\caption{Same as Fig.~\ref{Fig:HOD_ratio_1} but for galaxies selected by their SFR.} 
\label{Fig:HOD_ratio_2}
\end{figure*}

The full redshift evolution of the occupancy variation is captured in 
Fig.~\ref{Fig:HOD_ratio_1} and Fig.~\ref{Fig:HOD_ratio_2}, where we show the 
ratios of the HODs (as in the bottom subpanels of 
Fig.~\ref{Fig:HOD_single}-\ref{Fig:HOD_2}) for halos selected by age and
concentration at all redshifts explored, for galaxies selected by stellar mass 
and SFR, respectively. Here we corroborate that, for galaxies selected by stellar mass, the magnitude of the central galaxies occupancy variation 
is constant with redshift for haloes selected by age and it significantly 
decreases with increasing redshift (nearly diminishing by $z=3$) for haloes 
selected by concentration. The occupancy variation for the satellites part of 
the HOD progressively decreases for either age or concentration.
The overall shift of the ratios toward lower halo mass with increasing redshift
reflects the expected redshift evolution of the HOD (as studied for example 
by \citealt{C17}; see their Fig.~5).
For SFR selected galaxies (Fig.~\ref{Fig:HOD_ratio_2}), the occupancy variations decrease for both age and concentration, with a more pronounced trend for the latter.

The different evolution of the occupancy variation with age and concentration
indicates a different origin for these two effects. Even though they appear
similar at $z=0$, they evolve differently. We will further investigate their 
nature and origin in future work (Zehavi et al., in prep.). 
It is interesting to note that the evolution of the occupancy variation shows 
different trends compared to the evolution of the halo assembly bias found 
in Section~\ref{Sec:HAB}, where the signal decreased for the halo samples 
selected by age but not for those selected by concentration as in the occupancy
variation case. Both effects will influence the evolution of the galaxy 
assembly bias signal, as we will now show.

\section{The evolution of galaxy assembly bias}
\label{Sec:GAB}
In this section we show the effect of assembly bias on the galaxy correlation 
function at different redshifts. As we did in Section~\ref{Sec:HAB}, we 
measure the auto CF for the full galaxy sample as well as the cross CF of the
full sample with the given subsample (e.g. early/late formed haloes).

\begin{figure}
\includegraphics[width=0.48\textwidth]{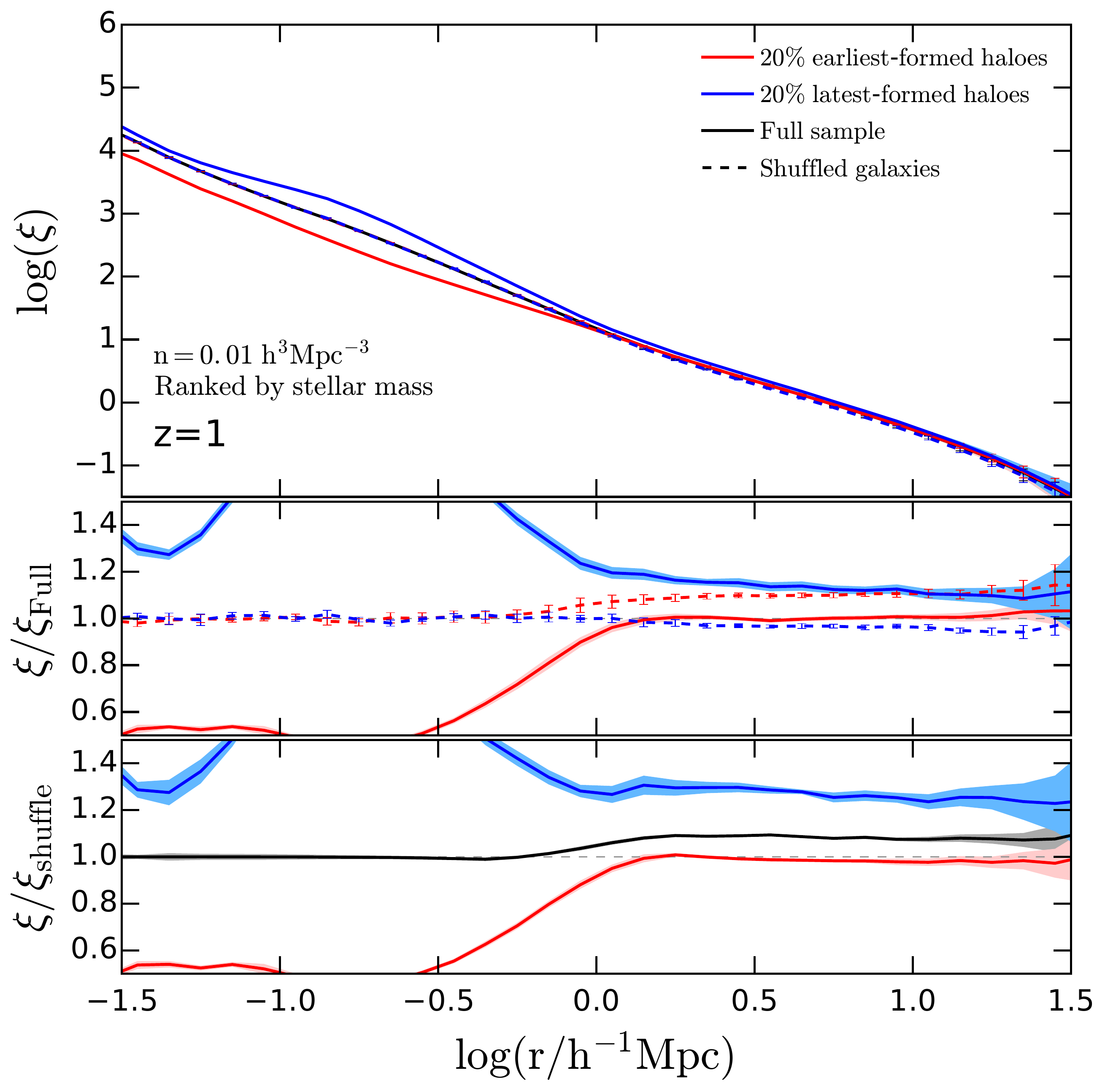}
\caption{The auto-correlation and cross-correlation functions at $z=1$ for a
stellar mass selected galaxy sample with number density of
$0.01\ h^{3}\, {\rm Mpc}^{-3}$.  Dashed lines correspond to the shuffled galaxy
sample (i.e., without occupancy variation) and solid lines represent the
original sample. 
The red (blue) lines show the cross-correlation function between the galaxies
that populate the $20$  per cent earliest (latest) forming haloes and the full
galaxy sample. The black lines show the auto-correlation function for the full
sample. The top panel shows the different correlation functions, the middle
panel shows the cross-correlation functions (red and blue lines) divided by
the auto-correlation function (black), and the bottom panel shows the ratio
between different correlation functions from the model and the ones from the
shuffled sample. The shaded region and errorbars represent the jackknife errors
calculated using 10 subsamples.}
\label{Fig:single_CF}
\end{figure}

To study the impact of assembly bias on the CF we shuffle galaxies among 
haloes of the same mass, following the approach of \cite{Croton:2007} and Z18. 
This consists of taking all haloes in a given bin of halo mass (0.1 dex wide 
in our case; we also tested using a bin width of 0.05 and 0.2 dex and found 
no major difference in our results) and randomly reassigning the galaxy 
population between these haloes. Central galaxies are located at the position 
of the central galaxies they replace (except if there is no galaxy in a halo; 
in which case the new galaxy is located at the potential minimum of the halo). 
The satellite galaxies are moved together with their original central galaxy 
and retain the same relative positions to it in their new halo. 
The shuffling 
removes any potential connection to the assembly history of the halos, and 
effectively transforms 
the HOD of any halo subsample (e.g., for a range of halo formation times 
or concentrations) to be the same as the total HOD 
(making, e.g., the red and blue lines of Fig.~\ref{Fig:HOD_single} be the 
same as the black line). 
This new galaxy sample will have, by construction, no occupancy variation.

The CFs for a stellar mass selected sample with number density
$n = 0.01\ h^{3}\, {\rm Mpc}^{-3}$ at $z=1$ is shown in top panel of
Fig.~\ref{Fig:single_CF}. The auto CF of the full sample is shown in black, and
the red and blue lines are the cross CF for the $20$ per cent earliest and
latest forming haloes, respectively.  The dashed lines show the CF for the
shuffled samples. The shaded region and errorbars represent the jackknife
errors calculated using 10 realisations for the real and the shuffled samples.
The middle panel shows the ratios between the cross CF of the subsamples and
the auto CF of the full sample, for both original (solid) and shuffled (dashed)
galaxy samples.  The bottom panel shows the ratios between the different CFs
measured for the original (unshuffled) galaxy samples and the corresponding
shuffled ones.  

A value above unity for the black line in the bottom panel of Fig.~\ref{Fig:single_CF} indicates 
that the original sample has a larger CF than that measured for the shuffled sample. 
These differences are the manifestation of galaxy assembly bias \citep{Croton:2007}. 
As explained in Z18, this arises from the combined effect 
of the occupancy variation and halo assembly bias. The central galaxies 
occupancy variation indicates a preferential occupancy of early-formed halos. 
These halos are more clustered, thus leading to a stronger clustering signal 
on large scales.  The significant clustering differences on small scales come 
about from the satellites occupancy variation, where the increased number of 
satellites in late-forming halos gives rise to a stronger clustering in the 
1-halo regime. 

It is interesting to note that the clustering of galaxies in the late-forming 
haloes is stronger on large scales than that for galaxies in early-formed 
haloes, as can be seen in the middle panel of Fig.~\ref{Fig:single_CF}.  This 
is the opposite to the results found by Z18 at $z=0$, implying that trend 
evolves with redshift.  This again arises from the inter-related  effects of 
halo assembly bias and the occupancy variation.
The dashed lines in the middle panel correspond to the shuffled galaxy samples,
and reflect the same halo assembly bias trends seen in Fig.~\ref{Fig:CF_Ev}, modulated 
by satellite galaxies.
The central occupancy variation acts to slightly increase this ratio for the 
galaxies in late-forming haloes and decrease it for the galaxies in 
early-formed halos (see corresponding discussion in Z18, specifically 
their \S5.3), thus likely resulting in the reversed clustering trend seen. 

\begin{figure*}
\includegraphics[width=0.49\textwidth]{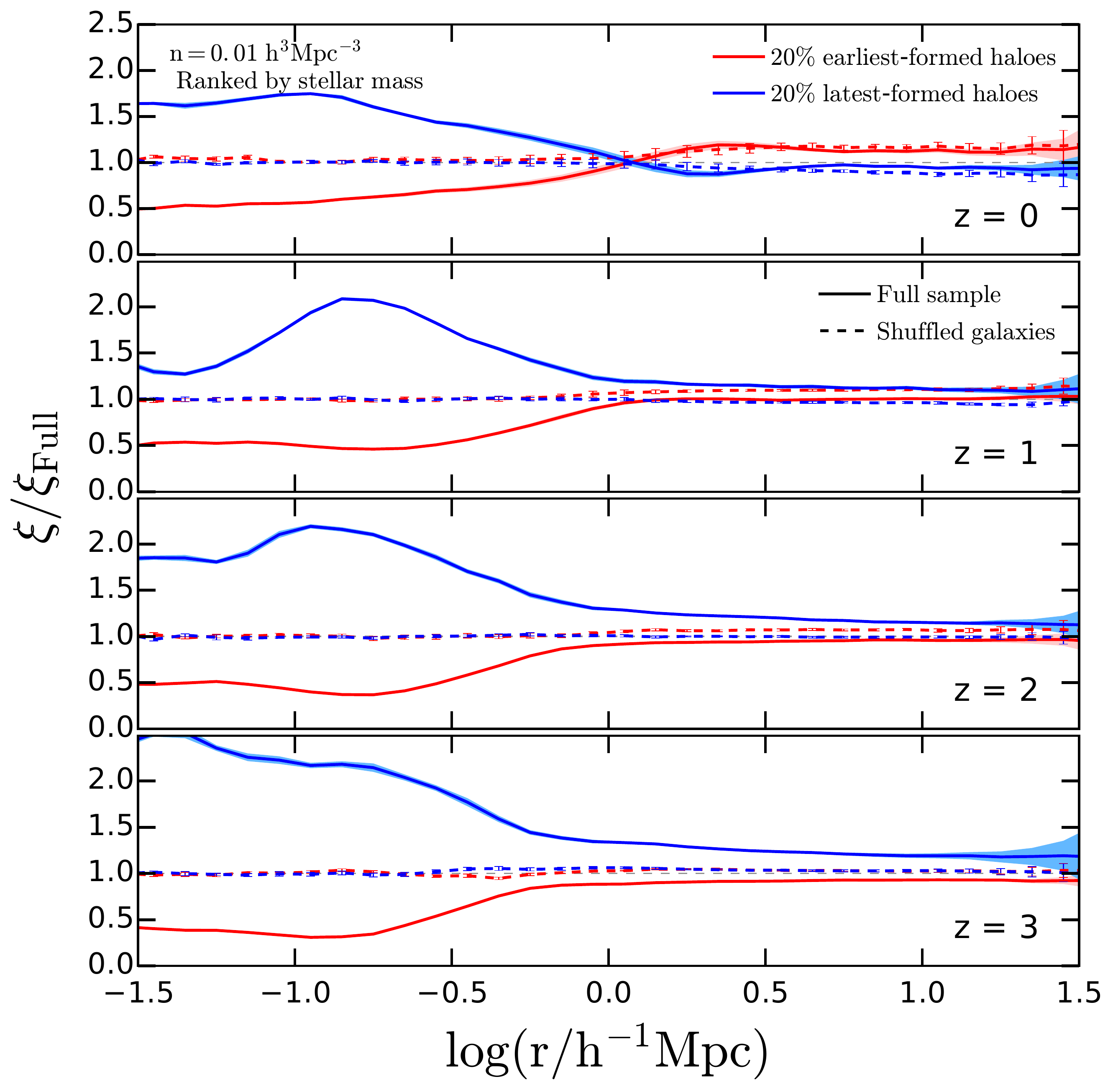}
\includegraphics[width=0.49\textwidth]{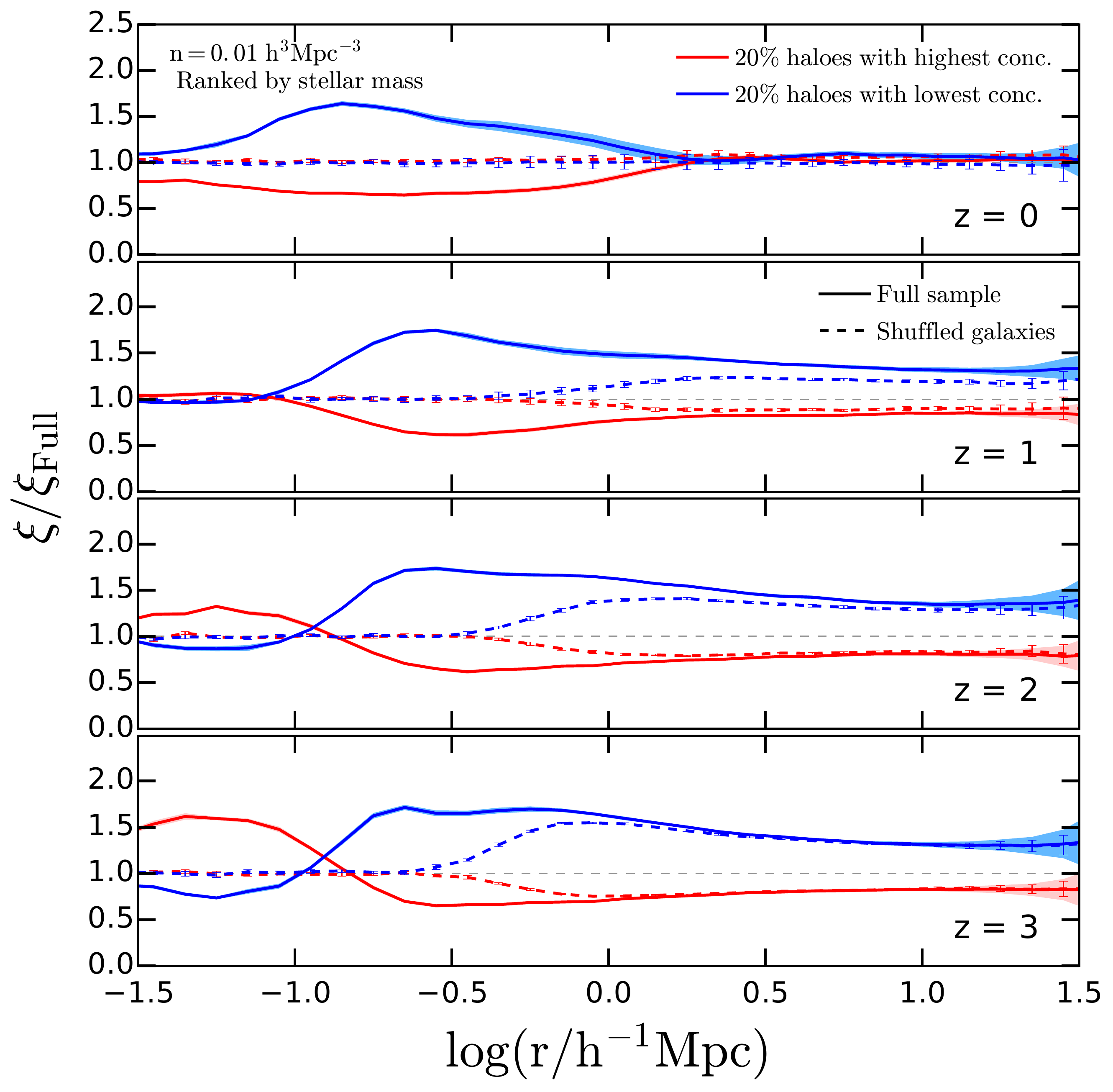}

\caption{The same as the middle panel of Fig.~\ref{Fig:single_CF}, for galaxies
  selected by their stellar mass with a number density of $0.01\ h^3\, {\rm Mpc}^{-3}$, at $z= 0, 1, 2$ and $3$. Halo samples selected by their formation redshift
  are shown on the left hand side and by their concentration on the  right.
  Please note that a larger y-axis range is shown here versus the one in  
Fig.~\ref{Fig:single_CF}.}
\label{Fig:CF_Ratio1}
\end{figure*}

The evolution of this ratio is individually presented Fig.~\ref{Fig:CF_Ratio1}. 
The left panels show the ratio between the CF measured for galaxies in haloes 
with the 20 per cent earliest/latest formation times and the CF of the full 
sample, for $z=0,1,2$ and $3$. The panels on right show the same but for 
haloes selected by concentration.  Solid lines show the original SAM galaxies 
and dashed lines show the shuffled sample results (i.e., with no occupancy
variation).
For the shuffled samples, we can see that the difference between the CFs of
galaxies in the earliest/latest-forming haloes decreases with increasing
redshift while the difference for haloes with the highest/lowest concentrations
is reversed, with increasingly stronger clustering found for the haloes with
a lower concentration. 
These trends are consistent with the evolution of the halo assembly bias signal shown in Section~\ref{Sec:HAB}, 
where we found a decrease (flip) of the of the difference in clustering on large scales for haloes selected by age (concentration).
At high redshift, the differences in the clustering of the SAM galaxies 
in early and late formed haloes come mostly from the occupancy variation. This 
is opposite to the situation of galaxies that live in high and low concentrated 
haloes, where the differences in their clustering come mostly from halo assembly bias.

For the galaxy population predicted by the SAM, galaxies in haloes with late
formation times or low concentrations show stronger large-scale clustering at
higher redshifts. For the concentration case, the clustering signal becomes
identical to that measured for the shuffled galaxies. This is expected since,
as shown in Section~\ref{Sec:OV}, the occupancy variation of haloes with 
concentration decreases strongly with increasing redshift. For galaxies in
haloes selected by formation time, the clustering of the galaxies in the
latest forming haloes is stronger than the shuffled galaxies, while for the
galaxies in the earliest-forming haloes it is lower than for the shuffled
case. This is again consistent with what we found in \S~\ref{Sec:OV}, with
the occupancy variation with age persisting to higher redshifts, with (at
each redshift) late (early) formation effectively shifting the occupation
toward higher (lower) halo masses, thus changing the clustering.

On small scales, the dashed lines in Fig.~\ref{Fig:CF_Ratio1} are identical
for both age and concentration, since there is no halo assembly bias in that
regime (1-halo scales). The SAM galaxies in late-forming haloes are more
correlated than those in early-forming haloes on small scales, at all redshifts.
This is due to the increased number of satellites in early versus late forming
halos, which persists at all redshifts (as seen in Fig.~\ref{Fig:HOD_ratio_1}).
Galaxies selected by halo concentration exhibit a similar behavior -- galaxies
in low concentration haloes are more clustered than those in high concentration
ones -- at small-to-intermediate scales. On very small scales (below $\sim 0.1
h^{-1}\, {\rm Mpc}$), though, this trend flips. 
One might have expected the same small-scale behaviour with concentration due
to the similar satellite occupancy variation.  However, the concentration 
differences impact the clustering as well. E.g., for the low-concentration 
sample, even with more satellite galaxies, they are likely less concentrated 
(since they trace the dark matter distribution) and as a consequence, less 
clustered on very small scales.

\begin{figure*}
\includegraphics[width=0.49\textwidth]{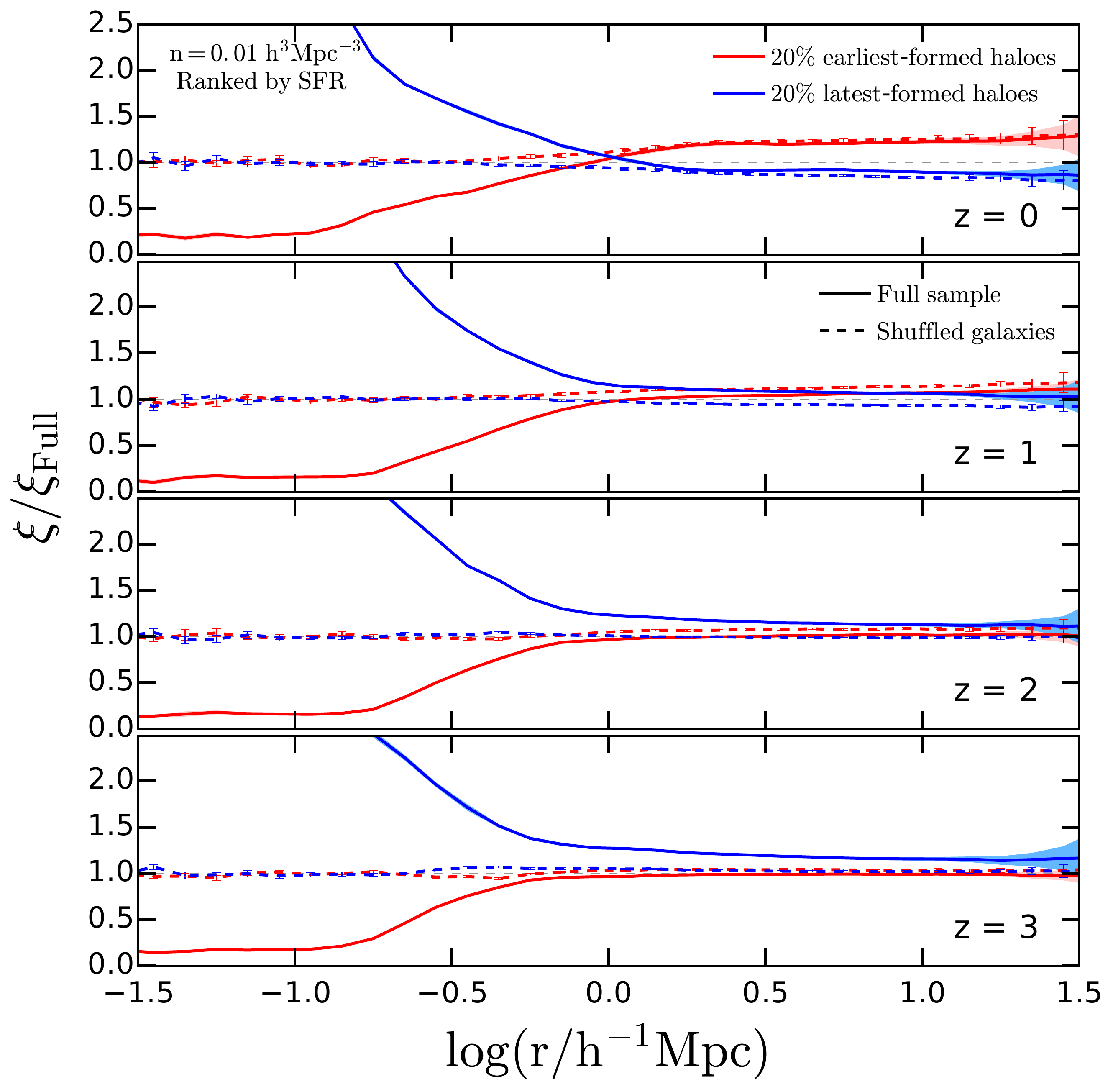}
\includegraphics[width=0.49\textwidth]{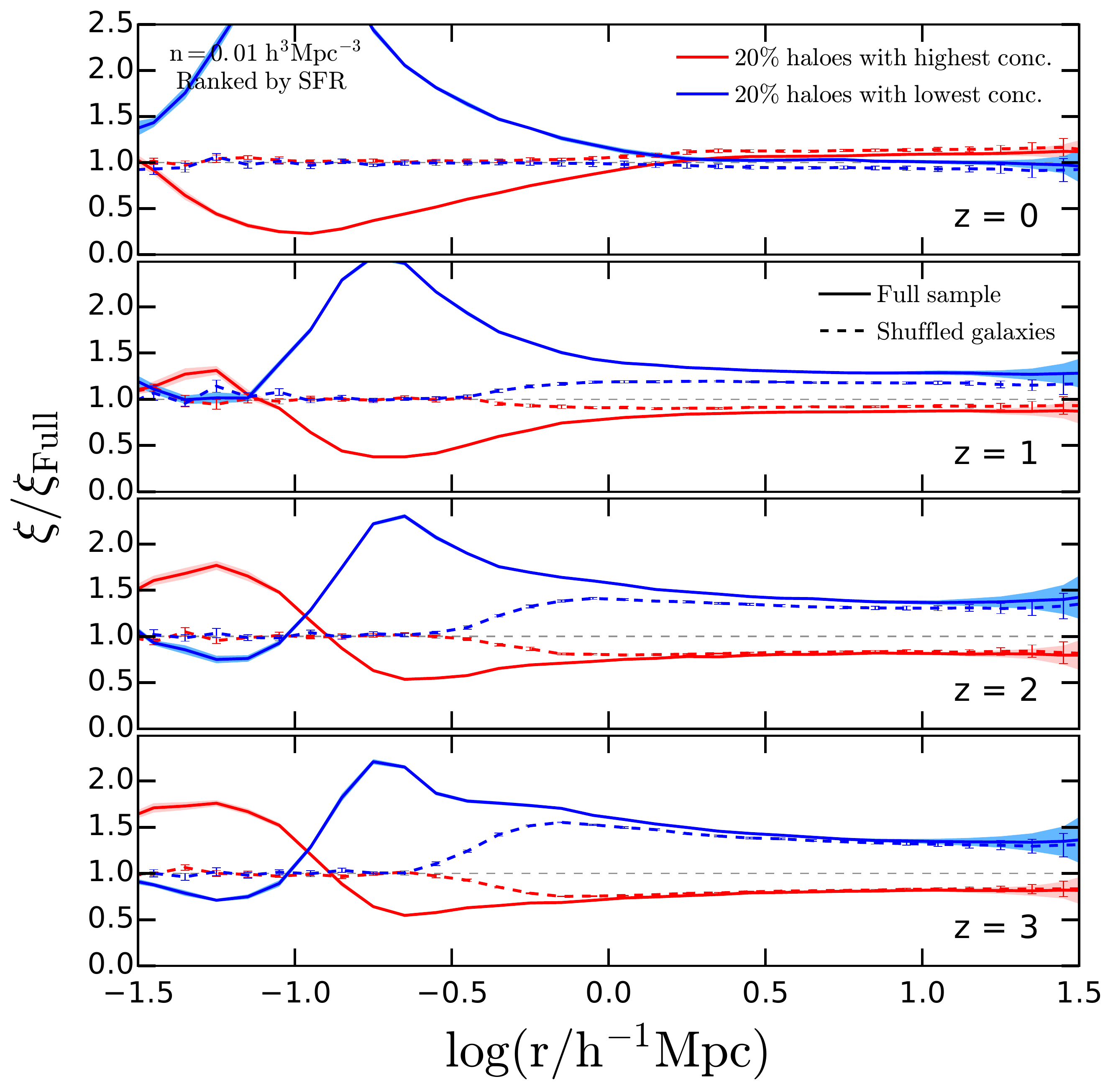}
\caption{The same as Fig.~\ref{Fig:CF_Ratio1}, but now for galaxies selected
  by their SFR instead of stellar mass.}
\label{Fig:CF_Ratio2}
\end{figure*}

In Fig.~\ref{Fig:CF_Ratio2} we show, for completeness, the corresponding
evolution of the CFs for galaxies selected by their SFR. We obtain the same
trends found for galaxies selected by stellar mass. We also analysed other
number density samples and found similar results for the evolution of the
galaxy CFs.

\begin{figure*}
\includegraphics[width=0.49\textwidth]{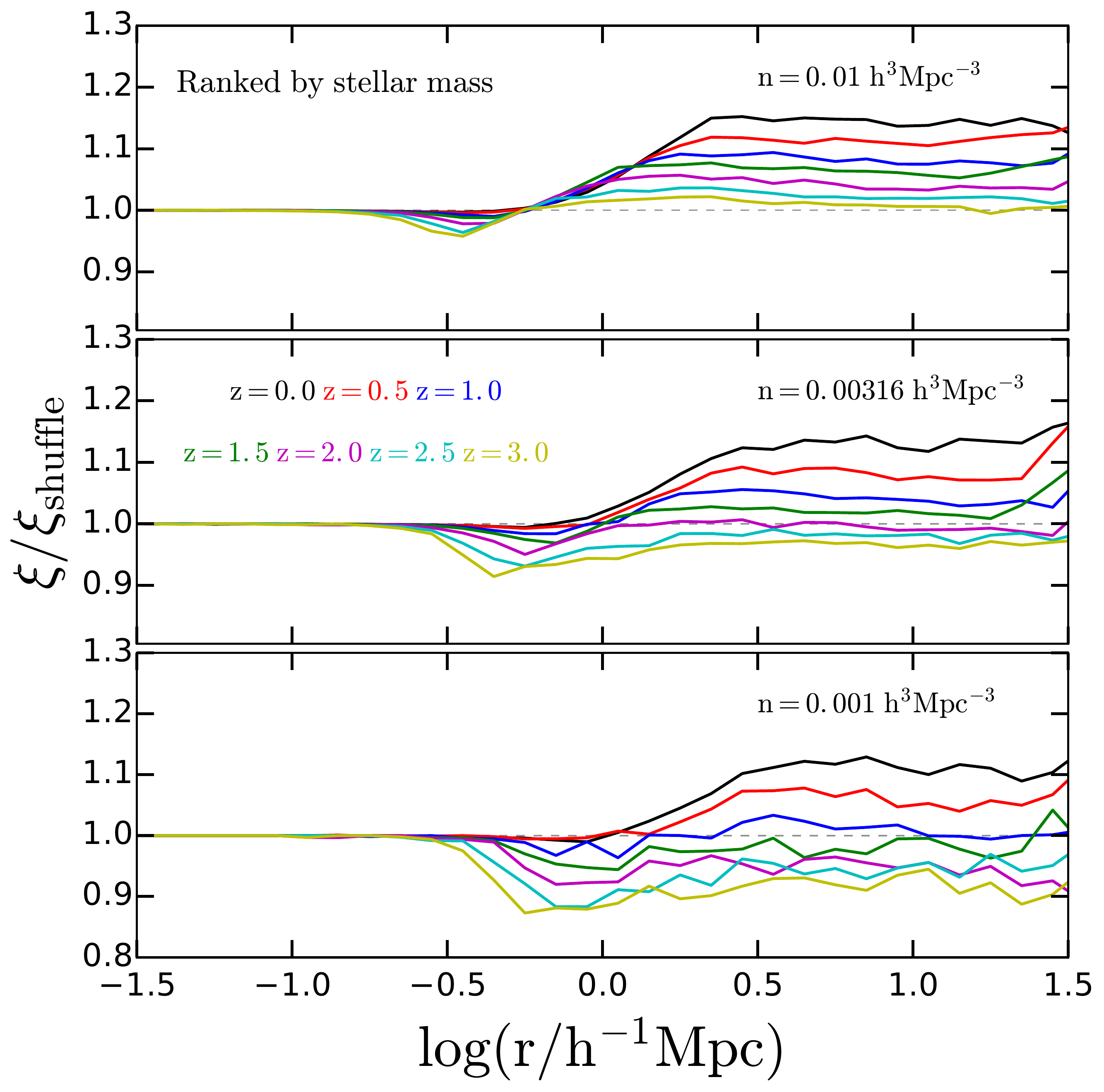}
\includegraphics[width=0.49\textwidth]{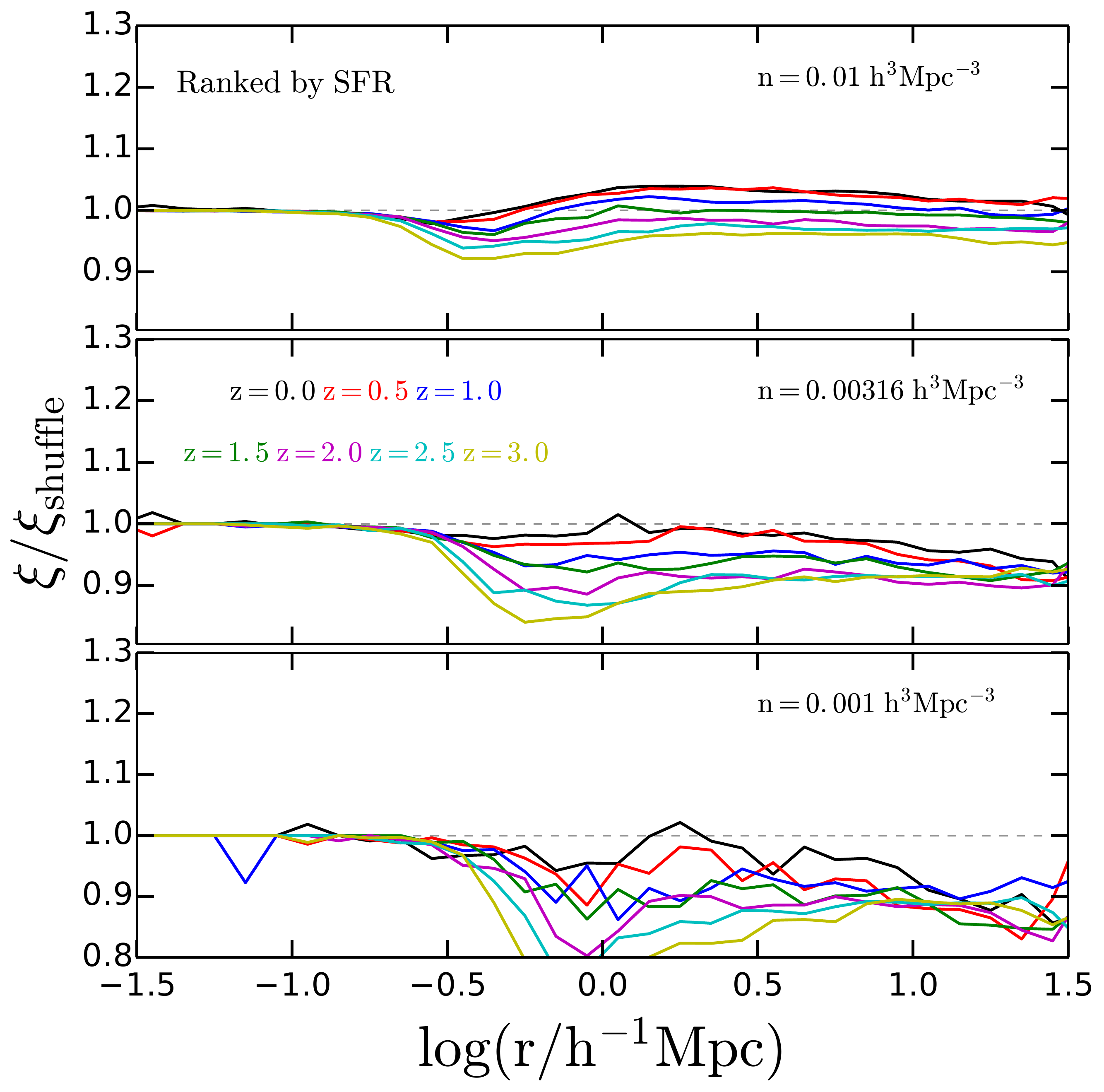}
\caption{The ratio between the correlation function of the full galaxy sample and the corresponding shuffled sample (equivalent to the black line in the bottom panel of Fig.~\ref{Fig:single_CF}) for three different number densities of $0.01\ h^3\, {\rm Mpc}^{-3}$ (top), $0.00316\ h^3\, {\rm Mpc}^{-3}$ (middle) and  $0.001\ h^3\, {\rm Mpc}^{-3}$ (bottom), and for $z=0, 0.5, 1, 1.5, 2, 2.5$ and $3$ as labelled.  Galaxies selected by stellar mass are shown on the left hand side while galaxies selected by SFR are shown on the right.}
\label{Fig:CF_Ratio4}
\end{figure*}

Finally, we consider the evolution of the galaxy assembly bias signal. As 
previously mentioned, galaxy assembly bias is quantified in terms of the ratio 
between the CF of a galaxy sample and that of a shuffled sample, where the
relation to halo assembly has been erased, as shown by the black line in the 
bottom panel of Fig.~\ref{Fig:single_CF}. Fig.~\ref{Fig:CF_Ratio4} presents
our measurements for three different number densities ($n= 0.01$, $0.00316$ 
and $0.001 \, h^{-3}\, {\rm Mpc}^3$) over a range of redshifts, 
for galaxies selected by stellar mass and SFR.  We find that this clustering 
ratio generally decreases for higher redshifts, and for lower number densities. 
Interestingly, this decrease can be large enough in some of these cases so that 
the original sample becomes less clustered than the corresponding shuffled sample, 
and the clustering {\it difference} changes sign and continues growing in magnitude.
This typically occurs for lower number densities and at high 
redshifts. This ratio is overall lower for the galaxy samples selected by 
their SFR rather than stellar mass. Nonetheless, the trends with redshift
and number density persist for these SFR selected samples, 

Our results are in qualitative agreement with those found by 
\citeauthor{Hearin:2016} (\citeyear{Hearin:2016}; their Fig.~8) over the 
limited redshift range they explore ($0<z<1$). However, we note that 
\citeauthor{Hearin:2016} compare samples with the same stellar mass thresholds 
at the different redshifts, not accounting for any stellar mass evolution. 
Effectively, this amounts to probing more massive galaxies (lower number 
densities) at higher redshifts, and thus it is impossible to separate the 
evolution they find from the expected number density dependence.

Again, the impact of assembly bias on galaxy clustering arises from the
combined contributions of the occupancy variation and halo assembly bias. 
At relatively low redshift for the stellar mass selected samples, these 
typically combine to produce an increased clustering (see the discussion 
following Fig.~\ref{Fig:CF_Ratio1} and in Z18). For example, using our results 
in the previous sections for $n=0.01 h^{3}\, {\rm Mpc}^{-3}$, we can explain the
behaviour exhibited in the top-left panel of Fig.~\ref{Fig:CF_Ratio4}. As we
saw earlier for the halo age case, the level of halo assembly bias decreases 
while the level of the central galaxies occupancy variation remains similar, 
leading to a diminishing galaxy assembly bias effect. For lower number 
densities at high redshifts, halo assembly bias reverses sense (in a similar 
manner to concentration) such that early-formed halos become less clustered
than the late-formed ones, and this gives rise to the reversed sense of galaxy 
assembly bias in those cases. 

One could have a-priory envisioned a scenario in which the stochasticity 
involved in the galaxy formation processes would serve to weaken galaxy
assembly bias over time.  Alternatively, one might have expected 
the signature to grow with time (i.e., diminish as one goes to higher 
redshift), due to the hierarchical growth of structure. However, it seems 
that the evolution of assembly bias is far more intricate. The overall trend
we find is that the CFs ratio monotonically increases with time (or decreases 
with increasing redshift; Fig.~\ref{Fig:CF_Ratio4}). This leads to a change
in the sign of the effect, i.e., whether the clustering of the galaxy sample
is stronger or weaker due to assembly bias effects, as well as a shift in
whether the magnitude of this clustering difference decreases or increases
with time. 
This gets more complex as the amplitude of the clustering ratio varies with
the specifics of the galaxy selection (e.g., stellar mass or SFR) and number 
density, thus it is non-trivial to predict which galaxy sample would show 
negligible or extreme assembly bias properties and at which redshift.

\section{Summary \& Conclusions}

We use a state-of-the-art semi-analytic model of galaxy formation, the G13 SAM 
model, to study the origin and evolution of assembly bias in the galaxy 
distribution. We identify two separate contributions to this effect: halo 
assembly bias, which refers to the different clustering of haloes with
different `secondary property', and occupancy variation, the dependence of
the number of galaxies in haloes of the same mass on a second property of
the haloes. We isolate the evolution of these two effects for haloes selected
by their concentration and formation redshift, two of the most common secondary
properties used to measure assembly bias. 
The galaxy samples correspond to different number densities based on either 
ranked stellar mass or SFR. Our key results are shown in 
Figures~\ref{Fig:CF_Ev}, \ref{Fig:HOD_ratio_1} and \ref{Fig:CF_Ratio4}. 
We now summarise our main findings:
\begin{itemize}
 \item At $z=0$ the concentration of dark matter haloes correlates with 
formation time. This correlation weakens at higher redshifts.
\item Haloes at $z=0$ with high concentrations or early formation times are 
more clustered than those with low concentrations or late formation times. At 
high redshift, there are no differences in the CF measured for haloes with 
different formation times, but low concentration haloes are more correlated 
than high concentration ones.
\item Haloes ranked to have an extreme concentration or formation time at a 
given redshift do not necessarily have the same ranking at other redshifts. We 
found that the main progenitors of $z=0$ haloes display clustering similar to 
that measured for their descendants. This means that the evolution of the halo assembly bias signal is not caused 
because a set of haloes (e.g., high concentration haloes) change their 
clustering over time, but because haloes change their ranking in terms of a 
secondary property. 
\item At $z=0$, haloes with early formation times or high concentrations are 
populated by galaxies starting at lower halo masses (for a fixed cut in stellar
mass) but they have fewer satellite galaxies for a fixed mass compared to 
haloes with late formation times or low concentrations. 
\item For galaxies selected by SFR we generally find similar occupancy 
variation trends to those found for galaxies selected by stellar mass 
(though different shape of the HOD). Haloes with early formation times or high 
concentrations  are first populated by galaxies at a lower mass and have fewer 
satellite galaxies at a given mass compared to haloes with late formation times
or low concentrations. The one difference is that at higher halo masses, 
where the central galaxies occupation drops, there are less centrals in haloes 
with early formation times or high concentrations than for those with either 
late formation times or low concentrations.  
\item The occupancy variation for central galaxies in haloes with different 
formation times stay roughly constant as a function of redshift for a fixed 
galaxy number density and for galaxies selected by either stellar mass or SFR. 
The corresponding satellite galaxies occupancy variation decreases somewhat 
with increasing redshift.  
\item The occupancy variation for galaxies in haloes with different concentrations 
diminishes for the central galaxies and satellites with increasing redshift, 
for both stellar mass or SFR selected galaxy samples.  
\item The evolution of the CF of galaxy samples without occupancy variation 
(i.e., the shuffled samples) reflects the same trends on large scales as the 
evolution of halo assembly bias for haloes selected by age or concentration; 
the CF differences for galaxies in haloes with early and late formation times 
decreases with look back time, while the CF of galaxies in low-concentration 
haloes increases relative to the CF of galaxies in high-concentration haloes 
when going to higher redshifts.  
\item The CF of galaxies hosted by haloes with late formation times or low 
concentration increases relative to the CF of galaxies in haloes with early 
formation times or high concentrations, respectively, with increasing redshift. 
\item The occupancy variation tends to increase the amplitude of the CF of 
galaxies that live in haloes with either late formation times or low 
concentrations, and decrease it for galaxies that live in haloes with early 
formation times or high concentrations.  
\item Galaxy assembly bias as measured by the ratio between the CF of the 
model galaxies and that of the shuffled galaxies decreases with redshift, going
below 1 in some cases.  This CFs ratio is generally smaller for lower number 
densities and for SFR-selected samples.
\end{itemize}

The different evolution of halo assembly bias and the occupancy variation
with age and concentration likely points to a different origin for the
dependence on these two secondary parameters. This is further corroborated
by their lack of correlation at high redshift. 
In general, we find similar trends in the evolution of assembly bias,
for both the occupancy variation and galaxy assembly bias, for galaxies 
selected by SFR versus stellar mass. This is quite impressive considering that 
galaxy samples selected by stellar mass and by SFR exhibit quite different 
behaviours in the SAMs \citep{C13,C15}, and may be relevant for upcoming surveys.

The results shown here will help to inform theoretical models of assembly bias
and the development of observational tests to detect its existence (or absence)
in the Universe. They can also be used to construct improved mock galaxy 
catalogues incorporating assembly bias (as standard HOD mocks do not include
this effect). For these purposes we are releasing all the HODs and occupancy
variation measures obtained in this work, as well as parametrised fits for 
them (see Appendix~A for more details).

\section*{Acknowledgements}
We thank the anonymous referee for insightful comments that helped improve the presentation of this manuscript.
We thank Peder Norberg for useful discussions.
This work was made possible by the efforts of Gerard Lemson and
colleagues at the German Astronomical Virtual Observatory in setting
up the Millennium Simulation database in Garching.
SC, IZ, NP, and EJ acknowledge the hospitality of the ICC at Durham University.
SC \& NP acknowledge support from a STFC/Newton-CONICYT Fund award (ST/M007995/1 - DPI20140114) and Anillo ACT-1417.
SC is also supported by the European Research Council through grant ERC-StG/716151.
IZ acknowledges support by NSF grant AST-1612085 and by a CWRU Faculty Seed Grant.
NP \& EJ are further supported by ``Centro de Astronom\'ia y Tecnolog\'ias Afines'' BASAL PFB-06 and by Fondecyt Regular 1150300. 
This project has received funding from the European Union's Horizon 2020 Research and Innovation Programme under the Marie Sk\l{}odowska-Curie grant agreement No 734374. 
The calculations for this paper were performed on the 
ICC Cosmology Machine, which is part of the DiRAC-2 
Facility jointly funded by STFC, the Large Facilities 
Capital Fund of BIS, and Durham University and on the Geryon computer at the Center for 
Astro-Engineering UC, part of the BASAL PFB-06, which received additional
funding from QUIMAL 130008 and Fondequip AIC-57 for upgrades.

\bibliography{Biblio}

\begin{thebibliography}{100}
\expandafter\ifx\csname natexlab\endcsname\relax\def\natexlab#1{#1}\fi

\bibitem[{{Abbas} \& {Sheth}(2006)}]{Abbas:2006}
{Abbas} U., {Sheth} R.~K., 2006, \mnras, 372, 1749

\bibitem[{{Artale} {et~al}\mbox{.}(2018){Artale}, {Zehavi}, {Contreras}, \&
  {Norberg}}]{Artale:2018}
{Artale} M.~C., {Zehavi} I., {Contreras} S., {Norberg} P., 2018, \mnras,\ in\
  press

\bibitem[{{Baugh}(2006)}]{Baugh:2006}
{Baugh} C.~M., 2006, Reports on Progress in Physics, 69, 3101

\bibitem[{{Baugh} {et~al}\mbox{.}(2018){Baugh}, {Gonzalez-Perez}, {del P
  Lagos}, {Lacey}, {Helly}, {Jenkins}, {Frenk}, {Benson}, {Bower}, \&
  {Cole}}]{Baugh:2019}
{Baugh} C.~M. {et~al.}, 2018, \mnras

\bibitem[{{Benson}(2010)}]{Benson:2010}
{Benson} A.~J., 2010, \physrep, 495, 33

\bibitem[{{Benson}(2012)}]{Benson:2012}
{Benson} A.~J., 2012, \na, 17, 175

\bibitem[{{Berlind} {et~al}\mbox{.}(2006){Berlind}, {Frieman}, {Weinberg},
  {Blanton}, {Warren}, {Abazajian}, {Scranton}, {Hogg}, {Scoccimarro},
  {Bahcall}, {Brinkmann}, {Gott}, {Kleinman}, {Krzesinski}, {Lee}, {Miller},
  {Nitta}, {Schneider}, {Tucker}, {Zehavi}, \& {SDSS
  Collaboration}}]{Berlind:2006}
{Berlind} A.~A. {et~al.}, 2006, \apjs, 167, 1

\bibitem[{{Berlind} \& {Weinberg}(2002)}]{Berlind:2002}
{Berlind} A.~A., {Weinberg} D.~H., 2002, \apj, 575, 587

\bibitem[{{Berlind} {et~al}\mbox{.}(2003){Berlind}, {Weinberg}, {Benson},
  {Baugh}, {Cole}, {Dav{\'e}}, {Frenk}, {Jenkins}, {Katz}, \&
  {Lacey}}]{Berlind:2003}
{Berlind} A.~A. {et~al.}, 2003, \apj, 593, 1

\bibitem[{{Bertone}, {De Lucia} \& {Thomas}(2007){Bertone}, {De Lucia}, \&
  {Thomas}}]{Bertone:2007}
{Bertone} S., {De Lucia} G., {Thomas} P.~A., 2007, \mnras, 379, 1143

\bibitem[{{Blanton} \& {Berlind}(2007)}]{Blanton:2007}
{Blanton} M.~R., {Berlind} A.~A., 2007, \apj, 664, 791

\bibitem[{{Bond} {et~al}\mbox{.}(1991){Bond}, {Cole}, {Efstathiou}, \&
  {Kaiser}}]{Bond:1991}
{Bond} J.~R., {Cole} S., {Efstathiou} G., {Kaiser} N., 1991, \apj, 379, 440

\bibitem[{{Bower} {et~al}\mbox{.}(2006){Bower}, {Benson}, {Malbon}, {Helly},
  {Frenk}, {Baugh}, {Cole}, \& {Lacey}}]{Bower:2006}
{Bower} R.~G., {Benson} A.~J., {Malbon} R., {Helly} J.~C., {Frenk} C.~S.,
  {Baugh} C.~M., {Cole} S., {Lacey} C.~G., 2006, \mnras, 370, 645

\bibitem[{{Bullock} {et~al}\mbox{.}(2001){Bullock}, {Dekel}, {Kolatt},
  {Kravtsov}, {Klypin}, {Porciani}, \& {Primack}}]{Bullock:2001}
{Bullock} J.~S., {Dekel} A., {Kolatt} T.~S., {Kravtsov} A.~V., {Klypin} A.~A.,
  {Porciani} C., {Primack} J.~R., 2001, \apj, 555, 240

\bibitem[{{Busch} \& {White}(2017)}]{Busch:2017}
{Busch} P., {White} S.~D.~M., 2017, ArXiv e-prints

\bibitem[{{Campbell} {et~al}\mbox{.}(2015){Campbell}, {van den Bosch},
  {Hearin}, {Padmanabhan}, {Berlind}, {Mo}, {Tinker}, \&
  {Yang}}]{Campbell:2015b}
{Campbell} D., {van den Bosch} F.~C., {Hearin} A., {Padmanabhan} N., {Berlind}
  A., {Mo} H.~J., {Tinker} J., {Yang} X., 2015, \mnras, 452, 444

\bibitem[{{Chaves-Montero} {et~al}\mbox{.}(2016){Chaves-Montero}, {Angulo},
  {Schaye}, {Schaller}, {Crain}, {Furlong}, \& {Theuns}}]{Chaves-Montero:2016}
{Chaves-Montero} J., {Angulo} R.~E., {Schaye} J., {Schaller} M., {Crain} R.~A.,
  {Furlong} M., {Theuns} T., 2016, \mnras

\bibitem[{{Cole} {et~al}\mbox{.}(1994){Cole}, {Aragon-Salamanca}, {Frenk},
  {Navarro}, \& {Zepf}}]{Cole:1994}
{Cole} S., {Aragon-Salamanca} A., {Frenk} C.~S., {Navarro} J.~F., {Zepf} S.~E.,
  1994, \mnras, 271, 781

\bibitem[{{Contreras} {et~al}\mbox{.}(2013){Contreras}, {Baugh}, {Norberg}, \&
  {Padilla}}]{C13}
{Contreras} S., {Baugh} C.~M., {Norberg} P., {Padilla} N., 2013, \mnras, 432,
  2717

\bibitem[{{Contreras} {et~al}\mbox{.}(2015){Contreras}, {Baugh}, {Norberg}, \&
  {Padilla}}]{C15}
{Contreras} S., {Baugh} C.~M., {Norberg} P., {Padilla} N., 2015, \mnras, 452,
  1861

\bibitem[{{Contreras} {et~al}\mbox{.}(2017){Contreras}, {Zehavi}, {Baugh},
  {Padilla}, \& {Norberg}}]{C17}
{Contreras} S., {Zehavi} I., {Baugh} C.~M., {Padilla} N., {Norberg} P., 2017,
  \mnras, 465, 2833

\bibitem[{{Cooper} {et~al}\mbox{.}(2010){Cooper}, {Gallazzi}, {Newman}, \&
  {Yan}}]{Cooper:2010}
{Cooper} M.~C., {Gallazzi} A., {Newman} J.~A., {Yan} R., 2010, \mnras, 402,
  1942

\bibitem[{{Cooray} \& {Sheth}(2002)}]{Cooray:2002}
{Cooray} A., {Sheth} R., 2002, \physrep, 372, 1

\bibitem[{{Coupon} {et~al}\mbox{.}(2012){Coupon}, {Kilbinger}, {McCracken},
  {Ilbert}, {Arnouts}, {Mellier}, {Abbas}, {de la Torre}, {Goranova},
  {Hudelot}, {Kneib}, \& {Le F{\`e}vre}}]{Coupon:2012}
{Coupon} J. {et~al.}, 2012, \aap, 542, A5

\bibitem[{{Croton}, {Gao} \& {White}(2007){Croton}, {Gao}, \&
  {White}}]{Croton:2007}
{Croton} D.~J., {Gao} L., {White} S.~D.~M., 2007, \mnras, 374, 1303

\bibitem[{{Croton} {et~al}\mbox{.}(2006){Croton}, {Springel}, {White}, {De
  Lucia}, {Frenk}, {Gao}, {Jenkins}, {Kauffmann}, {Navarro}, \&
  {Yoshida}}]{Croton:2006}
{Croton} D.~J. {et~al.}, 2006, \mnras, 365, 11

\bibitem[{{Croton} {et~al}\mbox{.}(2016){Croton}, {Stevens}, {Tonini}, {Garel},
  {Bernyk}, {Bibiano}, {Hodkinson}, {Mutch}, {Poole}, \&
  {Shattow}}]{Croton:2016}
{Croton} D.~J. {et~al.}, 2016, \apjs, 222, 22

\bibitem[{{Davis} {et~al}\mbox{.}(1985){Davis}, {Efstathiou}, {Frenk}, \&
  {White}}]{Davis:1985}
{Davis} M., {Efstathiou} G., {Frenk} C.~S., {White} S. D.~M., 1985, \apj, 292,
  371

\bibitem[{{De Lucia} \& {Blaizot}(2007)}]{DeLucia:2007}
{De Lucia} G., {Blaizot} J., 2007, \mnras, 375, 2

\bibitem[{{De Lucia}, {Kauffmann} \& {White}(2004){De Lucia}, {Kauffmann}, \&
  {White}}]{DeLucia:2004}
{De Lucia} G., {Kauffmann} G., {White} S.~D.~M., 2004, \mnras, 349, 1101

\bibitem[{{Dvornik} {et~al}\mbox{.}(2017){Dvornik}, {Cacciato}, {Kuijken},
  {Viola}, {Hoekstra}, {Nakajima}, {van Uitert}, {Brouwer}, {Choi}, {Erben},
  {Fenech Conti}, {Farrow}, {Herbonnet}, {Heymans}, {Hildebrandt}, {Hopkins},
  {McFarland}, {Norberg}, {Schneider}, {Sif{\'o}n}, {Valentijn}, \&
  {Wang}}]{Dvornik:2017}
{Dvornik} A. {et~al.}, 2017, \mnras, 468, 3251

\bibitem[{{Gao}, {Springel} \& {White}(2005){Gao}, {Springel}, \&
  {White}}]{Gao:2005}
{Gao} L., {Springel} V., {White} S.~D.~M., 2005, \mnras, 363, L66

\bibitem[{{Gao} \& {White}(2007)}]{Gao:2007}
{Gao} L., {White} S.~D.~M., 2007, \mnras, 377, L5

\bibitem[{{Gao} {et~al}\mbox{.}(2004){Gao}, {White}, {Jenkins}, {Stoehr}, \&
  {Springel}}]{Gao:2004}
{Gao} L., {White} S.~D.~M., {Jenkins} A., {Stoehr} F., {Springel} V., 2004,
  \mnras, 355, 819

\bibitem[{{Geach} {et~al}\mbox{.}(2012){Geach}, {Sobral}, {Hickox}, {Wake},
  {Smail}, {Best}, {Baugh}, \& {Stott}}]{Geach:2012}
{Geach} J.~E., {Sobral} D., {Hickox} R.~C., {Wake} D.~A., {Smail} I., {Best}
  P.~N., {Baugh} C.~M., {Stott} J.~P., 2012, \mnras, 426, 679

\bibitem[{{Gonzalez-Perez} {et~al}\mbox{.}(2018){Gonzalez-Perez}, {Comparat},
  {Norberg}, {Baugh}, {Contreras}, {Lacey}, {McCullagh}, {Orsi}, {Helly}, \&
  {Humphries}}]{Gonzalez:2018}
{Gonzalez-Perez} V. {et~al.}, 2018, \mnras, 474, 4024

\bibitem[{{Gonzalez-Perez} {et~al}\mbox{.}(2014){Gonzalez-Perez}, {Lacey},
  {Baugh}, {Lagos}, {Helly}, {Campbell}, \& {Mitchell}}]{Gonzalez:2014}
{Gonzalez-Perez} V., {Lacey} C.~G., {Baugh} C.~M., {Lagos} C.~D.~P., {Helly}
  J., {Campbell} D.~J.~R., {Mitchell} P.~D., 2014, \mnras, 439, 264

\bibitem[{{Guo} {et~al}\mbox{.}(2016){Guo}, {Gonzalez-Perez}, {Guo},
  {Schaller}, {Furlong}, {Bower}, {Cole}, {Crain}, {Frenk}, {Helly}, {Lacey},
  {Lagos}, {Mitchell}, {Schaye}, \& {Theuns}}]{Guo:2016b}
{Guo} Q. {et~al.}, 2016, \mnras, 461, 3457

\bibitem[{{Guo} {et~al}\mbox{.}(2013){Guo}, {White}, {Angulo}, {Henriques},
  {Lemson}, {Boylan-Kolchin}, {Thomas}, \& {Short}}]{Guo:2013a}
{Guo} Q., {White} S., {Angulo} R.~E., {Henriques} B., {Lemson} G.,
  {Boylan-Kolchin} M., {Thomas} P., {Short} C., 2013, \mnras, 428, 1351

\bibitem[{{Guo} {et~al}\mbox{.}(2011){Guo}, {White}, {Boylan-Kolchin}, {De
  Lucia}, {Kauffmann}, {Lemson}, {Li}, {Springel}, \& {Weinmann}}]{Guo:2011}
{Guo} Q. {et~al.}, 2011, \mnras, 413, 101

\bibitem[{{Hearin}, {Watson} \& {van den Bosch}(2015){Hearin}, {Watson}, \&
  {van den Bosch}}]{Hearin:2015}
{Hearin} A.~P., {Watson} D.~F., {van den Bosch} F.~C., 2015, \mnras, 452, 1958

\bibitem[{{Hearin} {et~al}\mbox{.}(2016){Hearin}, {Zentner}, {van den Bosch},
  {Campbell}, \& {Tollerud}}]{Hearin:2016}
{Hearin} A.~P., {Zentner} A.~R., {van den Bosch} F.~C., {Campbell} D.,
  {Tollerud} E., 2016, \mnras

\bibitem[{{Henriques} {et~al}\mbox{.}(2015){Henriques}, {White}, {Thomas},
  {Angulo}, {Guo}, {Lemson}, {Springel}, \& {Overzier}}]{Henriques:2015}
{Henriques} B.~M.~B., {White} S.~D.~M., {Thomas} P.~A., {Angulo} R., {Guo} Q.,
  {Lemson} G., {Springel} V., {Overzier} R., 2015, \mnras, 451, 2663

\bibitem[{{Henriques} {et~al}\mbox{.}(2013){Henriques}, {White}, {Thomas},
  {Angulo}, {Guo}, {Lemson}, \& {Springel}}]{Henriques:2013}
{Henriques} B.~M.~B., {White} S.~D.~M., {Thomas} P.~A., {Angulo} R.~E., {Guo}
  Q., {Lemson} G., {Springel} V., 2013, \mnras, 431, 3373

\bibitem[{{Jiang} {et~al}\mbox{.}(2014){Jiang}, {Helly}, {Cole}, \&
  {Frenk}}]{Jiang:2014}
{Jiang} L., {Helly} J.~C., {Cole} S., {Frenk} C.~S., 2014, \mnras, 440, 2115

\bibitem[{{Jing}, {Suto} \& {Mo}(2007){Jing}, {Suto}, \& {Mo}}]{Jing:2007}
{Jing} Y.~P., {Suto} Y., {Mo} H.~J., 2007, \apj, 657, 664

\bibitem[{{Kauffmann} {et~al}\mbox{.}(1999){Kauffmann}, {Colberg}, {Diaferio},
  \& {White}}]{Kauffmann:1999}
{Kauffmann} G., {Colberg} J.~M., {Diaferio} A., {White} S.~D.~M., 1999, \mnras,
  303, 188

\bibitem[{{Kauffmann} \& {White}(1993)}]{Kauffmann:1993}
{Kauffmann} G., {White} S.~D.~M., 1993, \mnras, 261

\bibitem[{{Kravtsov} {et~al}\mbox{.}(2004){Kravtsov}, {Berlind}, {Wechsler},
  {Klypin}, {Gottl{\"o}ber}, {Allgood}, \& {Primack}}]{Kravtsov:2004}
{Kravtsov} A.~V., {Berlind} A.~A., {Wechsler} R.~H., {Klypin} A.~A.,
  {Gottl{\"o}ber} S., {Allgood} B., {Primack} J.~R., 2004, \apj, 609, 35

\bibitem[{{Lacerna} {et~al}\mbox{.}(2018){Lacerna}, {Contreras},
  {Gonz{\'a}lez}, {Padilla}, \& {Gonzalez-Perez}}]{Lacerna:2018}
{Lacerna} I., {Contreras} S., {Gonz{\'a}lez} R.~E., {Padilla} N.,
  {Gonzalez-Perez} V., 2018, \mnras, 475, 1177

\bibitem[{{Lacerna} \& {Padilla}(2011)}]{Lacerna:2011}
{Lacerna} I., {Padilla} N., 2011, \mnras, 412, 1283

\bibitem[{{Lacerna} \& {Padilla}(2012)}]{Lacerna:2012}
{Lacerna} I., {Padilla} N., 2012, \mnras, 426, L26

\bibitem[{{Lacerna}, {Padilla} \& {Stasyszyn}(2014){Lacerna}, {Padilla}, \&
  {Stasyszyn}}]{Lacerna:2014}
{Lacerna} I., {Padilla} N., {Stasyszyn} F., 2014, \mnras, 443, 3107

\bibitem[{{Lacey} \& {Cole}(1993)}]{Lacey:1993}
{Lacey} C., {Cole} S., 1993, \mnras, 262, 627

\bibitem[{{Lacey} {et~al}\mbox{.}(2016){Lacey}, {Baugh}, {Frenk}, {Benson},
  {Bower}, {Cole}, {Gonzalez-Perez}, {Helly}, {Lagos}, \&
  {Mitchell}}]{Lacey:2016}
{Lacey} C.~G. {et~al.}, 2016, \mnras, 462, 3854

\bibitem[{{Lagos}, {Cora} \& {Padilla}(2008){Lagos}, {Cora}, \&
  {Padilla}}]{Lagos:2008}
{Lagos} C.~D.~P., {Cora} S.~A., {Padilla} N.~D., 2008, \mnras, 388, 587

\bibitem[{{Lagos} {et~al}\mbox{.}(2018){Lagos}, {Tobar}, {Robotham},
  {Obreschkow}, {Mitchell}, {Power}, \& {Elahi}}]{Lagos:2018}
{Lagos} C.~d.~P., {Tobar} R.~J., {Robotham} A.~S.~G., {Obreschkow} D.,
  {Mitchell} P.~D., {Power} C., {Elahi} P.~J., 2018, \mnras, 481, 3573

\bibitem[{{Leja} {et~al}\mbox{.}(2013){Leja}, {van Dokkum}, {Momcheva},
  {Brammer}, {Skelton}, {Whitaker}, {Andrews}, {Franx}, {Kriek}, {van der Wel},
  {Bezanson}, {Conroy}, {F{\"o}rster Schreiber}, {Nelson}, \&
  {Patel}}]{Leja:2013}
{Leja} J. {et~al.}, 2013, \apjl, 778, L24

\bibitem[{{Lemson} \& {Kauffmann}(1999)}]{Lemson:1999}
{Lemson} G., {Kauffmann} G., 1999, \mnras, 302, 111

\bibitem[{{Lin} {et~al}\mbox{.}(2016){Lin}, {Mandelbaum}, {Huang}, {Huang},
  {Dalal}, {Diemer}, {Jian}, \& {Kravtsov}}]{Lin:2016}
{Lin} Y.-T., {Mandelbaum} R., {Huang} Y.-H., {Huang} H.-J., {Dalal} N.,
  {Diemer} B., {Jian} H.-Y., {Kravtsov} A., 2016, \apj, 819, 119

\bibitem[{{Manera} {et~al}\mbox{.}(2015){Manera}, {Samushia}, {Tojeiro},
  {Howlett}, {Ross}, {Percival}, {Gil-Mar{\'{\i}}n}, {Brownstein}, {Burden}, \&
  {Montesano}}]{Manera:2015}
{Manera} M. {et~al.}, 2015, \mnras, 447, 437

\bibitem[{{Mao}, {Zentner} \& {Wechsler}(2018){Mao}, {Zentner}, \&
  {Wechsler}}]{Mao:2018}
{Mao} Y.-Y., {Zentner} A.~R., {Wechsler} R.~H., 2018, \mnras, 474, 5143

\bibitem[{{Miyatake} {et~al}\mbox{.}(2016){Miyatake}, {More}, {Takada},
  {Spergel}, {Mandelbaum}, {Rykoff}, \& {Rozo}}]{Miyatake:2016}
{Miyatake} H., {More} S., {Takada} M., {Spergel} D.~N., {Mandelbaum} R.,
  {Rykoff} E.~S., {Rozo} E., 2016, Physical Review Letters, 116, 041301

\bibitem[{{Mundy}, {Conselice} \& {Ownsworth}(2015){Mundy}, {Conselice}, \&
  {Ownsworth}}]{Mundy:2015}
{Mundy} C.~J., {Conselice} C.~J., {Ownsworth} J.~R., 2015, \mnras, 450, 3696

\bibitem[{{Navarro}, {Frenk} \& {White}(1996){Navarro}, {Frenk}, \&
  {White}}]{NFW:1996}
{Navarro} J.~F., {Frenk} C.~S., {White} S.~D.~M., 1996, \apj, 462, 563

\bibitem[{{Orsi} {et~al}\mbox{.}(2008){Orsi}, {Lacey}, {Baugh}, \&
  {Infante}}]{Orsi:2008}
{Orsi} A., {Lacey} C.~G., {Baugh} C.~M., {Infante} L., 2008, \mnras, 391, 1589

\bibitem[{{Padilla} {et~al}\mbox{.}(2010){Padilla}, {Christlein}, {Gawiser},
  {Gonz{\'a}lez}, {Guaita}, \& {Infante}}]{Padilla:2010}
{Padilla} N.~D., {Christlein} D., {Gawiser} E., {Gonz{\'a}lez} R.~E., {Guaita}
  L., {Infante} L., 2010, \mnras, 409, 184

\bibitem[{{Paranjape} \& {Padmanabhan}(2017)}]{Paranjape:2017}
{Paranjape} A., {Padmanabhan} N., 2017, \mnras, 468, 2984

\bibitem[{{Peacock} \& {Smith}(2000)}]{Peacock:2000}
{Peacock} J.~A., {Smith} R.~E., 2000, \mnras, 318, 1144

\bibitem[{{Press} \& {Schechter}(1974)}]{Press:1974}
{Press} W.~H., {Schechter} P., 1974, \apj, 187, 425

\bibitem[{{Saito} {et~al}\mbox{.}(2016){Saito}, {Leauthaud}, {Hearin}, {Bundy},
  {Zentner}, {Behroozi}, {Reid}, {Sinha}, {Coupon}, {Tinker}, {White}, \&
  {Schneider}}]{Saito:2016}
{Saito} S. {et~al.}, 2016, \mnras, 460, 1457

\bibitem[{{Sheth} \& {Tormen}(2004)}]{Sheth:2004}
{Sheth} R.~K., {Tormen} G., 2004, \mnras, 350, 1385

\bibitem[{{Sin}, {Lilly} \& {Henriques}(2017){Sin}, {Lilly}, \&
  {Henriques}}]{Sin:2017}
{Sin} L.~P.~T., {Lilly} S.~J., {Henriques} B.~M.~B., 2017, ArXiv e-prints

\bibitem[{{Springel} {et~al}\mbox{.}(2005){Springel}, {White}, {Jenkins},
  {Frenk}, {Yoshida}, {Gao}, {Navarro}, {Thacker}, {Croton}, {Helly},
  {Peacock}, \& {Cole}}]{Springel:2005}
{Springel} V. {et~al.}, 2005, \nat, 435, 629

\bibitem[{{Springel} {et~al}\mbox{.}(2001){Springel}, {White}, {Tormen}, \&
  {Kauffmann}}]{Springel:2001}
{Springel} V., {White} S. D.~M., {Tormen} G., {Kauffmann} G., 2001, \mnras,
  328, 726

\bibitem[{{Stevens} {et~al}\mbox{.}(2018){Stevens}, {Lagos}, {Obreschkow}, \&
  {Sinha}}]{Stevens:2018}
{Stevens} A.~R.~H., {Lagos} C.~d.~P., {Obreschkow} D., {Sinha} M., 2018,
  \mnras, 481, 5543

\bibitem[{{Tinker}, {Wetzel} \& {Conroy}(2011){Tinker}, {Wetzel}, \&
  {Conroy}}]{Tinker:2011}
{Tinker} J., {Wetzel} A., {Conroy} C., 2011, ArXiv e-prints

\bibitem[{{Tinker} {et~al}\mbox{.}(2008){Tinker}, {Conroy}, {Norberg},
  {Patiri}, {Weinberg}, \& {Warren}}]{Tinker:2008}
{Tinker} J.~L., {Conroy} C., {Norberg} P., {Patiri} S.~G., {Weinberg} D.~H.,
  {Warren} M.~S., 2008, \apj, 686, 53

\bibitem[{{Tinker} {et~al}\mbox{.}(2017){Tinker}, {Hahn}, {Mao}, {Wetzel}, \&
  {Conroy}}]{Tinker:2017a}
{Tinker} J.~L., {Hahn} C., {Mao} Y.-Y., {Wetzel} A.~R., {Conroy} C., 2017,
  ArXiv e-prints

\bibitem[{{Torrey} {et~al}\mbox{.}(2015){Torrey}, {Wellons}, {Machado},
  {Griffen}, {Nelson}, {Rodriguez-Gomez}, {McKinnon}, {Pillepich}, {Ma},
  {Vogelsberger}, {Springel}, \& {Hernquist}}]{Torrey:2015}
{Torrey} P. {et~al.}, 2015, \mnras, 454, 2770

\bibitem[{{Trenti} {et~al}\mbox{.}(2010){Trenti}, {Smith}, {Hallman},
  {Skillman}, \& {Shull}}]{Trenti:2010}
{Trenti} M., {Smith} B.~D., {Hallman} E.~J., {Skillman} S.~W., {Shull} J.~M.,
  2010, \apj, 711, 1198

\bibitem[{{Villarreal} {et~al}\mbox{.}(2017){Villarreal}, {Zentner}, {Mao},
  {Purcell}, {van den Bosch}, {Diemer}, {Lange}, {Wang}, \&
  {Campbell}}]{Villarreal:2017}
{Villarreal} A.~S. {et~al.}, 2017, \mnras, 472, 1088

\bibitem[{{Wang} {et~al}\mbox{.}(2013){Wang}, {Weinmann}, {De Lucia}, \&
  {Yang}}]{Wang:2013b}
{Wang} L., {Weinmann} S.~M., {De Lucia} G., {Yang} X., 2013, \mnras, 433, 515

\bibitem[{{Wechsler} {et~al}\mbox{.}(2006){Wechsler}, {Zentner}, {Bullock},
  {Kravtsov}, \& {Allgood}}]{Wechsler:2006}
{Wechsler} R.~H., {Zentner} A.~R., {Bullock} J.~S., {Kravtsov} A.~V., {Allgood}
  B., 2006, \apj, 652, 71

\bibitem[{{White}(1999)}]{White:1999}
{White} S.~D.~M., 1999, \apss, 267, 355

\bibitem[{{White} \& {Rees}(1978)}]{WhiteRees:1978}
{White} S. D.~M., {Rees} M.~J., 1978, \mnras, 183, 341

\bibitem[{{Xu} \& {Zheng}(2017)}]{Xu:2017}
{Xu} X., {Zheng} Z., 2017, ArXiv e-prints

\bibitem[{{Xu} \& {Zheng}(2018)}]{Xu:2018}
{Xu} X., {Zheng} Z., 2018, \mnras, 479, 1579

\bibitem[{{Yang} {et~al}\mbox{.}(2005){Yang}, {Mo}, {Jing}, \& {van den
  Bosch}}]{Yang:2006}
{Yang} X., {Mo} H.~J., {Jing} Y.~P., {van den Bosch} F.~C., 2005, \mnras, 358,
  217

\bibitem[{{Zehavi} {et~al}\mbox{.}(2018){Zehavi}, {Contreras}, {Padilla},
  {Smith}, {Baugh}, \& {Norberg}}]{Zehavi:2017}
{Zehavi} I., {Contreras} S., {Padilla} N., {Smith} N.~J., {Baugh} C.~M.,
  {Norberg} P., 2018, \apj, 853, 84

\bibitem[{{Zehavi} {et~al}\mbox{.}(2011){Zehavi}, {Zheng}, {Weinberg},
  {Blanton}, {Bahcall}, {Berlind}, {Brinkmann}, {Frieman}, {Gunn}, {Lupton},
  {Nichol}, {Percival}, {Schneider}, {Skibba}, {Strauss}, {Tegmark}, \&
  {York}}]{Zehavi:2011}
{Zehavi} I. {et~al.}, 2011, \apj, 736, 59

\bibitem[{{Zehavi} {et~al}\mbox{.}(2005){Zehavi}, {Zheng}, {Weinberg},
  {Frieman}, {Berlind}, {Blanton}, {Scoccimarro}, {Sheth}, {Strauss}, {Kayo},
  {Suto}, {Fukugita}, {Nakamura}, {Bahcall}, {Brinkmann}, {Gunn}, {Hennessy},
  {Ivezi{\'c}}, {Knapp}, {Loveday}, {Meiksin}, {Schlegel}, {Schneider},
  {Szapudi}, {Tegmark}, {Vogeley}, {York}, \& {SDSS
  Collaboration}}]{Zehavi:2005}
{Zehavi} I. {et~al.}, 2005, \apj, 630, 1

\bibitem[{{Zentner}, {Hearin} \& {van den Bosch}(2014){Zentner}, {Hearin}, \&
  {van den Bosch}}]{Zentner:2014}
{Zentner} A.~R., {Hearin} A.~P., {van den Bosch} F.~C., 2014, \mnras, 443, 3044

\bibitem[{{Zheng} {et~al}\mbox{.}(2005){Zheng}, {Berlind}, {Weinberg},
  {Benson}, {Baugh}, {Cole}, {Dav{\'e}}, {Frenk}, {Katz}, \&
  {Lacey}}]{Zheng:2005}
{Zheng} Z. {et~al.}, 2005, \apj, 633, 791

\bibitem[{{Zheng} \& {Guo}(2016)}]{Zheng:2016}
{Zheng} Z., {Guo} H., 2016, \mnras, 458, 4015

\bibitem[{{Zhu} {et~al}\mbox{.}(2006){Zhu}, {Zheng}, {Lin}, {Jing}, {Kang}, \&
  {Gao}}]{Zhu:2006}
{Zhu} G., {Zheng} Z., {Lin} W.~P., {Jing} Y.~P., {Kang} X., {Gao} L., 2006,
  \apjl, 639, L5

\bibitem[{{Zu} \& {Mandelbaum}(2016)}]{Zu:2016a}
{Zu} Y., {Mandelbaum} R., 2016, \mnras, 457, 4360

\bibitem[{{Zu} \& {Mandelbaum}(2017)}]{Zu:2017}
{Zu} Y., {Mandelbaum} R., 2017, ArXiv e-prints

\bibitem[{{Zu} {et~al}\mbox{.}(2016){Zu}, {Mandelbaum}, {Simet}, {Rozo}, \&
  {Rykoff}}]{Zu:2016b}
{Zu} Y., {Mandelbaum} R., {Simet} M., {Rozo} E., {Rykoff} E.~S., 2016, ArXiv
  e-prints

\bibitem[{{Zu} {et~al}\mbox{.}(2008){Zu}, {Zheng}, {Zhu}, \& {Jing}}]{Zu:2008}
{Zu} Y., {Zheng} Z., {Zhu} G., {Jing} Y.~P., 2008, \apj, 686, 41

\end{thebibliography}

\appendix

\section{HOD catalogues}
To facilitate the creation of mock galaxy catalogues with occupancy variation, 
which can be used for the creation of mocks with galaxy assembly bias, we are 
making public the HODs calculated in this work. The HODs are 
calculated for the following number densities, $n = 0.0316, 0.01, 0.00316, 
0.001, 0.000316$ and $0.0001 \, h^3\, {\rm Mpc}^{-3}$, for galaxies ranked either by stellar mass 
or SFR.  The following redshifts are used, $z=0,0.5,1,1.5,2,2.5$ and $ 3$,  and the haloes are selected in 10 bins 
of ranked age and concentration. This yields more than 1,800 HODs in total. This material can be found at 
\url{https://github.com/hantke/-HOD_Extractor2}

Additionally, we provide in this same repository the HOD fitting parameters for all galaxy 
samples selected by stellar mass, for the commonly used 5-parameter model 
introduced by \citealt{Zheng:2005}. The HOD parameters of the galaxies 
selected by SFR can not be well represented by this standard parametrization 
(see, e.g., \citealt{C13}) and will be investigated further in future work.  
The 5-parameters model is given by
\begin{equation}
 \langle N_{\rm gal}(M_{\rm h})\rangle = \langle N_{\rm cen}(M_{\rm h})\rangle + \langle N_{\rm sat}(M_{\rm h})\rangle,
\label{Eq:Cen_HOD}
\end{equation}
with 
\begin{equation}
 \langle N_{\rm cen}(M_{\rm h})\rangle = \frac{1}{2}\left[ 1 + {\rm erf} \left( \frac{\log M_{\rm h} - \log M_{\rm min}}{\sigma_{\log M}}  \right) \right]
\label{Eq:Cen_HOD2}
\end{equation}
and
\begin{equation}
 \langle N_{\rm sat}(M_{\rm h})\rangle = \left( \frac{M_{\rm h}-M_{\rm cut}}{M^*_1}\right)^\alpha,
\label{Eq:Sat_HOD}
\end{equation}
where $M_{\rm h}$ is the host halo mass and 
$ {\rm erf}(x)$ is the error function,
\begin{equation}
 {\rm erf}(x) = \frac{2}{\sqrt{\pi}} \int_{0}^{x} e^{-t^2} {\rm d}t.
\end{equation}
$M_{\rm min}$ is the mass where, on average, half of the haloes are occupied by 
a central galaxy (i.e., $\langle N_{\rm cen}(M_{\rm min})\rangle = 0.5$);   
$\sigma_{\log M}$ characterises the width  of the transition from zero to one 
central galaxy per halo, where $\sigma_{\log M}=0$  represents a vertical 
step-function transition;  $\alpha$ is the slope of the satellite HOD
power-law; $M_{\rm cut}$ is the minimum halo mass at which haloes can host a 
satellite galaxy and $M^*_1$ is the normalization. Note that we provide instead
the value of a related parameter, $M_{1}$, the halo mass where on average there
is one satellite per halo (i.e., $\langle N_{\rm sat}(M_{1})\rangle = 1$) and is 
equal to $ M_{\rm cut} + M^*_1$. 

As an example, we provide in Table~\ref{Table:HOD} the HOD parameters for 
the galaxy samples with $n=0.01\ h^3\, {\rm Mpc}^{-3}$ at $z=0,1,2,3$ and for 
the $10$ per cent oldest haloes, the $10$ per cent youngest haloes, and the full halo sample.
The full set of fitted parameters is provided in our public release website.
Together with the HODs and their parameters (stored in two {\tt HDF5} files),
we are also releasing tools to read and plot the HODs to facilitate their analysis.

\begin{table}
 \begin{tabular}{c c c c c c } 
  \hline
 & & $z=0$ & & & \\
 \hline
   & $M_{\rm min}$ & $\sigma_{\log M}$ & $\alpha$ & $M_{\rm cut}$ & $M_{1}$ \\ 
  10$ \% $ oldest haloes & 11.44 & 0.10 & 1.15 & 11.99 & 12.82\\
  All haloes                   & 11.62 & 0.21 & 0.99 & 11.83 & 12.57\\
  10$ \% $ youngest haloes & 11.92 & 0.30 & 0.84 & 11.74 & 12.31\\
 \hline
  & & $z=1$ & & & \\
 \hline
   & $M_{\rm min}$ & $\sigma_{\log M}$ & $\alpha$ & $M_{\rm cut}$ & $M_{1}$ \\ 
  10$ \% $ oldest haloes & 11.36 & 0.16 & 1.07 & 11.97 & 12.57\\
  All haloes                   & 11.59 & 0.28 & 0.92 & 11.81 & 12.34\\
  10$ \% $ youngest haloes & 11.94 & 0.37 & 0.85 & 11.70 & 12.18 \\
 \hline
  & & $z=2$ & & & \\
 \hline
   & $M_{\rm min}$ & $\sigma_{\log M}$ & $\alpha$ & $M_{\rm cut}$ & $M_{1}$ \\ 
  10$ \% $ oldest haloes  & 11.15 & 0.11 & 1.08 & 11.66 & 12.33\\
  All haloes                    & 11.41 & 0.30 & 0.92 & 11.60 & 12.13\\
  10$ \% $ youngest haloes   & 11.74 & 0.34 & 0.86 & 11.46 & 11.97\\
  \hline
  & & $z=3$ & & & \\
 \hline
   & $M_{\rm min}$ & $\sigma_{\log M}$ & $\alpha$ & $M_{\rm cut}$ & $M_{1}$ \\ 
  10$ \% $ oldest haloes & 10.89 & 0.13 & 1.01 & 11.47 & 12.07 \\
  All haloes                   & 11.17 & 0.32 & 0.93 & 11.34 & 11.91 \\
  10$ \% $ youngest haloes  & 11.51 & 0.33 & 0.85 & 11.24 & 11.70 \\
  \hline
  \end{tabular}
\caption{The HOD parameters described in Eqns.~A1 to A4 for galaxy samples 
corresponding to a number density of $n=0.01\ h^{3}\, {\rm Mpc}^{-3}$ selected 
by stellar mass. From top to bottom we show the parameters for 
$z = 0,\ 1,\ 2\ \&\ 3$, for the $10$ per cent oldest haloes, for the full halo sample 
and for the $10$ per cent youngest haloes.} 
\label{Table:HOD}
\end{table}

\label{lastpage}
\end{document}